\documentclass[multphys,vecphys]{svmult}

\usepackage{epsfig}
\usepackage{makeidx}         
\usepackage{multicol}        
\usepackage[bottom]{footmisc}
\usepackage{amsmath}
\usepackage{bm}

\makeindex             

\newcommand{\rem}[1]{}

\newcommand{\up}{\uparrow}
\newcommand{\down}{\downarrow}

\newcommand{\Id}[1] {\int \! \! d^3 #1}

\newcommand{\vr} {{\bf r}}

\begin{document}

\setcounter{chapter}{5}

\title{Theoretical aspects of highly correlated fullerides:
        metal-insulator transition}
\titlerunning{Theory of MI transition}

\author{Nicola Manini\inst{1,2}
\and
Erio Tosatti\inst{2,3,4}}
\institute{Dipartimento di Fisica, Universit\`a di Milano,\\
Via Celoria 16, 20133 Milano, Italy\\
\and International School for Advanced Studies (SISSA),\\
Via Beirut 4, 34014 Trieste, Italy
\and INFM Democritos National Simulation Center,\\
Via Beirut 4, 34014 Trieste, Italy
\and International Centre for Theoretical Physics (ICTP),\\
P.O. Box 586, I-34014 Trieste, Italy
}
%
%

\maketitle

\abstract{
We review some theoretical aspects connected with the
interplay of strong electron correlations and vibron phenomena in negative
fullerene ions and in 
solid 
fulleride conductors, superconductors and
insulators.
The first part covers molecular ions, their intra-molecular Jahn-Teller
effect, Coulomb (Hund's rule) exchange interactions, molecular vibrons and
multiplet splittings.
The second part addresses electron propagation in molecular fullerides,
with special emphasis given to trivalent cases such as K$_3$C$_{60}$ and
NH$_3$\,K$_3$C$_{60}$, where metallic, superconducting and Mott insulating
phases are at play.
Dynamical mean field theory approaches to a simplified Hamiltonian for this
system are discussed in the light of some of the observed phenomenology.
It is argued in particular that NH$_3$\,K$_3$C$_{60}$ is a Mott-Jahn-Teller
insulator, which under pressure turns into a strongly correlated
superconductor, thus sharing some important elements with the high-$T_c$
cuprates.
}

\section{Introduction}
\label{sec:1}

\rem{
moderately-high-$T_c$ superconductivity
in metallo-fullerene compounds\cite{Hebard,Rosseinsky,Holczer,Tanigaki}.

superconductors, research on fullerene-based compounds in the last few
years demonstrated many unexpected features, such as organic ferromagnetism
in TDAE C$_{60}$ \cite{Allemand}, insulating behavior of ''open-shell''
compounds A$_4$C$_{60}$ \cite{Kiefl}, an orientational transition between a
high-temperature free rotating state and a low-temperature disordered
frozen state in pure \cite{David} and doped \cite{Stephens} C$_{60}$, and
polymerization \cite{Koller} in RbC$_{60}$.

\cite{Goldoni01}
} 

The wealth of experimental data on the alkali fullerides collected through
the 1990s and early 2000s has raised several exciting theoretical issues
which, in turn, have attracted 
a large
research effort.
Different phases are realized when temperature, alkali concentration,
lattice parameter and lattice structure are varied.
Metallic, insulating, and superconducting phases are obtained in solid
compounds characterized by the partial occupancy -- by an average $n$
electrons per C$_{60}$ with $n$ ranging between 0 and 6 -- of the narrow
electronic band originating from the $t_{1u}$ LUMO molecular orbital.
When this band is completely empty (pure C$_{60}$) or completely filled
(K$_6$C$_{60}$) the solid is a band insulator (although the lattice
structure in the two solids are different).
Band-structure calculations of compounds characterized by an incomplete
filling of this band consistently predict metallicity
\cite{Erwin,Satpathy92,Erwin94}.
However, experimentally some of these compounds (e.g.\ K$_4$C$_{60}$,
Rb$_4$C$_{60}$, NH$_3$\,K$_3$C$_{60}$) are found 
to be
insulators
\cite{Benning,Martin,Rosseinsky93,Iwasa95,Takenobu00,Obu00,Tou2}.
Among the metallic compounds, many with $n = 3$ turn superconducting at low
temperature, with transition temperatures as high as $30-40$~K, while several
others remain metallic at all experimentally investigated temperatures.

For even $n$, as in K$_4$C$_{60}$, electron-electron correlations and the JT
coupling stabilize a {\em correlated insulating state} of a lattice of
evenly-charged C$_{60}$ molecules. In this type of insulator, fluctuations
about $\langle n_i\rangle =n$ are suppressed, and a gap 
opens in the electronic spectrum.
This state is non-magnetic, very much like a regular band insulator.
However, the mostly intra-molecular electron correlations responsible for
band narrowing and gap opening are largely {\em coulombic} and {\em
vibronic} in origin.
We have suggested that the (body-centered tetragonal \cite{Fleming})
structure of K$_4$C$_{60}$ and Rb$_4$C$_{60}$ may be a realization of this
state where electronic and vibronic interactions play an important role.
Similarly, the insulating state observed in strongly correlated
NH$_3$\,K$_3$C$_{60}$ 
(and (NH$_3$)$_6$\,Li$_3$C$_{60}$ \cite{Durand03}) as well as 
the insulator-superconductor transition obtained
under pressure are likely to have the same origin.

A satisfactory understanding of how similar physical parameters lead to
very different ground states (insulat\-or/metal/sup\-erconductor) has long
eluded the research community.
A novel picture relating the transitions between metallic, insulating, and
superconducting phases to the interplay of strong electron-electron
correlation with electron-phonon coupling in the LUMO band has emerged in
recent years.
The scenario is now in our view better understood, even though 
many
quantitative details 
still escape the full grasp of theory.
The present Chapter reviews and illustrates this picture.

\section{C$_{60}$ molecules and molecular ions}
\label{single:sec}

We review for a start some basic aspects of the physics of C$_{60}$,
concerning electron-electron correlation and electron-vibration coupling in
C$_{60}$ molecular ions and ionic fullerides. For more complete reviews
we suggest consulting Refs.\
\cite{Ramirez,Gelfand,Reed00review,GunnarssonBook}.

In C$_{60}$ all the 60 carbon atoms of fullerene are arranged as a regular,
icosahedral, roughly spherical cage, of $\sim 0.7$~nm in diameter
\cite{Schoenherr}. This cluster may be thought as a piece of graphene sheet,
wrapped up to a spherical shape.
The regular hexagonal structure of graphite is distorted, with 12
five-membered rings intercalating 20 six-membered ones, thus introducing
the Eulerian $4\pi$ solid angle necessary to yield a closed surface. The
order is such that all carbon atoms remain equivalent, each sitting at the
corner of one 5-member ring and two 6-member ones.
30 chemical bonds, the so-called double bonds, are shared by two hexagonal
rings only; the remaining 60 bonds,
shared by pentagonal and hexagonal rings, are
about 5\% longer \cite{David}.
The $\sigma $-bonding sp$^2$ graphite orbitals constitute the backbone of the
molecule.
Spherical curvature alters their character to an average
''sp$^{2.28}$''\cite{Haddon}.  In energy, the $\sigma $-bonding orbitals
range from several eV to a few tens of eV below the vacuum, the antibonding
states lying $+10$~eV and higher above vacuum zero \cite{Satpathy}.
The chemically active electronic states are those derived from the
''p$_z$'' carbon orbitals (actually of hybrid ''s$^{0.09}$p'' nature
\cite{Haddon}) that are directed radially, supporting a half filled
$\pi$-electron system.

Although theoretically metastable compared to graphite and diamond,
C$_{60}$ is nevertheless an empirically very stable and long lived
allothropic form of carbon both as an isolated molecule and as an fcc, or
Pa3, solid.
The reason for stability is the substantial amount of electronic energy
gained in the delocalized MO's, making in some sense C$_{60}$ the spherical
counterpart of the aromatic benzene ring.
The molecular orbitals at the origin of the conduction bands of
solid fullerides are all of $\pi$-bonding/antibonding nature.

The overall molecular symmetry group is the icosahedral group ${\cal I}_h$,
the largest point group in 3D (except for axial groups).
\rem{
, composed by 120
elements. ${\cal I}_h$ is the product group of ${\cal I}\times i$, where
$i$ is the group composed by the identity and space inversion, and the 60
operations in ${\cal I}$ are organized in the following
classes \cite{Wilson55}: the identity, 12$C_5$, 12$C_5^2$, 20$C_3$, and
15$C_2$. The great richness of the group is directly related to the large
degree of symmetry carried by this unique molecule, appearing in particular
in the complete equivalence of all its 60 atoms.
}
Symmetry implies a large degeneracy of the group's irreducible
representations \cite{Wilson55}: $A_{g/u}$ (1-dimensional), $T_{1\,g/u}$,
$T_{2\,g/u}$ (3-dimensional), $G_{g/u}$ (4-dimensional), and $H_{g/u}$
(5-dimensional).
Large degeneracies are accordingly very common to all electronic,
vibrational,vibronic molecular states.
This makes C$_{60}$ a rich playground of novel vibronic structures, where
Berry phases and entanglement plays a fascinating role
\cite{AMT,MTA,AssaPRL,Delos96,Paris97,ManiniAErice,noberry,hbyh,Moate96,Moate97,Manini05}.

Icosahedral symmetry is of course restricted to the ideal situation of a
molecule or molecular ion in vacuum. In compounds and/or in the solid state
it will be affected by crystal fields of lower symmetry.
Even in vacuum, isotope substitution of one carbon atom is enough to reduce
${\cal I}_h$ to simple bilateral reflection $C_v$.
Due to 1.10\% isotope abundance of $^{13}$C in natural carbon, only about
50\% of C$_{60}$ is pure $^{12}$C$_{60}$.  However, isotope substitution
induces only small splittings (about 1\%) of the vibron frequencies, and
can be safely neglected for many purposes.
For a discussion of the intricacies of isotope shifts in solid-state
properties of the fullerides, see Ref.~\cite{GunnarssonBook}.

As mentioned, immersion of the icosahedral molecule in a solid-state
environment reduces its symmetry to that of the local crystal field. For
example, solid C$_{60}$ has electronic bands and optical phonons compatible
with the local cubic field induced by the fcc lattice.
In most (but not all) solid state compounds, intermolecular interactions
are relatively loose, each molecule retaining its close structure.
Accordingly, splittings of the molecular vibrations (now optical phonons in 
the solid) due to reduced symmetry are small and hardly observed at all \cite{Bethune91}.

\subsection{Molecular electronic states}
\label{ElecStruc:GenIntro}

Electrons in the $\pi$ orbitals of C$_{60}$ represent the chemically
relevant region of the spectrum.
These orbitals provide the basic one-electron picture, neglecting first e-e
correlation effects, which we shall address in Sec.~\ref{coulomb:sec}.

Many approaches have been taken to the electronic structure of C$_{60}$,
from simple Huckel tight-binding with one orbital per atom (already
yielding the correct order of molecular energies and gaps)
\cite{Satpathy,Haymet,Haddon}, to more extended bases \cite{Satpathy,Negri}
to microscopic DFT-LDA (density functional theory in the local density
approximation) calculations on localized \cite{Satpathy,Green96} and
extended bases \cite{Erwin,Andreoni}.

\begin{figure}[t]
\centering
\epsfig{file=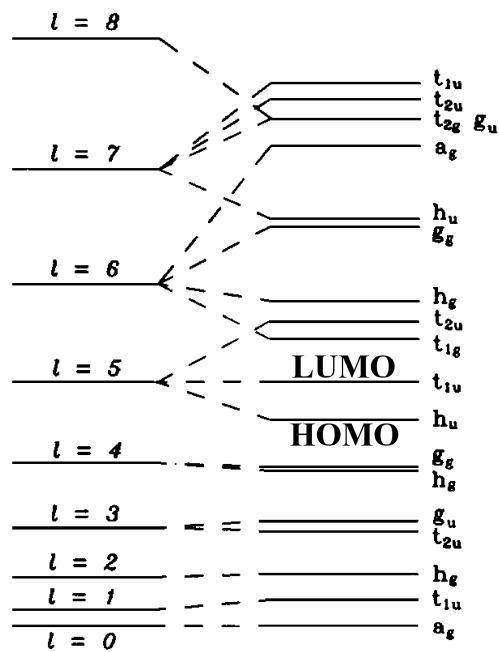,width=6.5cm}
\caption{\label{C60levels:fig}
A particle-on-a-sphere schematic representation of the electronic levels of
C$_{60}$. The HOMO and LUMO levels, originating from the $L=5$ orbital, are
indicated.  (Adapted from Ref.\ \cite{Savina}.)
}
\end{figure}

These approaches provide an increasing degree of quantitative accuracy in
the description of the electronic spectrum.
Simple intuitive models, such as the particle-on-a-sphere model
\cite{Savina}, are often sufficient for a qualitative and synthetic
understanding of the electronic structure.
In this approximation the idea is treating the $\pi$ electrons as
though moving 
on a spherical shell of radius $R$, mimicking
the attractive potential generated by the carbon ions.  The precise
localization of these ions produces then a weak icosahedral perturbation.
As long as this icosahedral perturbation can be neglected, the
single-electron angular wavefunctions are atomic-like spherical harmonics
$Y_{LM}$, with energy
\begin{equation}
\label{SphericalE}E(L)=\frac{L(L+1)\hbar ^2}{2m_eR^2}\ ,
\end{equation}
where $m_e$ is the electron mass. 50 out of 60 p$_z$ electrons of the
neutral molecule fill completely the MO up to $L=4$. The lowest $L=0,1,2$
orbitals coincide with icosahedral states labelled $a_g$, $t_{1u}$, $h_g$
respectively.
All higher $L$ values are split into icosahedral representations by
the icosahedral field generated by the cage.
After filling all states including $L= 4$, 
10 electrons are left in the $L=5$ shell, which is therefore only partly
filled.
As it happens, the icosahedral splitting ($L=5\to h_u+t_{1u}+t_{2u}$) of
this 11-fold degenerate orbital generates a closed-shell configuration, as
shown in Fig.~\ref{C60levels:fig}.
In accord with microscopic calculations and with experiment, the
completely-filled HOMO has $h_u$ symmetry, and the LUMO is $t_{1u}$.
The HOMO-LUMO gap is therefore caused by the icosahedral perturbation in
the $L=5$ shell, and is experimentally $\sim 1$~eV for molecules in vacuum
\cite{Tomita05}.
%
%
A $t_{2g}$ LUMO+1 state, originated from the $L=6$ shell, is found
approximately $1$~eV above the $t_{1u}$ LUMO.
\rem{
As Fig.\ \ref{C60levels:fig} shows, the icosahedral splitting is most
effective on the largely-degenerate states with largest $L$, due to the
greater number of nodes in the wavefunction.
This model, although approximate and qualitative, agrees in the
main qualitative features of the experiment, and provides easily the
spherical parentage of each orbital.
}

The electron affinity of C$_{60}$ is large (2.69~eV) \cite{Yeretzian,Wang99}
and experimental evidence has been found that C$_{60}^{-}$ \cite{Yang} and
even C$_{60}^{2-}$ \cite{Limbach} are stable ions in vacuum.
In solution, a wider spectrum of ionization states has been demonstrated
electrochemically, up to C$_{60}^{6-}$
\cite{Dubois,Heath,Fullagar,Baumgarten93}.
As an adsorbate on a metal surface, the electronegative C$_{60}$ molecule
naturally picks up electrons \cite{Erwin,Burstein}, and evidence has been
provided of charge transfer as large as $n$=6 \cite{Modesti}.
In the solid state, compounds have been synthesized, covering a wide range
of charge transfers, from $n$=1, as in TDAE-C$_{60}$
\cite{Allemand,Denisov} or Rb$_1$C$_{60}$ \cite{Benning}, $n$=3, as in
K$_3$C$_{60}$ or Rb$_3$C$_{60}$ \cite{Hebard}, $n$=4 as
in K$_4$C$_{60}$ \cite{Kiefl}, $n$=6 as in Rb$_6$C$_{60}$ \cite{Fisher}, or
even higher as in Li$_{12}$C$_{60}$ \cite{Chabre,koh}.
More recently, also positive C$_{60}$ ions were produced in gas phase
\cite{Tomita01}, in liquid solution \cite{Reed00,Bruno03}, in Ar matrix
\cite{Gasyna}, and in the solid state
\cite{Datars95,Datars96,Panich02,Panich03}.
Since the LUMO can hold up to 6 electrons, the negatively charged ions
C$_{60}^{n-}$ with $1\leq n\leq 5$ are open shells.
Likewise, the cations C$_{60}^{n+}$ with $1\leq n\leq 9$ are also open
shells due to the fivefold degeneracy of the HOMO orbital.

\subsection{Molecular vibrations}

\begin{figure}[t]
\centering
\epsfig{file=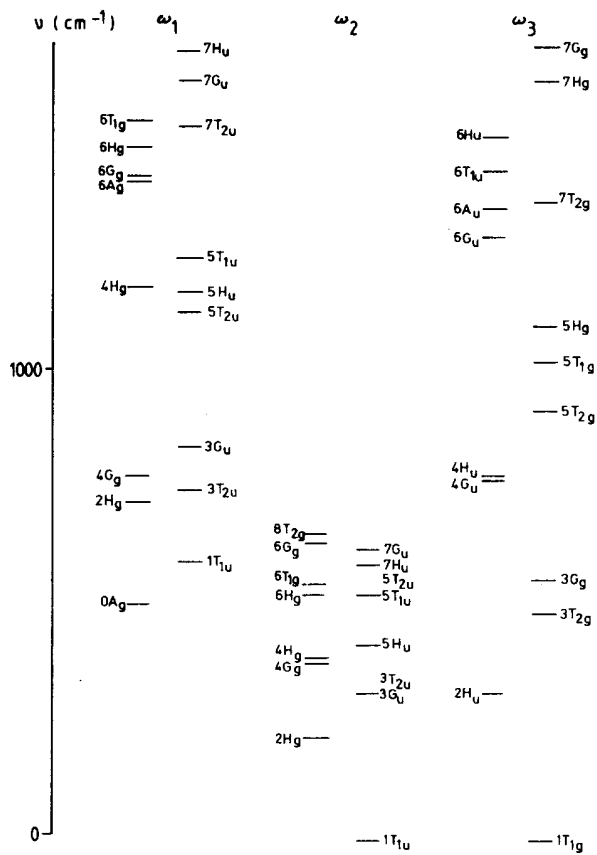,width=8cm}
\caption{\label{C60modes:fig}
Spherical resolution of the vibrational spectrum of C$_{60}$. The
modes are organized according to the three series, indicated with
$\omega_{1/2/3}$. The spherical parent $L$ is indicated before the
icosahedral label.  (From Ref.~\cite{CeulemansIII}).
}
\end{figure}

C$_{60}$ has 174 vibrational degrees of freedom, but thanks to
symmetry-induced degeneracy and selection rules, the vibrational spectra
show relatively few peaks, with clear identification.
In particular, only the 4 $T_{1u}$ dipolar modes are 
infrared active, and
only the 2 $A_g$ and 8 $H_g$ modes are Raman active.
Neutron experiments \cite{Prassides} are sensitive to all modes, including
the silent ones, but with low resolution.
Calculations are therefore crucial, in order to get a global picture of the
vibrational spectrum \cite{CeulemansIII}.

Like for the electronic states, many routes to the calculation of
vibrational eigenfrequencies and normal modes have been pursued, both based
on force field fits to the experimentally accessible data
\cite{Prassides,Zhou,Onida}, or based on {\it ab initio} calculations of
the molecular structure \cite{Negri,Andreoni,Jiang,Manini01}.
The agreement among different calculations is somewhat worse than for the
electronic levels: typical discrepancies on the order of several meV are
well above nowadays' experimental resolution.

A qualitative understanding of the vibrational structure of C$_{60}$ is
provided by the the analogy of the C$_{60}$ cage with a hollow elastic
sphere proposed in Ref.\ \cite{CeulemansIII} and illustrated in
Fig.~\ref{C60modes:fig}.
The eigenmodes of a homogeneous spherical membrane with a
stretching and a bending rigidity are collected in three classes: mainly
radial, mainly tangential and purely tangential.
The first class contains levels of parity $(-1)^L$, with $L=0,1,2,3,...$;
the levels in the second class have the same parity, but start off at
$L=1$; in the third series the parity is reversed $(-1)^{L+1}$, and they
start at $L=1 $.
The modes of rigid translation and rotation of the sphere are identified in
the $L=1$ states at zero frequency in series 2 and 3.
Of course, when the homogeneous sphere is replaced by the discrete 60-atoms
molecule, the infinite set of spherically symmetric eigenmodes goes into a
finite number of modes, now labeled by ${\cal I}_h$
representations. $L=0,1,2$ states have icosahedral counterparts in
$A_{g/u}$, $T_{1u/g}$, $H_{g/u}$ respectively: the first {\it
gerade/ungerade} label corresponds to series 1 and 2, the second labels the
third series, which has inverted parity. States with $L>2$ are split,
according to the rules given in Table~III of Ref.\ \cite{CeulemansIII};
for example a $L=3$ vibration becomes $T_{2u}\oplus G_u$, if its parity is
odd.
The explicit eigenmodes, computed for example by force-field methods, can
be easily analyzed in terms of the spherical basis, to obtain their
parentage in terms of hollow sphere modes \cite{CeulemansIII}: the three
series are readily identified in Fig.~\ref{C60modes:fig}.
Inter-mode mixing is present, but most vibrations hold a well-defined
spherical parentage,
so that the spherical picture remains generally valid and useful.

\subsection{Jahn-Teller coupling between electrons and vibrations: 
molecular vibronic states and energetics}
\label{vibronic:sec}

The electronic state of most C$_{60}^{n\pm}$ ions is orbitally degenerate.
According to the Jahn-Teller (JT) theorem, the highly symmetric ${\cal I}_h$
geometry is unstable toward symmetry reducing molecular distortions.
Theoretically, such distortion can be evaluated by direct calculation of
the relaxed geometry of the molecular ions, and are very well defined.
However, direct experimental evidence of the JT distortions is rather poor,
and limited to few compounds, mainly ``discrete'' salts, as discussed in
detail in Ref.~\cite{Reed00review}.
The reason for this is twofold: (i) the atomic displacents from the ${\cal
I}_h$ positions of neutral C$_{60}$ are very small -- of the order of few
pm, and (ii) quantum tunneling and/or thermal hopping of the molecule between equivalent JT
distortions generally restores, on average, the original ${\cal I}_h$ symmetry.

Despite the small absolute atomic displacements, the JT physics involves
serious energetics, since the phonon frequencies are high (owing to the large
stiffness of the fullerene cage and the light carbon mass) and the JT couplings
are large.
At zero temperature, a significant fraction of the JT energy is associated 
to nonadiabatic effects related to the quantum motion of carbon nuclei, and 
cannot be obtained by standard adiabatic calculations based on {\em classical} atomic positions.
However, quantitative insight in the coupled electron-phonon dynamics is
obtained by means of the study of formally simple models.
To the extent that these distortions are not too large, the distortion
mechanism is well described within a model where the distortion is expanded
on the basis of the normal modes of vibration: these are treated as
harmonic oscillators, and the coupling between the degenerate electron
level and the JT-active distortions can be assumed to be linear in the
phonon coordinate.
These simplifying approximations remain valid only as long as the
distortions are relatively small, as in C$_{60}$ ions, and could not be
applied to ionic states of e.g.\ SiH$_4$, whose molecular shape is
radically modified upon charging \cite{Balamurugan04}.

The basic model Hamiltonian for the JT dynamics of an icosahedral
molecule has the following standard structure \cite{ChoB97}:
\begin{equation}
\label{Hschemat}\hat{H}=\hat{H}_0+\hat{H}_{\rm e-v} \,,
\end{equation}
where $\hat{H}_0$ describes the free (uncoupled) electrons and vibrations and
$\hat{H}_{\rm e-v}$ the linear JT interaction.
To begin with, consider a single vibrational mode of energy $\hbar \omega$
and symmetry label $\Lambda$: the Hamiltonian are written
\begin{eqnarray}
\label{H0:eq}
\hat{H}_0 & = & \hbar \omega \sum_{\mu \in \Lambda}
\left(\hat{b}_\mu^{\dagger }\hat{b}_\mu+\frac 12\right)
+(\epsilon -\mu) \sum_{m \in \lambda}\sum_{\sigma =\up,\down }
\hat{c}_{m,\sigma}^{\dagger }\hat{c}_{m,\sigma }\,, \\
\label{Hev:eq}
\hat{H}_{\rm e-v} & = & g\hbar \omega \frac{\sqrt{3}}2
\sum_{m_1,m_2\in \lambda\ }
\sum_{\mu\in \Lambda\, \sigma}
C^{\Lambda,\mu}_{\lambda m_1\, \lambda m_2} 
\left( \hat{b}_{\mu}^{\dagger}+ \hat{b}_{\mu}\right)
\hat{c}_{m_1\,\sigma }^\dagger\,\hat{c}_{-m_2\,\sigma}\,.
\end{eqnarray}
Here, the harmonic oscillator is represented by the boson operators
$b_\mu^{\dagger}$: the indexes $\mu$ span the degenerate representation
$\Lambda$.
The distortion coordinate appears in the coupling term (\ref{Hev:eq}) as
$\hat{Q}_\mu= 2^{-1/2}\left( \hat{b}_{\mu}^{\dagger}+
\hat{b}_{\mu}\right)$.\ \
$\lambda=t_{1u}$ is the symmetry label of the LUMO electronic state, span
by indexes $m_i$ and generated by the fermion operators
$\hat{c}_{m\,\sigma}^{\dagger}$.\ \ 
$g$ is the dimensionless parameter measuring the electron-vibration
coupling strength.
$C^{\Lambda,\mu}_{\lambda m_1\, \lambda m_2}$ is a Clebsch-Gordan
coefficient for the icosahedral-group \cite{butler81}, which recouples the
fermion tensors of $\lambda$ symmetry with the boson tensor to an
icosahedral scalar.
In practice, this recoupling is possible only for modes of symmetry
$\Lambda=A_g,\ H_g$.  All other $\Lambda$'s yield vanishing Clebsch-Gordan
coefficients, which means that all modes of non-$A_g\, /\,H_g$ symmetry
are not linearly coupled to the $t_{1u}$ LUMO.

It is interesting to note that the ${\cal I}_h$ Clebsch-Gordan coefficients
for the coupling of two $t_{1u}$ representations to $\Lambda=A_g / H_g$
happen to coincide numerically with the corresponding $C^{L,\mu}_{l m_1\, l
m_2}$ Clebsch-Gordan coefficients for the full rotations group SO(3).
The $t_{1u}$ label maps to angular momentum $l=1$, while $A_g$ maps to
$L=0$, and $H_g$ maps to $L=2$, in accord with the spherical picture of
Figs.\ \ref{C60levels:fig} and \ref{C60modes:fig}.
This means physically that linear JT coupling of the $t_{1u}$ orbital does
not distinguish between a soccer-ball molecule and a perfectly spherically
shaped one: the $t_{1u}$ level behaves as an atomic-like p state perturbed
by $A_g / H_g$ distortions of monopolar and quadrupolar nature
\cite{ChoB97,ob69,ob71}.
The implications of this extra symmetry are striking: Hamiltonian
(\ref{Hschemat}) is actually a scalar with respect to the full rotation
group SO(3), its eigenstates being representations of this larger group.
SO(3) states are to be expected in the outcoming spectrum, which means that
vibronic states belonging to different ${\cal I}_h$ representations
collapse to degenerate SO(3) multiplets (labeled by some angular momentum
$L$, rather than by ${\cal I}_h$ labels), as on the left side of
Fig.~\ref{C60levels:fig}.
In actual C$_{60}$, the linear electron-vibration coupling term is just
the leading term in an infinite expansion, where only higher-order
terms introduce
the actual icosahedral symmetry, thus splitting the highly degenerate
spherical vibronic states not unlike Fig.\ \ref{C60levels:fig}.
Eventually, the effect of these higher-order terms is small, yielding
very small splittings of the vibronic states: neglect of all these
higher-order effects is a good approximation for C$_{60}^{n-}$.

Electron coupling to the $\Lambda=A_g$ mode represents a trivial shifted
oscillator, since the nondegenerate mode does not split the electronic
degeneracy, and the distortion is purely proportional to the total charge
in the coupled level.  Henceforth we shall mostly concentrate on coupling
to degenerate $\Lambda=H_g$ modes.

There is no loss of generality in choosing the energy reference so that
$\epsilon = \mu$ in Eq.~(\ref{H0:eq}). We are then left with a two-parameters
Hamiltonian operator. The value $\hbar\omega$ of the frequency of the
harmonic oscillator sets the energy scale.
$\hat{H}_0$ and $\hat{H}_{\rm e-v}$ are written with a common factor
$\hbar\omega$, which sets the energy scale of the model.
The dimensionless parameter $g$ tunes the intensity of the JT coupling,
thus of the tendency of the system to distort.
This is seen as follows: express each pair of vibrational operators
$(\hat{b}_{\mu}^{\dagger} , \ \hat{b}_{\mu})$ as a dimensionless coordinate
$\hat{Q}_\mu$ and conjugate momentum $\hat{P}_\mu$.
In this notation, the adiabatic potential is obtained by ignoring the
$\frac 12 \sum \hat{P}_\mu^2 $ phonon kinetic term in $\hat{H}_0$ and
treating the $\hat{Q}_\mu$ operators as classical variables $Q_\mu$
\cite{Bersuker}: one readily finds that $Q_0= g$ (all other $Q_\mu=0$) is a
minimum of the sum of the competing $Q_\mu$-linear term ($\hat{H}_{\rm
e-v}$) and $Q_\mu$-quadratic term in $\hat{H}_0$.
We see here that $g$ measures the amount of distortion (in dimensionless
oscillator coordinates) along the normal mode.
For small $g\ll 1$ (weak-coupling regime) distorsions are small.
In this regime quantum fluctuations dominate: the correlated vibrational and
electronic dynamics involves all electronic states in the multiplet at the
same time, in a profoundly nonadiabatic fashion.
Despite this intricacy, as a small parameter can be identified, the JT
problem can be dealt with by treating $\hat{H}_{\rm e-v}$ by perturbation
theory.
For intermediate $g\simeq 1$, the distortion is sizeable, but quantum
kinetic energy has a relevant role in promoting tunneling among equivalent
minima.  The vibronic spectrum in this region shows nontrivial structures.
Eventually, when the coupling becomes very large ($g\gg 1$), the JT
distortion is so large that individual JT minima remain well separate so
that tunneling is efficiently suppressed: the system freezes in one of the
equivalent local minima.

Here we should note a peculiarity of the so-called linear $t\otimes H$ JT
problem Eq.~(\ref{H0:eq})-(\ref{Hev:eq}) at hand, and precisely that the
minimization of the lowest adiabatic potential sheet leads to a continuous
manifold of equivalent minima (a {\em trough}) rather than isolated minima
\cite{AMT,MTA,ChoB97}.
This is a consequence of the extra spherical symmetry of this special
icosahedral linear JT system.
However, at strong coupling higher-than-linear terms, neglected in
Eq.~(\ref{Hev:eq}) become relevant. They produce a warping of the JT
trough, leading to isolated minima.
The molecule distorts to a minimum in a set of static JT reduced-symmetry
configurations separated by saddle points of the adiabatic potential
surface.
For the intermediate coupling case of C$_{60}^-$, detailed Hartree-Fock
calculations have shown, for example, that the total energy lowering in
going from the ${\cal I}_h$-symmetric molecular configuration to static JT
distortions of $D_{5d}$, $D_{3d}$, $D_{2h}$ symmetries, are in fact
identical to within 1\% \cite{Koga}.
Therefore, the linear JT coupling approximation Eq.~(\ref{Hev:eq}) is very
well justified for C$_{60}$ ions.


\begin{figure}[t]
\centering
\epsfig{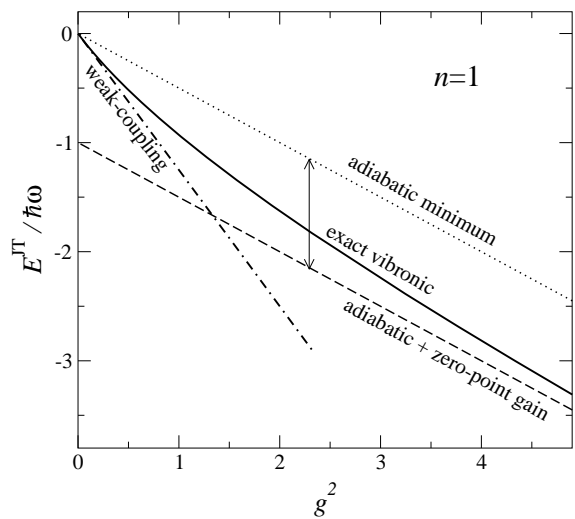}
\caption{\label{Softening:fig}
Ground-state energy lowering for a $t_{1u}$ electron coupled to an $H_g$
mode as a function of coupling.
At weak coupling (dot-dashed), the ground-state energy (solid) drops $2.5$
times more rapidly than the adiabatic minimum lowering (dotted) in order to
reach the correct strong-coupling semiclassical limit (dashed), which takes
into account the softening of two (for $n=1$) harmonic modes for a total
zero-point gain $\hbar\omega$ (indicated by a vertical arrow).
(Adapted from Ref.~\cite{ManiniPhD}.)
}
\end{figure}

Analytical expression for the $t\otimes H$ JT energies can be obtained in
the simple weak-coupling limit, by means of perturbation theory \cite{MTA}.
The main outcome of the weak-coupling regime is that the ground-state
energy gain $E^{\rm e-v}$ in this JT system is $5/2$ times larger in
perturbation theory than the simple adiabatic lowering \cite{Yabana}.
This point is illustrated in Fig.~\ref{Softening:fig} for $n=1$.
For a simple nondegenerate non-JT coupling, such as that to the $A_g$ mode,
the two quantities would instead coincide.
The physical origin of this enhancement is a substantial reduction in
zero-point energy.
The ground-state energy drops faster at small $g$ because the system
transforms rapidly from 5 harmonic oscillators (zero-point energy $=\frac
52 \hbar\omega$) at $g=0$ to a Mexican-hat potential well, which is more
``square-well''-like than the original harmonic potential, and, in
particular, carries several soft (``pseudorotational'') modes along the
trough.
Perturbative expressions for the ground-state energy lowering at weak
coupling for $t_{1u}^n \otimes H$ in all possible spin states are collected
in Table~\ref{JTen:table}.
The perturbative expressions are essentially exact for $g\leq0.3$.
For larger (intermediate) coupling, the actual JT energy lowering is
located between the perturbative ``antiadiabatic'' estimate (which
systematically overestimates the JT energy lowering) and the adiabatic
energy lowering (which consistently underestimates the energy lowering).

\begin{table}[t]
\centering
\caption{
Perturbative (weak coupling, to order $g^2$ \cite{MTA}) and semiclassical
(strong coupling, to order $g^{-2}$ \cite{ChoB97}) ground-state energies
for an $H_g$ mode coupled coupled to $n$ electrons in a $t_{1u}$ orbital,
for the allowed values of the total spin $S$.
%
Here, only electron-phonon coupling is considered, while electron-electron
repulsion is neglected.
%
}
\label{JTen:table}
\begin{tabular}{l@{\quad}l@{\quad}l@{\quad}l@{\quad}l@{\quad}l}
\hline
\hline
C$_{60}^{n-}$	&spin	&$E^{\rm e-v}_n/(\hbar\omega)$
			&$ E^{\rm e-v}_n/(\hbar\omega)$
				&Berry phase
					&ground-state\\
$n$	&$S$	& [for $g\to0$]
			& [for $g\to\infty$]
				& $\gamma$
					&symmetry\\
\hline
0	&0	&0	&0	&0	&$A_g$\\
1	&$\frac12$&$-\frac54 g^2$&$-\frac12 g^2 -1 + \frac 13 g^{-2}$ &$\pi$&$T_{1u}$\\
2	&0	 &$-5 g^2$&$-2 g^2 -1  + \frac 1{12}g^{-2}$ &$0$	&$A_g$\\
2	&1	 &$-\frac54 g^2$&$-\frac12 g^2 -1 + \frac 13 g^{-2}$ &$\pi$&$T_{1g}$\\
3	&$\frac12$&$-\frac{15}4 g^2$&$-\frac32 g^2 -\frac 32 + \frac 38 g^{-2}$ &$\pi$&$T_{1u}$\\
3	&$\frac32$&$0$	&$0$	&$0$	&$A_u$\\
\hline
\hline
\end{tabular}
\end{table}

In the strong-coupling limit, $E^{\rm e-v}_n$ concides with the classical
adiabatic energy lowering (proportional to $g^2\hbar\omega$), with the
addition of a negative zero-point softening contribution (of order
$\hbar\omega$).
As was the case at weak coupling, in multi-electron contexts, JT
distortions leads to a larger energy gain when charge concentrates in as
few orbital as possible, thus favoring larger distortions.
Due to this, low-spin configurations end generally lower in energy than
high spin ones, giving rise to a kind of reversed (first) Hund's rule
exchange behavior (here we ignore the actual Hund rule, which is however
important, and will be introduced in Sec.~\ref{coulomb:sec}).
To describe correctly the quantum dynamics in the JT trough, a detailed
analysis of the coupled electron-phonon system is needed.
A crucial ingredient in this problem is the value 0 or $\pi$ of an
electronic Berry phase \cite{AMT,AssaPRL,Berry,Levi,Ihm} $\gamma$ in the
semiclassic motion: it imposes the boundary condition on the soft-mode
pseudorotational dynamics in the trough, thus selecting vibronic states of
a given parity $\exp(i\gamma)$ \cite{AMT,ob71}, not unlike similar vibronic
contexts \cite{Wilczek,lh,Mead,Wolf}.
This boundary condition determines the leading corrections to the adiabatic
energetics, of order ${\hbar\omega}/{g^2}$, reported in
Table~\ref{JTen:table} \cite{ChoB97}.

The model Hamiltonian (\ref{Hschemat}) is readily extended to include all
vibrational modes of real C$_{60}$.
Each mode adds one harmonic term of the type (\ref{H0:eq}) and a coupling
term of the type (\ref{Hev:eq}), characterized by an individual coupling
amplitude $g_k$ and frequency $\omega_k$.
In the evaluation of the total electron-phonon coupling, one should also
include two $A_g$ vibrons which also couple linearly to the LUMO electrons,
even though they do not split its degeneracy.
This is a simple polaron problem, which is exactly solvable \cite{Mahan}
and independent from the $t_{1u}\otimes 8H_g$ problem.
The amount of $A_g$-related electron-phonon energy is simply $-n^2
g^2\hbar\omega/2$, independent of the total spin $S$.

Coming to the eight $H_g$ modes, generally speaking, even within the
harmonic-phonon linear-coupling approximation, a realistic description of
the dynamical JT state of a C$_{60}^{n-}$ ion is a significantly more
complicated affair than the single-mode problem.
Linear superposition of in the coupling of individual $H_g$ modes to a {\em
same} $t_{1u}$ orbital is valid only in the antiadiabatic perturbative
regime, but does not seriously apply to the coupling range generally
accepted for C$_{60}$: for a general coupling strength, there is in fact no
linear superposition.
The JT splitting of the LUMO increases with the cooperative action of the
single modes, so that even moderate coupling of many individual modes
builds up a rather strong total coupling.

\begin{table}[t]
\centering
\caption{
Dimensionless couplings $g_k$ for the eight $H_g$ modes to the LUMO of
C$_{60}$, according to several theoretical and experimental determinations.
In the literature, the electron-phonon coupling are often gauged in terms
of the weak-coupling superconductivity parameter $\lambda$: the relation
with the Hamiltonian coupling strengths $g_k$ is \cite{Schl}:
$g_k^2=\frac{6}{5}\lambda_k/N(\epsilon_{\rm F}) / \hbar\omega_k$ for the
$H_g$ modes (where $N(\epsilon_{\rm F})$ is the density of states per
C$_{60}$ at the Fermi energy $\epsilon_{\rm F}$; one would get
$g_k^2=3\lambda_k/N(\epsilon_{\rm F}) / \hbar\omega_k$ for the $A_g$ modes).
$g_{\rm eff}^2=\sum_k g_k^2$ measures the total interaction strength, and 
$w_{\rm eff}=\sum_k g_k^2 w_k / g_{\rm eff}^2$ provides and average phonon
frequency to estimate zero-pint effects.
The last line reports the total $\lambda/N(\epsilon_{\rm F})$ from $H_g$
modes, according to the different estimates.  }
\label{couplings:table}
\begin{tabular}{cc|ccccc|cc}
\hline
\hline
$H_g$	& $\hbar\omega_k$&	Varma	&	Schl\"uter&	Antropov&	Faulhaber&	Manini	&	Raman	&	PES	\\
mode	& \cite{Bethune91} & \cite{vzr}	&  \cite{Schl}	& \cite{Antropov}&\cite{Faulhaber}&\cite{Manini01}	& \cite{Winter96}	& \cite{Gunnarsson} \\
$k$	&[meV]	&\multicolumn{7}{c}{$g_k$ \qquad\qquad -- \qquad\qquad dimensionless}\\
\hline
1	&	34	&	0.326	&	0.533	&	0.326	&	0.188	&	0.421	&	1.300	&	0.821			\\
2	&	54	&	0.149	&	0.394	&	0.365	&	0.471	&	0.494	&	0.666	&	0.941			\\
3	&	88	&	0.117	&	0.234	&	0.202	&	0.117	&	0.350	&	0.202	&	0.421			\\
4	&	96	&	0	&	0.296	&	0.194	&	0.354	&	0.224	&	0.194	&	0.474			\\
5	&	136	&	0.230	&	0.094	&	0.163	&	0.133	&	0.188	&	0.094	&	0.325			\\
6	&	155	&	0	&	0.152	&	0.249	&	0.124	&	0.152	&	0.088	&	0.197			\\
7	&	177	&	0.480	&	0.297	&	0.368	&	0.319	&	0.319	&	0.165	&	0.339			\\
8	&	195	&	0.260	&	0.235	&	0.368	&	0.235	&	0.293	&	0.136	&	0.376			\\
\hline
tot $H_g$:&&&&&&&\\
$g_{\rm eff}^2$& $\sum_k g_k^2$&0.493	&	0.756	&	0.677	&	0.586	&	0.840	&	2.285	&	2.363	\\
$\hbar w_{\rm eff}$&[meV]&	136	&	83	&	121	&	102	&	93	&	44	&	75\\
$\lambda/N(\epsilon_{\rm F})$& [meV]&56	&	52	&	68	&	49	&	65	&	83	&	147		\\
\hline
\hline
\end{tabular}
\end{table}

For the detailed values of the electron-phonon couplings, several estimates
are available in the literature
\cite{Bethune91,Negri,Green96,Manini01,Schl,vzr,Antropov,Faulhaber,Winter96,Gunnarsson},
some of which are reported in Table~\ref{couplings:table}.
The main observation here is that individual mode couplings are indeed
intermediate to weak.
However, a significant spread in the values of different estimates is due
to the large sensitivity of the coupling to the individual phonon
eigenvector, which differ largely in different approximations.
Moreover, the total coupling of all calculations is significantly weaker
than all experimental estimates, such as those obtained by
fitting the vibronic features of the photoemission spectrum of C$_{60}^-$
\cite{Gunnarsson}.
This is rather surprising in view of the fact that a recent parameter-free
calculation \cite{ManiniComm03,Manini03,Gattari03} found instead good
accord with observed vibronic structures in photoemission data of neutral
C$_{60}$ \cite{Bruhwiler97,Canton02}, based on HOMO couplings computed
\cite{Manini01} with the same {\it ab-initio} techniques as those for the
LUMO reported in Table~\ref{couplings:table}.
Even though some ideas are being pursued to understand the
theory-experiment discrepancy for the LUMO couplings \cite{Bordoni04}, this
is still an unresolved issue.

As pointed out by Bergomi and Jolicoeur \cite{Bergomi}, experiments on
anions in matrix may yield relevant information to the vibronic effects.
Near-infrared and optical spectra of C$_{60}^{n-}$ ions in solution and
frozen Ar are available \cite{Heath,Langford99}.
A major $t_{1u}\to t_{1g}$ optical transition near $1$~eV is present for
all values of $n$.
It is accompanied by additional vibronic shake-up structures, typically
near 350, 750, 1400 and 1600~cm$^{-1}$, which involve the vibronic
couplings of both the $t_{1u}$ LUMO and $t_{1g}$ LUMO+1.
Also, a satisfactory vibronic assignement of a recently observed gas-phase
spectrum \cite{Tomita05} is not yet available.
The experimental information seems insufficient as yet for any relevant
comparison with our calculations.
Well-defined vibrational spectra are instead available for
chemisorbed C$_{60}^{n-}$ \cite{Modesti} and for A$_n$C$_{60}$ alkali
fullerides \cite{Fisher}.
In this case, however, interaction of the electronic $t_{1u}$ level with
surface states or with other $t_{1u}$ states of neighboring balls turns the
level into a broad band, and our treatment as it stands is invalid.
Rapid electron hopping from a molecule to another interferes substantially
with the dynamical JT process, in a way which is not known in detail
\cite{Bersuker,Gehring75,Kaplan95}.
The spectra of negatively charged C$_{60}$ adsorbates and solids, in any
case, do not present any clear evidence of vibronic splittings, but rather
of the gradual continuous shift most likely due to a gradual overall change
of geometry, also suggested by DFT-LDA calculations \cite{Andreoni}.

The full quantum-mechanical problem of a threefold degenerate electronic
state linearly coupled to eight fivefold degenerate harmonic oscillators is
conceptually simple but in practice, for arbitrary couplings, rather hard
to solve exactly.
On one side, it is straightforward to compute the Hamiltonian matrix
elements on a truncated oscillator basis.
Also, this matrix is sparse, involving nonzero terms only between states
whose numbers of vibrons differ by exactly one.
Moreover, the inclusion of states with up to $N$ vibrons, because of linear
e-v coupling, implies decay as $\exp (-N)$ of components with $N$-vibron
states, for large enough $N$.
For this reason, a truncated basis set including all states up to $N^{\rm
max}$ vibrons gives a variational estimate of the lowest eigenvalues, with
good convergence with increasing $N^{\rm max}$ \cite{AMT,Delos96,Manini03}.
The structure of the problem makes it especially suitable for a Lanczos
algorithm.
By this numerical technique, not only the ground-state energy but also
several low-lying excitations can be computed \cite{Reno}.

To get an estimate of the accuracy of the adiabatic and antiadiabatic
approximations applied to C$_{60}^-$, the couplings of Ref.~\cite{Manini01}
(see Table~\ref{couplings:table}) can be plugged in the strong and
weak-coupling formulas of Table~\ref{JTen:table} and obtain the following
estimates for the ground-state energy: $E^{\rm e-v}_{\rm ad\ 1}=-38$~meV,
$E^{\rm e-v}_{\rm ad+zp\ 1}=-131$~meV, and $E^{\rm e-v}_{\rm
antiad\ 1}=-97$~meV.
These estimates should be compared to the exact ground-state energy
lowering obtained by exact Lanczos diagonalization: $E^{\rm e-v}_{\rm
exact\ 1}=-76$~meV \cite{Lueders03}.
The better accuracy of the perturbative antiadiabatic estimate is due to
the relatively small-to-intermediate value of the total $\sum_k g_k^2=0.84$
(Fig.~\ref{Softening:fig} indicates that the zero-point corrected adiabatic
energy $E^{\rm e-v}_{\rm ad+zp\ 1}$ becomes fairly accurate only beyond
$\sum_k g_k^2\geq 1.5$).
For $n=2$ $S=0$, and even for $n=1$ if the larger couplings from PES
\cite{Gunnarsson}  are taken, the adiabatic estimate comes in much
better agreement, especially after a zero-point softening correction is
included \cite{Lueders03,Leuven02}. 
In effect, this adiabatic-plus-zero-point-correction approximation should
be quite generally applicable to all strongly coupled JT problems.
The relative difference of the classical adiabatic energy from the exact
energy shows the relative importance of phonon quantum effects in
C$_{60}^{n-}$.
So, while for $n=1$ -- a weaker coupling case -- the exact energy lies 
closer to the antiadiabatic
expression, the exact result for $n=3$ and even more for $n=2$  is much closer to the
adiabatic (zero-point corrected) estimate, because in C$_{60}^{2-}$ and
C$_{60}^{3-}$ JT coupling is effectively stronger (see
Table~\ref{JTen:table}).

Even with the uncertainties on the exact value of the couplings, fullerene
ions are generally taken as intermediate- to strong-coupled JT systems, the
relatively strong coupling realized as the the cooperative effect of
several moderate individual couplings.
This fact is more general: any JT system with a
large enough number of vibron modes weakly-coupled to the {\em same}
electronic level behaves as if strongly coupled, since the contributions to
the splitting of the electronic multiplet add cooperatively.
In fact, for intermediate coupling, the JT energy gains of the individual
modes do {\em not add} algebraically, as they would in the limiting
perturbative and adiabatic regimes.
To further understand this, we may wish to see the splitting as if
effectively due to the coupling to a single mode.
In this picture, the addition of more coupled modes effectively acts like
an increase in $g$, thus a rightward shift along the solid line of
Fig.~\ref{Softening:fig}.
Since the curvature is upward, the sum of the individual energy gains for
each mode coupled separately is always larger than what one obtains from
diagonalizing with all the modes included together.

\begin{table}[t]
\centering
\caption{
Perturbative (weak coupling, to order $g^2$) and semiclassical (strong
coupling, to order $g^{-2}$) electron-phonon pair energies for an $H_g$
mode coupled coupled to an average $n$ electrons in a $t_{1u}$ orbital,
assuming for all $n$ a low-spin ground state.
The numerical pairing energies are computed for the C$_{60}^-$ couplings of
Ref.~\cite{Manini01} (see Table~\ref{couplings:table}) in the
anti-adiabatic approximation.
The last column includes the contribution $-\sum g_k^2 \hbar\omega_k =
-93$~meV of the $A_g$ modes \cite{Agerror:note}.
The strong-coupling expressions are rigorously valid only in the limit
where the intra-shell electron-electron repulsion can be totally neglected.
}
\label{Ueff:table}
\begin{tabular}{l@{\quad}l@{\quad}l@{\quad}l@{\quad}l}
\hline
\hline
C$_{60}^{n-}$
	&${\cal U}_n^{\rm e-v}/(\hbar\omega)$
		&${\cal U}_n^{\rm e-v}/(\hbar\omega)$
			&${\cal U}_n^{\rm e-v}$ [meV]
				&${\cal U}_n^{\rm e-v}$ [meV]\\
$n$	& antiadiab
		& adiabatic
			& [$H_g$ modes only]
				& [$A_g$ and $H_g$ modes]\\
	& [$g\to0$]
		& [$g\to\infty$]
			& (antiadiab)
				& \\
\hline
1	&$-\frac52 g^2$&$ - g^2 +1 - \frac 7{12}g^{-2}$  &$-195$ & $-288$\\
2	&\ $5g^2$&\ $2 g^2 -\frac12  + \frac {13}{24}g^{-2}$&\ \ $390$
							&\ \  $297$\\
3	&$-\frac52 g^2$&$- g^2 +1 - \frac 7{12}g^{-2}$ 
	 &$-195$ & $-288$\\
\hline
\hline
\end{tabular}
\end{table}

In a solid-state context, electron hopping between molecules will alter
this picture as soon as it is large enough to disturb substantially the
local JT physics. However, as we will argue later on, the relevant hopping
is not so much the bare electron hopping, but rather the quasiparticle
effective hopping, a quantity which becomes much smaller near a Mott
transition.
So long as that effective hopping is small enough, the JT energetics as
described above is expected to hold even in the solid state.
Electronic correlation in a compound where conduction occurs through
electron hopping between C$_{60}$ ionic sites is mainly governed by the
extra electron-electron repulsion when the local charge fluctuates away
from its average filling of $n$ electrons.
The {\em pair energy}, or Hubbard $U$, defined by
\begin{equation}
\label{U_n}
{\cal U}_n=E_{n+1}+E_{n-1}-2E_n\,,
\end{equation}
(where $E_n$ are the fully relaxed ground-state energies of $n$ electrons)
measures precisely the strength of this local correlation.
As we shall see in detail below (Sec.~\ref{coulomb:sec}), ${\cal U}_n$ is
largely dominated by the positive Coulomb electron-electron repulsion.
However, also the JT electron-phonon energies $E^{\rm e-v}_n$ of
Table~\ref{JTen:table} show a strong nonlinear $n$-dependence, and thus
necessarily also contributes to ${\cal U}_n$.

In the standard many-body perturbation theory language, formally ${\cal
U}_n^{\rm e-v}$ is the real part of the two-electron vertex function at
zero frequency, including electron-phonon contributions in the
weak-coupling antiadiabatic limit.
The contribution of a nondegenerate $A_g$ mode ${\cal U}_n^{{\rm e-v}\,A_g}
\equiv - g^2 \hbar\omega < 0$, independent of $n$ and irrespective of spin
$S$.
If coupling to $A_g$ modes prevailed, the electron-phonon pair energy would
be negative, and thus electrons would lower their energy if, rather than
keeping an uniform occupancy $n$ of all sites, they separated into $(n-1)$
and $(n+1)$ at different molecules, thus creating a bipolaron, or a charge
density wave.
Even though charge disproportionation has been claimed to play some role in
K$_3$C$_{60}$ \cite{Ceulemans97}, in practice the large value of the on-site
Coulomb repulsive $U$ parte rules this possibility out in the fullerides.

\begin{figure}[t]
\centering
\epsfig{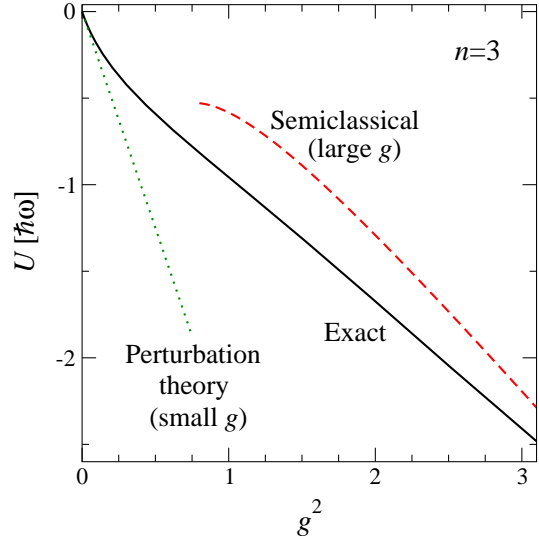}
\caption{\label{PairBinding:fig}
Exact single-mode electron-phonon pair-binding energy ${\cal U}_1^{\rm
e-v}$ (solid line), compared to weak-coupling perturbation theory for
$g\ll 1$ (dotted) and semiclassical theory for $g\gg 1$ (dashed).
${\cal U}_n^{\rm e-v}$ is similar for $n=1,3,5$ electrons.
The coupling strengths $g^2$ for the C$_{60}$ modes are all less than
$\approx 1$.
(Adapted from Ref.~\cite{ManiniPhD}.)
}
\end{figure}

Assuming a low-spin ground state for all $n$ (we shall discuss the validity
of this assumption in Sec.~\ref{coulomb:sec}), Table~\ref{JTen:table}
provides all the ingredients necessary to determine approximate expressions
for the $H_g$ modes contribution to ${\cal U}_n^{\rm e-v}$, both in the
antiadiabatic limit and in the adiabatic limit (neglecting retardation
completely).
These expressions are reported in Table~\ref{Ueff:table}.
For odd $n$, ${\cal U}_n^{\rm e-v}$ is {\em negative}, thus as customary
the electron-phonon coupling is attractive and favors pairing
\cite{Baskaran}, thus opposing 
Coulomb repulsion.
Interestingly, for even $n=2,4$, ${\cal U}_n^{\rm e-v}$ is instead {\em
positive}: here the electron-phonon coupling of the JT type re-enforces the
Coulomb repulsion and contributes to {\em suppress} local fluctuations away
from the average number $n$ of electrons at each site.
The sign change of the JT contribution to ${\cal U}_n$ is due to the
nonparabolic dependence of the ground-state energy on $n$, as reported in
Table~\ref{JTen:table}.
Figure~\ref{PairBinding:fig} shows a comparison of the approximate
expressions with the pair energy obtained by exact diagonalization for
$n=3$.
As for the total energy, the exact pair energy is bracketed between the
antiadiabatic estimate (overestimating) and the semiclassical adiabatic
expression (underestimating).

The couplings of Table~\ref{couplings:table} can be plugged into the
expressions of Table~\ref{Ueff:table}, to obtain an estimate of the phonon
contributions to the pair energy.
In Table~\ref{Ueff:table} we use the (probably underestimated) couplings of
Ref.~\cite{Manini01}, and the antiadiabatic approximation (which insteads
overestimates ${\cal U}_n$), to obtain our best realistic estimate of the
phonon contribution to the pair energies.
These values are relatively large, and even though they come a long way
from reversing the much larger Coulomb repulsion addressed in
Sec.~\ref{coulomb:sec} below, they represent a serious correction with
cannot be neglected.

\subsection{Intramolecular Coulomb repulsion and Hund's rule exchange}
\label{coulomb:sec}

We come now to the Coulomb electron-electron intra-molecular energetics.
As appropriate to an orbitally degenerate level, two distinct Coulomb 
enter: the ``Hubbard U'' term coupling to the total charge in
the degenerate shell, and an intramolecular exchange term.
The first term is strictly related to the pair energy of Eq.~(\ref{U_n}):
basically measuring the energy cost required to change electron occupancy
of $n$ of the molecular site. In a solid conductor, this term will oppose
all local fluctuations away from the average occupancy.
Exchange terms, of smaller value, act instead to split multiplets 
at any given fixed occupancy. They enforce in particular the first two
Hund rules, lowering in energy the otherwise degenerate multiplet states 
of highest spin and largest orbital momentum.

The structure of the Coulomb Hamiltonian for a $(t_{1u})^n$ configuration
is formally the same as that for an atomic $p^n$ in spherical symmetry, and
as such entirely determined by two parameters only, usually chosen as the
configuration-averaged $U$ and a Hund-rule intra-molecular exchange $J$
\cite{ChoB97,Martin93,Han00}.
For a detailed symmetry analysis of the structure of the Coulomb Hamiltonian
$\hat{H}^{\rm e-e}$, we refer to Ref.~\cite{Lueders02}.

The Coulomb Hamiltonian is a 2-body electronic operator which can be
written as
\begin{equation}
\label{Coulomb-hamiltonian}
\hat{H}^{\rm e-e} =
\frac{1}{2} \sum_{\sigma, \sigma'} \sum_{{m m'}\atop{n n'}} 
w_{\sigma,\sigma'}(m,m';n,n') \ 
\hat{c}^\dagger_{m \sigma} \hat{c}^\dagger_{m' \sigma'} \, 
\hat{c}_{n' \sigma'} \hat{c}_{n \sigma},
\end{equation}
where the Coulomb integrals
\begin{equation}
\label{Coulomb-ints}
w_{\sigma,\sigma'}(m,m';n,n') = \!\Id{r} \! \Id{r'} \, 
\varphi^{*}_{m \sigma}(\vr) \, 
\varphi^{*}_{m'\sigma'}(\vr') \,
u_{\sigma,\sigma'}(\vr,\vr') \, 
\varphi_{n\sigma}(\vr) \,
\varphi_{n'\sigma'}(\vr') 
\end{equation}
are expressed on the basis of the single-particle LUMO wave functions
$\varphi_{m\sigma}(\vr)$, associated to the fermion operators
$\hat{c}^\dagger_{m \sigma}$ introduced above.

These Coulomb integrals (\ref{Coulomb-ints}) can be evaluated
directly for the simple kernel
$u_{\sigma,\sigma'}(\vr,\vr')= q_e^2 / (4\pi \epsilon_0 |\vr-\vr'|)$, with
phenomenological molecular orbitals \cite{Nikolaev02}.
This approach neglects the screening due to the other electrons within the
same fullerene molecule.
Due to that, it yields very large exchange values, close to early
Hartree-Fock results \cite{Chang91}, which generally overestimate exchange
due precisely to lack of screening.
With that in mind, one can regard as substantially more realistic the
evaluation of the interaction Hamiltonian (\ref{Coulomb-hamiltonian})
obtained in Ref.~\cite{Lueders02} on the basis of DFT-LDA electronic
structure calculations, which account for the full polarization response of
the total charge density.
Since on the other hand LDA overestimates screening, the outcome of this
calculation represents in turn an underestimate of actual intramolecular
exchange.

Like in the atomic case, the symmetry of the Coulomb interaction plus the
molecular symmetry of the problem allow us to express all of the Coulomb
integrals in (\ref{Coulomb-ints}) as functions of a small set of physical
parameters.
Indeed, the Coulomb integrals are expressed as follows
\begin{equation}
w(m,m';n,n') = \sum_{\Lambda=A_g,H_g} F^{\Lambda}
\left( \sum_\mu C^{\Lambda \mu}_{m n} \, 
C^{\Lambda \mu}_{m' n'} \right) 
\label{Fdecomposition}
\end{equation}
in terms of a {\em minimal} set of independent parameters $F^{\Lambda}$,
defined as:
\begin{equation}\label{F-params}
F^{\Lambda}=
\frac{1}{|\Lambda|} \sum_{\mu}  \Id{r} \Id{r'} 
\Phi^{\Lambda}_{\mu}(\vr) \,
u(\vr,\vr') \,
\Phi^{\Lambda}_{\mu}(\vr') \,,
\end{equation}
with
\begin{equation}\label{phicoupled:eq}
\Phi^{\Lambda}_{\mu}(\vr)
=
\sum_{m,n} 
C^{\Lambda \mu}_{m n} \,
\varphi_{m}(\vr) \,
\varphi_{n}(\vr) \, .
\end{equation}
The two parameters governing $\hat{H}^{\rm e-e}$ are related to the $k=0$ and
$k=2$ Slater-Condon integrals $F^{(k)}$ for $p$ electrons in spherical
symmetry \cite{Cowan}.

With the decomposition (\ref{Fdecomposition}) in hand, it is convenient to
re-organize the interaction Hamiltonian (\ref{Coulomb-hamiltonian}) in
terms of number-conserving symmetry-adapted fermion operators:
\begin{equation}\label{He-ecombined}
\hat{H}^{\rm e-e} = \frac{1}{2} \sum_{\Lambda} F^{\Lambda} 
\left( \sum_\mu \hat{w}^{\Lambda \mu} \, \hat{w}^{\Lambda \mu} \right)
- \frac 16 \left(F^{A_g}-5 F^{H_g}\right) \hat{n} 
\end{equation}
where we defined the operators
\begin{equation}\label{woperators}
\hat{w}^{\Lambda \mu} := \sum_{\sigma} \sum_{m n} 
C^{\Lambda \mu}_{m n} \, \hat{c}^\dagger_{m\sigma} \hat{c}_{n\sigma} \,.
\end{equation}

Rather than the $F^\Lambda$ parameters, it is customary to use the
parameters $U=F^{A_g}/3 - F^{H_g}/3$, and $J= F^{H_g}/2$ (the notation $K$
is sometimes used for this quantity; it also equals $F^{(2)}/3$ of
spherical symmetry \cite{ChoB97,Cowan}).
With this definition, observe that
\begin{equation}\label{Eave}
E^{\rm e-e\ ave}(n)=
{\rm Tr}|_n ( \hat{H}^{\rm e-e} )= U \, \frac {n (n-1)}2 \, ,
\end{equation}
where ${\rm Tr}|_n$ is the trace restricted to the $n$-electrons states.
$U$ governs therefore the parabolic dependence of the Coulomb energy
averaged over all possible multiplet configurations on the total number of
electrons $n$.
It should be noted that $U$ involves average multiplet energies, not {\em
ground-state energies} like the pair binding energy of Eq.~(\ref{U_n}): the
two quantites therefore would only coincide in a hypothetical system where
multiplets were unsplit by JT and exchange terms.
With standard multiplet splitting instead, the pair-energy definition,
based on the ground-state energy, depends on $n$, which makes it rather
inconvenient to characterize the Coulomb repulsion in the $t_{1u}$ shell
with a single microscopical parameter.

\begin{table}[t]
\centering
\caption{
Coulomb energies for the $(t_{1u})^n$ or $p^n$ multiplets expressed in
terms of the two Coulomb parameters $U$ and $J$.
These multiplets are correct under the assumption that electron-phonon
coupling can be totally neglected.
The $n=4$ and $n=5$ multiplet have the same $J$ terms as $n=2$ and $n=1$,
with the center-of-mass ($U$) term given by Eq.~(\ref{Eave}): $6\,U$ and
$10\,U$ respectively.
}
\label{multiplets:table}
\begin{tabular}{l@{\quad}l@{\quad}l@{\quad}l@{\quad}l@{\quad}l}
\hline
\hline
C$_{60}^{n-}$
	&$^{(2S+1)}[L]$
		&$^{(2S+1)}\lambda$
			&$E^{\rm e-e}(n,S,L)$\\
\hline
1	&$^2$P	&$^2t_{1u}$	&0 \\
2	&$^1$S	&$^1a_g$	&$U + 4J$ \\
2	&$^1$D	&$^1h_g$	&$U + J$ \\
2	&$^3$P	&$^3t_{1g}$	&$U - J$ \\
3	&$^2$P	&$^2t_{1u}$	&$3U + 2  J$ \\
3	&$^2$D	&$^2h_{u}$	&$3U $ \\
3	&$^4$S	&$^4a_u$	&$3U -3 J$ \\
\hline
\hline
\end{tabular}
\end{table}

The $J$ parameter controls the multiplet exchange splittings, i.e.\ it
implements Hund's rules.
One can label all multiplet states with an orbital ``angular momentum'' $L$
(recall that the $t_{1u}$ orbitals behave effectively as $p$ orbitals), as
in Table~\ref{multiplets:table}.
In fact, it is possible to express the Coulomb energy of all multiplets
with the equation
\begin{equation}\label{multipletEn:eq}
E^{\rm e-e}(n,S,L) = U \frac{n (n-1)}{2} - J \left[2 S(S+1) +
  \frac12 L(L+1) + \frac12 n(n-6)\right],
\end{equation}
as a function of $n$, $S$ and $L$ \cite{Lueders02}.
These energies are also reported in Table~\ref{multiplets:table}.

\rem{





\begin{table}[t]
\begin{center}
\begin{tabular}{crrr}
\hline
\hline
Parameter & ($\hat{H}^{\rm e-e}$) &
($\hat{H}_{\rm e-v}$, 2$^{\rm nd}$ order) & Total: (e-e)+(e-v) \\
        &       [meV]   &       [meV]   &       [meV]   \\
\hline
$U$     & 3069          &         32    &       3101    \\
$J$     &   32          &        -57    &        -25    \\
\hline
\hline
\end{tabular}
\caption{
The Coulomb parameters for C$_{60}^{n-}$, as obtained from the LSDA
calculations, the effective parameters obtained from second-order treatment
of $\hat{H}_{\rm e-v}$ [based on the couplings of
(Manini {\it et al.}\ 2001)], and the sum of the two contributions.
\label{parameters-:table}
}
\end{center}
\end{table}

}

The Coulomb parameters were computed by ab-initio density-functional
calculations in Ref.~\cite{Lueders02}.
The value of the molecular Hubbard U term of isolated C$_{60}$ was
determined $U=3.07$~eV, in the same range as previous calculations
\cite{Martin93,Antropov92,Pedersen92,deCoulon92}.
This estimate can be compared to experiment.
The pair energy ${\cal U}_1=E_0+E_2-2E_1= (E_0-E_1) -(E_1-E_2)=E_{\rm
A}({\rm C}_{60}) - E_{\rm A}({\rm C}_{60}^-)$, where $E_{\rm A}$ indicates
the electron affinity.  These quantities are experimentally available for
molecular C$_{60}$ \cite{Yeretzian} and C$_{60}^-$ \cite{Limbach,Hettic91}
\begin{equation}
E_{\rm A}\left({\rm C}_{60}\right)-E_{\rm A}\left({\rm C}_{60}^-\right)=
2.7\,{\rm eV}-0.17\,{\rm eV} \approx 2.5\,{\rm eV}.
\end{equation}
This value is slightly reduced from pure Coulomb by electron-phonon
coupling and orbital relaxation (see also Table~\ref{Ueff:table}).
If we neglect these comparatively small contributions to the pair energy
${\cal U}_1\approx U$: a range $U\simeq 2.5\div 3$~eV provides a realistic
estimate of the actual $U$ parameter of gas-phase C$_{60}$.

\begin{table}[t]
\centering
\caption{
The energy (in meV) of the lowest state of $t_{1u}^n$ for each $n$ and $S$,
including the electron-electron and JT contributions from $\hat{H}_0 +
\hat{H}_{\rm e-v} + \hat{H}^{\rm e-e}$ (but excluding the $\left[\epsilon n
+ U n(n-1)/2\right]$ term).
Adiabatic approximation: the phonons are treated in the adiabatic
(strong-coupling) approximation, by full relaxation of the phonon modes to
the optimal classical JT distortion for each $n$ and $S$.
Anti-adiabatic: the electron-phonon behaves effectively as a negative
$J^{\rm e-v}=-\frac 34 \sum_k g_k^2 \hbar\omega_k$ \cite{Lueders02}: the
combined Hund and JT interaction yields levels split according to the
pattern of Table~\ref{multiplets:table}, with $J$ replaced by $J+J^{\rm
e-v}$. (From Ref.~\cite{Lueders02}.)
}
\label{e-v--e-eenergies}
\begin{tabular}{c@{\quad}c@{\quad}|@{\quad}l@{\quad}l}
\hline
\hline
$n$&$S$ & adiabatic & anti-adiabatic \\
\hline
$2$	&  0  &  ${\bf -92}$ & ${\bf -100}$\\
	&  1  &  $-71$ &  $-25$\\
& & \\
$3$	& 1/2 &  $-85$ & ${\bf -50}$\\
	& 3/2 &  ${\bf -97}$ & $+75$ \\
\hline
\hline
\end{tabular}
\end{table}

For the intra-orbital exchange interaction, the calculation of
Ref.~\cite{Lueders02} finds a value $J=32$~meV.
Calculations where screening is ignored or underestimated find much larger
values, e.g.\ $J=110$~meV (Hartree-Fock \cite{Chang91}) and $J=95$~meV
(calculation of the unscreened integrals in model orbitals
\cite{Nikolaev02}).
An early calculation including screening by the strongly polarizable
C$_{60}$ molecule found $J=25$~meV \cite{Martin93}.
Because DFT tends to overestimate screening, DFT values of $J$ may be
somewhat underestimated.
We believe that the actual Coulomb parameters of C$_{60}$ could lie
somewhere in between the DFT couplings assumed here and the ``bare'' ones
of Ref.~\cite{Nikolaev02,gosia03}, but most likely closer to the DFT ones,
due to the large polarizability of C$_{60}$.
An intermediate value of $J \simeq 50$~meV is probably realistic.

It should be noted in this respect that suggestions advanced in earlier
times that overscreening could reverse altogether the sign of this
frozen-molecule, purely electronic exchange $J$
\cite{Baskaran,Chakravarty92,Murthy95} appear incorrect.
This is confirmed by recent purely electronic quantum Monte Carlo
calculations \cite{Lin05} finding a spin gap between triplet (ground) and
singlet (excited) states consistent with $J\simeq 54$~meV.
What is emerging as the correct picture is that only after inclusion of
nuclear motion, and specifically of JT effects that are strongly
anti-exchange, the {\em effective} $J$ may change sign, from positive to
negative.
As the above results in fact show, the JT energetics is opposed to Hund's
rule, and favor low-spin states against intra-molecular exchange, as
evident from Table~\ref{JTen:table}.  The effects of the two interactions,
electronic exchange and JT, tend to cancel.
In fact the antiadiabatic expressions of the JT energies
(Table~\ref{JTen:table} and Ref.\ \cite{MTA}) are consistent with the
structure given by Eq.~(\ref{multipletEn:eq}), if we replace $J$ with a
negative local exchange $J^{\rm e-v}=-\frac 34 \sum_k g_k^2 \hbar\omega_k$.
Therefore when the antiadiabatic approximation holds -- for example when
the JT coupling is weak and the effective electron hopping is much smaller
than important $H_g$ vibration frequencies -- one can account
simultaneously for both JT and Coulomb exchange by replacing $J$ with an
effective $J_{\rm eff}=J+J^{\rm e-v}$.
The resulting energetics for C$_{60}^{n-}$ ions is summarized in
Table~\ref{e-v--e-eenergies}.
Likewise, an additional term $J^{\rm e-v} \frac 13 n (6 - n)$ must be
introduced to account for the shift of the center of mass and contributes
$U^{\rm e-v}=\frac 23 J^{\rm e-v}=-\frac 12 \sum_k g_k^2 \hbar\omega_k$ to
the multiplet center-mass Coulomb $U$.
However, as $U\gg J$, the cancellation of $U$ is much less effective than
for the exchange term, and the phonon's contribution leaves a strongly
repulsive total $U$.
We should stress again that for general couplings and general electron
hoppings the electron-vibration and electron-electron contributions do not
add so simply, and must be treated in full \cite{Sookhun03bis,Dunn05}.
Nevertheless these approximate results are quite useful in providing a
basic insight in this competition.

The balance between Hund's rule exchange and JT is much less definite for
C$_{60}^{n+}$, where, according to calculations \cite{Lueders03}, overall
Coulomb exchange marginally prevails, leading to regular Hund's rule
high-spin ground states. In that case, for example C$_{60}^{n+}$ should
have a $S=1$ ground state as an isolated ion, and might or might not retain
high spin in solution \cite{Bruno03} or in solid-state compounds
\cite{Datars95,Datars96,Panich02,Panich03}, depending on the environment.

Experiments do confirm a substantial screening of $J$.
Consider first C$_{60}^{4-}$, present in insulating A$_4$C$_{60}$ (A=K,
Rb).
NMR \cite{Kerkoud,ZimmerAll,Lukyanchuk95} has identified an important
singlet-triplet excitation around $100\div 140$~meV.
Taking this spin gap as representative of the $^1$S$\to ^3$P transition
between the singlet molecular ground state and the lowest triplet state
(separated by $5\,J_{\rm eff}$ according to Table~\ref{multiplets:table}),
NMR data are consistent with a small effective value of $J_{\rm eff}\simeq
-20 \div -28 $~meV.
Thus the electron-phonon coupling counteracts very effectively the Hund's
rule exchange, and eventually prevails, even if marginally, in these
electron-doped fullerene compounds. As we will see below, this is the
origin of the low-spin character of the Mott insulating phases (Mott JT
insulators), and of (singlet, s-wave) superconductivity in fullerides.
EPR and NMR of discrete salts indicate that also C$_{60}^{3-}$ ions have a
low-spin ground state \cite{Bhyrappa93,Bossard93}.
The case of C$_{60}^{2-}$ is more intriguing.
Here, it appears that the balance between Coulomb effects, matrix effects,
and JT energies is more critical, and the final ground state
can jump between $^3$P and $^1$S, depending on marginal effects.
Early reports suggested that when frozen in an organic glass, C$_{60}^{2-}$
is in a triplet state \cite{Dubois}, but this evidence has been questioned
\cite{Reed00review}.
In discrete salts, the singlet and triplet have been claimed to coincide
closely \cite{Chen95,Boyd}, whereas in solutions the singlet appears to be
stabilized by about 70~meV \cite{Reed00review,Trulove,Sun97}.
Evidence of a 10~meV spin gap attributed to long-lived fluctuating
C$_{60}^{2-}$ in cubic CsC$_{60}$  suggests almost complete compensation, with
marginal prevalence of the singlet state \cite{Brouet99}.
On the other hand, ferromagnetism of TDAE-C$_{60}$ \cite{Schilder94} has been
interpreted \cite{Arovas95} in terms of fluctuating $S=1$ local
C$_{60}^{2-}$ sites.
Structural evidence of a JT distortion is relatively rare.  When available
it is not especially clear in showing twice larger distortion in
C$_{60}^{2-}$ \cite{Paul94} relative to C$_{60}^{-}$, which is what is expected for low spin.
The only indirect evidence of a distortion in K$_4$C$_{60}$ is provided by
the splitting of the highest $T_{1u}$ vibrational mode demonstrated in
Ref.~\cite{Iwasa95}.

\rem{
In the antiadiabatic limit, the effective e-e interaction is simply the
superposition of the Coulomb repulsion and the phonon-mediated attraction.
The total net result as far as $U$ is concerned 
is still repulsive, the large Coulomb term only marginally
corrected by molecular distortions. All the
other intramolecular exchange terms are instead heavily reduced. However,
while in the case of C$_{60}^{n-}$ that leads to an effective
sign reversal from Hund to ``anti-Hund'', (and from repulsive 
to attractive for an electron pair in the singlet channel) 
the balance is much less definite for C$_{60}^{n+}$,
where the overall sign remains positive for $n$=2 and is uncertain
for higher $n$ values.
The efficiency of the JT effect in reversing Hund-rules 
couplings for C$_{60}^{n+}$ is even weaker when the adiabatic approximation is 
considered instead of the anti-adiabatic one. In the adiabatic
approximation, where ionic motion is classical, the molecular ground 
state of C$_{60}^{n+}$ turns out to be always high spin for all $n$ 
values, in contrast to C$_{60}^{n-}$ where it is always low spin. 
}

In summary, experimental and theoretical evidence agrees that (i) in the
energy comparison between different number occupancies $n$ (pair energy
${\cal U}_n$) the repulsive Coulomb energies dominate; (ii) in the energy
comparison between different multiplet (spin) states at fixed occupancy $n$
(exchange term $J$), the attractive anti-Hund JT energies tends to prevail
over slightly weaker competing Coulomb exchange, but there are exceptions
and high spin may prevail in particular environments.

\section{Strong correlation in solid alkali fullerides: general}
\label{correlations:sec}

\begin{figure}[t]
\centering
\epsfig{file=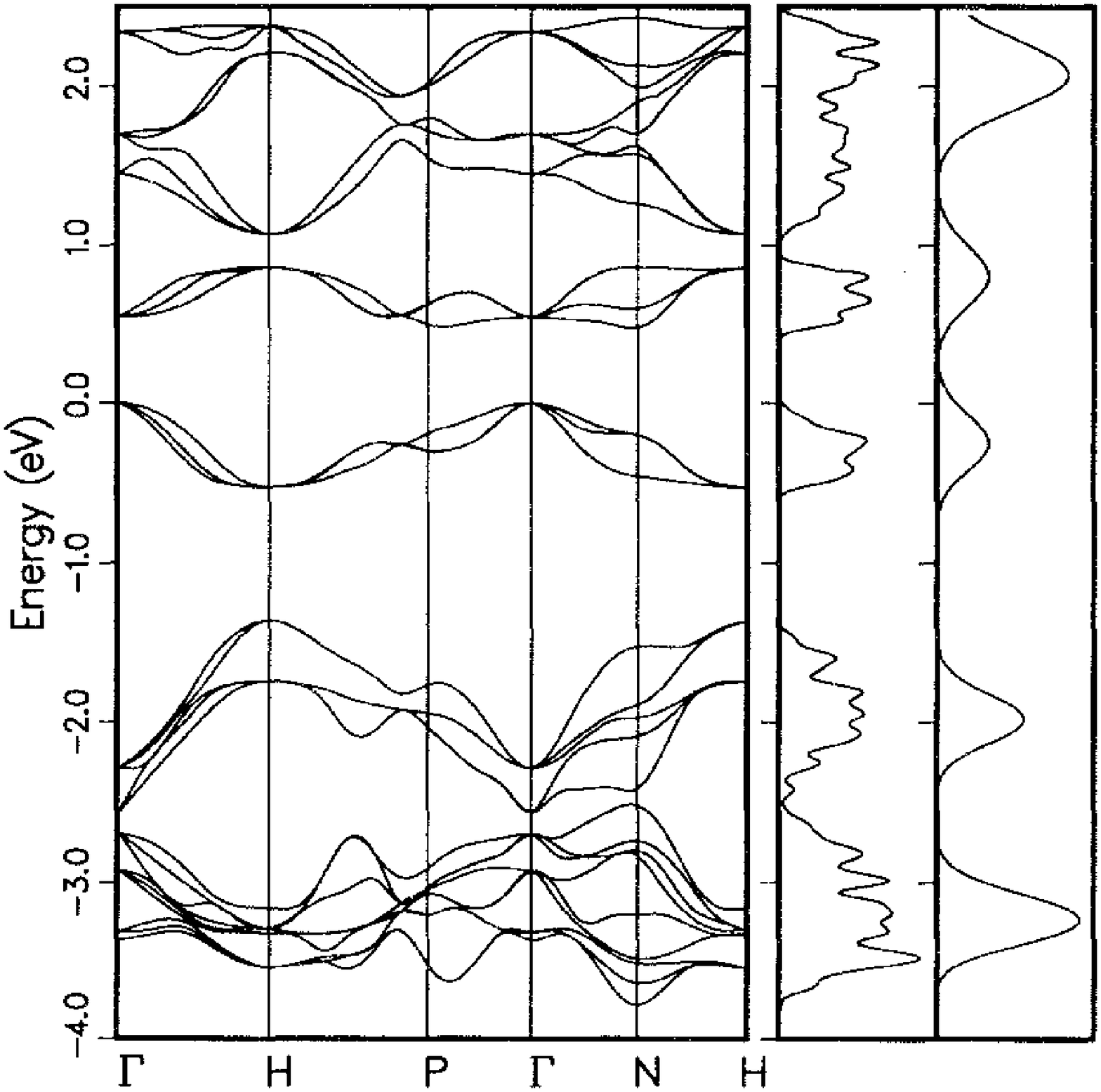,height=62mm,clip=}\quad\epsfig{file=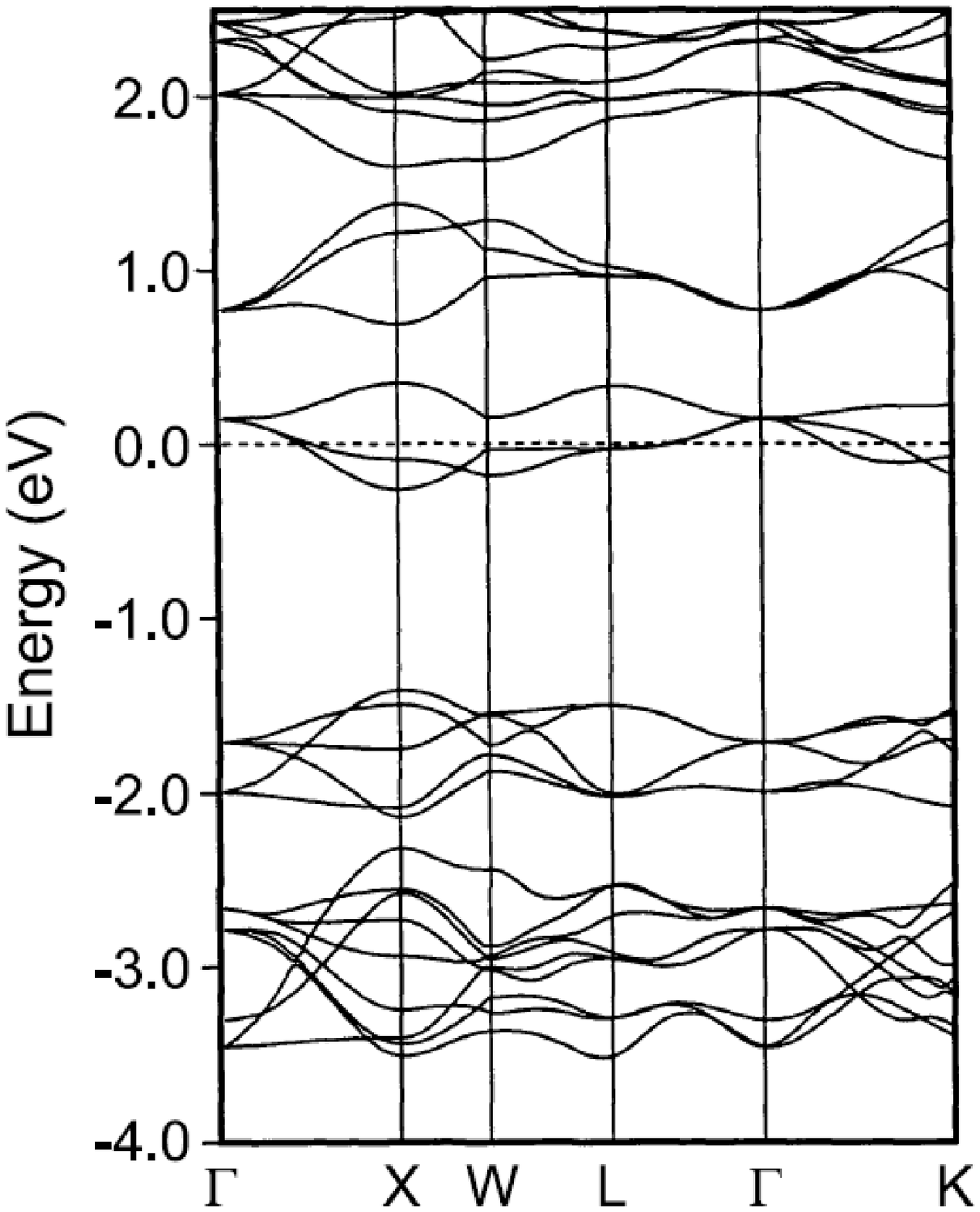,height=62mm,clip=}
{\bf\small
(a)\hfill
\quad\qquad\qquad(b)\hfill
}
\caption{\label{bands:fig}
(a) Band structure of K$_6$C$_{60}$.  Energy is referred to the $t_{1u}$
band maximum.  The solid-state density of states is compared to
Gaussian-broadened single-molecule levels.
(From Ref.~\cite{Erwin91}.)
(b) Band structure of K$_3$C$_{60}$.  Energy is referred to the Fermi
level.  (From Ref.~\cite{Erwin}.)
}
\end{figure}

\begin{figure}[t]
\centering
\epsfig{file=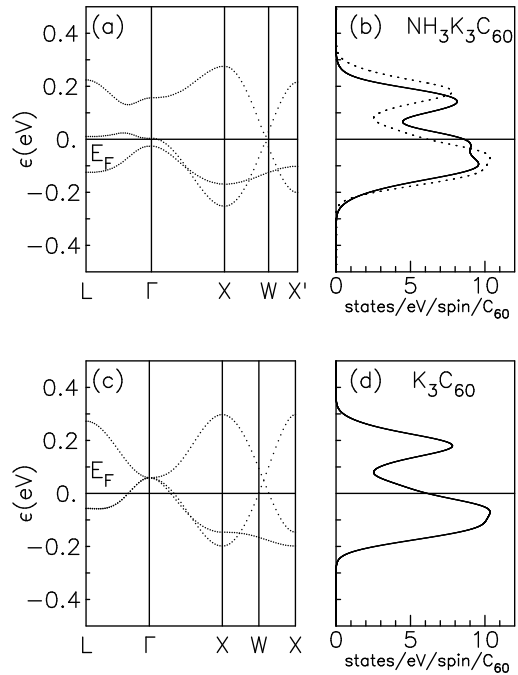,height=9cm}
\caption{\label{bands_NH3:fig}
LUMO band structure of NH$_3$\,K$_3$C$_{60}$ (a), obtained by an DFT-LDA
calculation, compared to that of K$_3$C$_{60}$ (c).
The corresponding densities of states are given in panels (b) and (d).  In
panel (b) the density of states of K$_3$C$_{60}$ (dashed line) is re-drawn
for direct comparison with that of NH$_3$K$_3$C$_{60}$ (solid line).
(From Ref.~\cite{splitDMFT}.)
}
\end{figure}

After this long preliminary on the properties of molecular ions
we now come to the solid state compounds, that constitute our main 
concern.
The starting point for the discussion of their electronic behavior is 
as always the standard electron band structure.
The band structure of solid alkali fullerides has been computed by many
authors \cite{Erwin,Erwin94,Erwin91,splitDMFT,Gunnarsson98}, generally based
on DFT-LDA methods.
As illustrated in Figs.~\ref{bands:fig} and \ref{bands_NH3:fig}, the
bandwidth $W$ of the LUMO band is generally obtained in the $500\div
600$~meV region.
Differences for different lattice structures are of course present, as well
as a strong dependence upon the relative rotational ordering of the
C$_{60}$ balls \cite{Shirley93}.
In particular, smaller bandwidths are obtained in correspondence of larger
lattice spacings, associated, e.g., to larger intercalated cations such as
Cs$^+$ in the place of K$^+$.
The main observation is that the electron bandwidths in the solid are
generally a fraction of the separation between neighboring molecular
levels, the latter giving rise to well defined rigid bands.  In particular, 
the threefold LUMO band remains well separated from
the LUMO+1 and HOMO bands through the Brillouin zone.
The metal-insulator transition in the C$_{60}^{n-}$-based compounds can
then be interpreted in terms of the localization of the electrons within
the partly filled $t_{1u}$ LUMO band.
Self-energy effects are predicted to increase $W$ by about 30\% in the
usually more accurate GW approximation over that calculated by LDA
\cite{Shirley93}.
%

A direct experimental determination of the bands in solid C$_{60}$
compounds is complicated by characteristic strong electron-phonon
satellites derived from the vibronic couplings of Sec.~\ref{vibronic:sec},
which tend to broaden and influence all spectra.
Angle-resolved photoemission, which could access direcly the band
dispersion is made additionally hard by the orientational disorder and the
small size of the Brillouin zone.
A recent experiment \cite{WangBrouet03} accessed directly the
dispersion in a K$_3$C$_{60}$ monolayer, obtaining an estimate $W\approx
250$~meV, significantly smaller than the DFT-LDA bandwidth of the
monolayer.
Newer experiments on K$_3$C$_{60}$ multilayers find an even smaller
$W\approx 150$~meV, about one third of the DFT bandwidth
\cite{GoldoniInSerenaBook}.
These small values do not deny the DFT estimates but most likely represent
a quasiparticle effective dispersion, whose magnitude relative to the bare
band dispersion is strongly affected, and nontrivially reduced by
correlations.
It should be noted that, even if there is so far no accurate prediction for
the full $\vec k$-dependent spectral function, this kind of observed
electron dispersion is not to be identified with the simple quasiparticle
bandwidth $zW$ either.  The effective observed bandwidth probably
corresponds to the intermediate energy scale discussed in recent DMFT
studies of a simplified model, where it is designated as $T_+$
\cite{DeLeo05,Capone04,FabrizioPriv}.
Indeed, the same angle-resolved photoemission measurement carried out 
on K$_6$C$_{60}$, where correlations
play no role, find \cite{GoldoniInSerenaBook} a band dispersion in very
good quantitative agreement with the DFT calculation of
Fig.~\ref{bands:fig} ($W\simeq 0.6$~eV) \cite{Erwin91}.

In the light of the discussion of Sec.~\ref{coulomb:sec}, it is clear that
the largest energy scale of for the LUMO-band electrons in the fullerides
is the Coulomb repulsion $U$.
It remains the largest energy scale, even with the extra screening
characteristic of the solid state, which appears to reduce the molecular
$U$ from about $2.5\div 3$~eV to a smaller $0.9\div 1.6$~eV
\cite{Antropov92,Lof,Pederson92,Bruhwiler93,Gunnarsson97}.
The smallest value in this range is probably closer to the correct bulk
estimate in the metallic doped materials, while the upper value better
characterizes insulating states and the molecules at the surface, where
screening may be less effective due to reduced coordination
\cite{GunnarssonBook,Antropov92}.
The ratio of the local Coulomb repulsion to the bandwidth can therefore be
estimated in the range $1.5 \leq U/W \leq 3$ for typical LUMO-band
electrons in standard fullerides.

\begin{figure}[t]
\centering
\epsfig{file=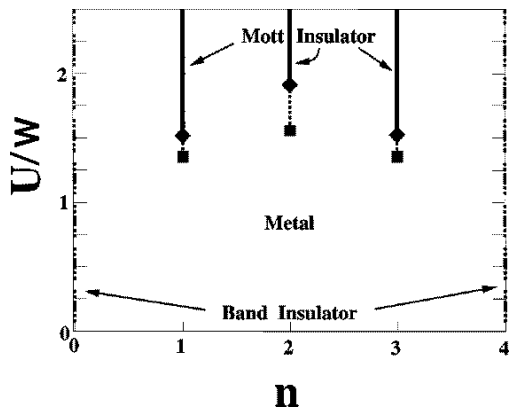,scale=.333333333}
\caption{\label{deg_phased}
DMFT zero-temperature phase diagram for the two-band Hubbard model as a
function of filling and of the ratio $U/W$ of the local Coulomb repulsion
to the half width.
In this calculation molecular Coulomb exchange $J=0$, and the possibilities
of magnetism and/or charge order are ignored.
The $T=0$ Mott transitions at integer filling are located
between the pairs of finite-temperature estimates indicated by squares and
diamonds.
(Adapted from Ref.~\cite{Rozenberg97}.)
}
\end{figure}

A local Coulomb repulsion larger than the bandwidth is the signature of
strong electron correlation in a solid.
Situations of this kind are conventionally described in the language of
Hubbard models.
The single-band half-filled 3D Hubbard model is believed to undergo a
metal-insulator transition of the Mott type, as soon as the ratio $U/W$
exceeds some critical $U_{\rm c}/W$ of order unity \cite{Noack99}.
The conventional small-$U$ metal transforms to a large-$U$ Mott-insulator
where electrons remain essentially frozen, one at each site, so that
the occurrence of zero and double occupancies is strongly even if
not totally suppressed.
The residual effect of intersite hopping is to induce a generally
antiferromagnetic correlation between the spins of electron at neighboring
sites.
Away from half filling the model has a metallic (and possibly
superconducting) ground state even for large $U/W$.

The single-band Hubbard model, however, can hardly be applied to the
fullerides, where there are $d=3$ relevant bands, derived from the three
degenerate molecular LUMO orbitals.
The rough physics of the multi-band Hubbard model is summarized in the
phase diagram of Fig.~\ref{deg_phased}, for the simpler case of only $d=2$
bands.
At integer-filling, lines of strongly correlated Mott insulators extending
above $U_{\rm c}/W$ are surrounded by the metallic phase encompassing all
noninteger fillings.
The same qualitative picture should hold for $d=3$, where metal-insulator
transitions should occur for stoichiometric phases, in particular $n=3$ (as
in K$_3$C$_{60}$) and $n=4$ (as in K$_4$C$_{60}$).

\begin{figure}[t]
\centering
\epsfig{file=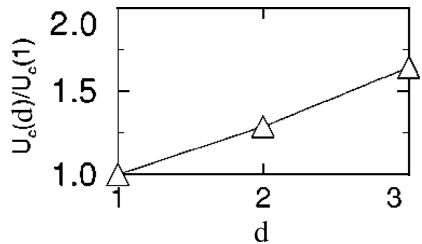,scale=.16}
\caption{\label{Han98:fig}
The critical $U_{\rm c}/W$ as a function of orbital degeneracy $d$ at half
filling $n=d$.
For $d=1$ band $U_{\rm c}^{\rm 1-band} \approx 1.46\,W$.
In these DMFT calculations the local Coulomb exchange is neglected ($J=0$).
The possibility of magnetic and/or charge and/or superconductive order
is also ignored.
(Adapted from Ref.~\cite{Han98}.)
}
\end{figure}

Recently, the metal-insulator transition has been attacked theoretically
within the dynamical mean field theory (DMFT) \cite{Georges96}.  This
method, which becomes exact in the limit of large space dimensions, has
provided quantitative estimates of the $U_{\rm c}/W$ under various
conditions.
The lines of Fig.~\ref{deg_phased} are constructed based on the DMFT
\cite{Rozenberg97}, assuming a ``pure'' Mott transition, with no
complications induced by possible spin or/and charge order in the
insulating phase.
The ``paramagnetic'' Mott-Hubbard transition occurs at different critical
$U$ for different orbital degeneracy $d$ (number of bands) and fillings $n$.
Earlier work showed that the Mott transition in orbitally $d$-degenerate
lattice models takes place at larger values $U_{\rm c}/W$ for larger
degeneracy $d$, roughly proportionally to $\sqrt{d}$
\cite{Gunnarsson96,Han98,Koch99}, as illustrated in Fig.~\ref{Han98:fig}.
Specifically, the Mott transition in the half-filled one-band model ($d=1$,
$n=1$) has been calculated at $U_{\rm c}^{\rm 1-band}/W \approx 1.3$ in a
previous DMFT study based on the the quantum Monte Carlo method at the
relatively low temperature $T=W/32$ \cite{Rozenberg94}.
A more accurate $T=0$ estimate of $U_{\rm c}^{\rm 1-band}/W \approx 1.46$
is provided by iterated perturbation theory \cite{Georges96}.
The transition was found at $U_{\rm c}^{\rm 2-band}/W \approx 1.5$ for
$d=2$, $n$=1 \cite{Rozenberg97}, and $U_{\rm c}^{\rm 2-band}/W \approx 1.8$
for $d=2$, $n$=2 \cite{Rozenberg97,Han98}, with the same method, and at the
same temperature.
$T=0$ Lanczos diagonalization DMFT calculations push these values
slightly up, with the $n$=1 value close to $U_{\rm c}^{\rm 2-band}/W \approx
1.8$ \cite{splitDMFT}.
For $d=3$, $n$=3, the transition was located close to $U_{\rm c}^{\rm
3-band}/W \approx 2.3$ \cite{Han98}, although at a slightly higher
temperature.
The finite temperature 
appears to affect somewhat the numerical values:
the zero-temperature DMFT calculations based on Lanczos diagonalization
provide slightly larger values \cite{Caffarel94}, but confirm that
$U_{\rm c}^{\rm 3-band}>U_{\rm c}^{\rm 2-band}>U_{\rm c}^{\rm 1-band}$
\cite{splitDMFT}.

These values of $U_{\rm c}/W$ are significantly reduced when more
realistic DMFT calculations allow for intra-site couplings and the ensuing
magnetically/charge ordered phases, and when any kind of local exchange
term is included to break the local multiplet degeneracy.
Therefore one cannot compare directly the values of $U_{\rm c}/W$ from
paramagnetic DMFT calculations of the multi-band Hubbard model with the
metal-insulator transition in the actual compound, where it is affected by
the detailed band structure, nesting,
JT electron-vibration interaction and intra-site exchange.
Most of these effects tend to lower $U_c$.

In fact, the range $1.5 \leq U/W \leq 3$ typical of the alkali fullerides
should put all these materials above the metal-insulator transition: all
integer-filled compounds should then be insulators.
%
%
%
%
In practice, the alkali fullerides lie experimentally close to the Mott
transition, on both sides.
Different compounds exibit a variety of behavior, including unconventional
metals like cubic CsC$_{60}$ \cite{Brouet99}, superconductors of the
A$_3$C$_{60}$ family (A= K,Rb,Cs) \cite{Ramirez,Gunnarsson97} including
also NH$_3$Na$_2$CsC$_{60}$ \cite{Zhou93}, a ferromagnet -- TDAE-C$_{60}$
\cite{Allemand,Denisov,Schilder94,Arovas95,Blinc96,Blinc98}, and Mott
insulators.
Among the latter we can place A$_4$C$_{60}$ \cite{Benning}, possibly
Na$_2$C$_{60}$ \cite{Brouet01}, (NH$_3$)$_6$Li$_3$C$_{60}$ \cite{Durand03},
and the full class of compounds NH$_3$\,K$_3$C$_{60}$,
NH$_3$\,K$_2$RbC$_{60}$, NH$_3$\,KRb$_2$C$_{60}$, and
NH$_3$\,Rb$_3$C$_{60}$ \cite{Rosseinsky93,Takenobu00,Obu00}.
The fact that the $n=2,4$ compounds are insulators is not expecially
surprising in view of the increase in repulsive pair energy due to
electron-phonon contribution (Sec.~\ref{vibronic:sec}), and the probably
smaller $U_{\rm c}^{\rm 3-band} (n=2)<U_{\rm c}^{\rm 3-band} (n=3)$.
%
Altogether, this scenario indicates a reduction of $U$ in solid fullerides,
and a lowering of $U_c$ relative to its bare theoretical value of
Figs.~\ref{deg_phased} and \ref{Han98:fig}.

Magnetic susceptibility data \cite{Durand03,Robert98} in particular place
the metallic/superconducting C$_{60}^{3-}$ compounds close to the
metal-insulator transition.
Indeed, in the Hubbard model, the strongly-correlated metallic state near
$U_{\rm c}$ is expected to show an anomalously large uniform magnetic
susceptibility $\chi$ compared to the free-electron value
$\mu_{\rm B}^2 N(\epsilon_{\rm F})$
(where $N(\epsilon_{\rm F})$ is the density of states per C$_{60}$ at the
Fermi energy $\epsilon_{\rm F}$) \cite{Georges96}.
The magnetic susceptibility is observed to increase rapidly with
the lattice spacing in the C$_{60}^{3-}$ fullerides
\cite{Durand03,Robert98}.
When moving closer to the metal-insulator transition, the observed increase
cannot be interpreted purely in terms of band narrowing and a related
increase in the density of states at the Fermi level $N(\epsilon_{\rm
F})$.
The discrepancy between this increase due to bare band narrowing and
observation is huge, and can only by explained by strong correlations.

Further, if less direct, evidence of the strong electronic correlation in
K$_3$C$_{60}$ is provided by the electron spectral function at $E_{\rm F}$
which can be assimilated to a narrow, dispersionless Kondo-like resonance
by photoemission data \cite{Goldoni01}.
On the other hand, photoemission is on one hand strongly affected by
vibrational effects and also possibly more sensitive to surface molecules
\cite{Schiessling05}, where correlation is stronger due to a less screened
Hubbard $U$, thus caution is required before photoemission data should be
taken as representative of bulk K$_3$C$_{60}$ as magnetic susceptibility
data can.

\begin{figure}[t]
\centering
\epsfig{file=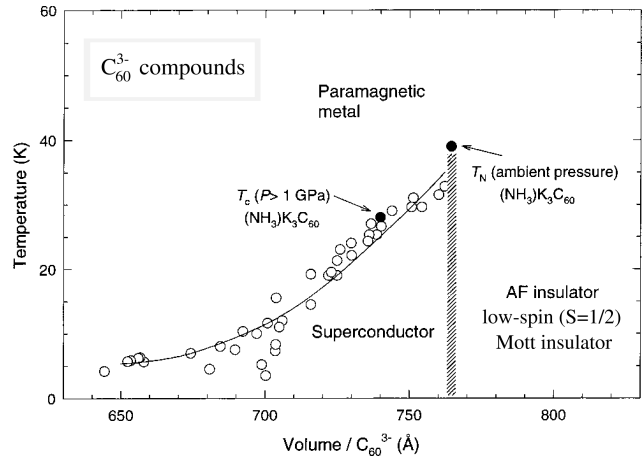,scale=1}
\caption{\label{Prassides99:fig}
Schematic electronic phase diagram of C$_{60}^{3-}$ compounds, showing the
approximate location of the metal (superconductor)-insulator phase
boundary.
The open symbols are literature values of $T_c$ for a variety of
superconducting fullerides, while the solid symbols mark $T_N$ (ambient
pressure) and $T_c$ ($>1$~GPa) of ND$_3$\,K$_3$C$_{60}$.
This expanded trivalent fulleride is only metallic and superconducting
under pressure, and reverts to an antiferromagnetic insulator at zero pressure,
in strong analogy with the cuprates (exchanging pressure with doping).
Moreover, the C$_{60}^{3-}$ is in a $S\!=\!1/2$, low-spin state, and not in
$S\!=\!3/2$ high-spin state.
This indicates an effective inversion of the effective Hund's rule $J$,
supporting the conclusion that ND$_3$\,K$_3$C$_{60}$ and related compounds
are Mott-JT insulators.
(Adapted from Ref.~\cite{Prassides99}.)
}
\end{figure}

In the ammoniated compounds, insertion of the electronically inert NH$_3$
molec\-ules expands the C$_{60}$ lattice, turning the cubic, metallic, and
superconducting state of K$_3$C$_{60}$ into an orthorhombic narrow-gap
$S=1/2$ antiferromagnetic insulator \cite{Rosseinsky93,Takenobu00,Tou1}.
%
Upon application of pressure to NH$_3$\,K$_3$C$_{60}$, the Mott insulating
state reverts to a fully metallic and superconducting -- while still
orthorhombic -- phase \cite{Zhou95,Margadonna01}.
A similar pressure-induced reversion from insulator to metal has been
observed in tetragonal Rb$_4$C$_{60}$ \cite{Kerkoud}: these experiments
rule out the possibility that metallic states of the fullerides should
entirely be attributed to nonstochiometry.
These transitions in the alkali fullerides are rare examples of
experimentally accessible Mott transitions.
We will take that of NH$_3$\,K$_3$C$_{60}$ as a paradigm where several
of the concepts illustrated above emerge most clearly.
It was suggested \cite{Prassides99} that the increase in volume per
C$_{60}$ molecule relative to K$_3$C$_{60}$, with its probable slight
decrease of electronic effective bandwidth $W$, should drive the Mott
transition (Fig.~\ref{Prassides99:fig}).

While the above is certainly relevant, in addition to the simple change in
volume, thus of $W$, a role of the crystal-structure anisotropy in this
pressure-driven transition from the insulator to the metal and
superconductor has been proposed \cite{Takenobu00}.
Crystal anisotropy affects hopping between neighboring fullerene balls,
making it different for pairs of molecules at different distances, along
different crystal directions, thus affecting the band structure.
The effect of anisotropy is most clearly seen in the the band splitting at
$\Gamma$, illustrated in Fig.~\ref{bands_NH3:fig}.
A strong enough ``crystal-field'' splitting of the threefold degenerate
$t_{1u}$ molecular orbital of C$_{60}$ caused by orthorhombic lattice
anisotropy could remove one or several bands away from the Fermi level,
effectively reducing the orbital degeneracy $d$, and shifting the Mott
transition to a smaller critical value $U_{\rm c}/W$.
Exploring this concept, the basic question is how large a splitting is
required to promote the effective reduction of degeneracy.
In a non-interacting system, that reduction would clearly require a 
splitting magnitude at least similar to the electron bandwidth $W$.
In a strongly-interacting system, it is important to understand whether the
effective degeneracy lifting, and the associated substantial displacement
of the metal-insulator transition, will again require anisotropic
splittings as large as the bandwidth, or else if some smaller energy scale
will emerge in its place.

In the final sections in this chapter we will adddress this issue within
DMFT, by studying the effects of a band splitting on the Mott transition of
an orbitally degenerate, strongly correlated metal.
The theory for the distorted ammoniated compounds underscores the central
role of a new low-energy scale characteristic of strongly correlated
itinerant electrons close to the Mott transition: the width $z\,W$ of the
Kondo peak.
We will argue about the importance of this concept in the context of both
C$_{60}^{3-}$ and C$_{60}^{4-}$ fullerides, and of strongly correlated
materials in general.

\section{Strong correlations in fullerides: theoretical models}
\label{models}

Models of alkali fullerides should be able to describe metals,  Mott
insulators, and superconductors.
The largest energy parameter being the intra-molecular repulsion $U$,
with the electron hopping only second, ab initio approaches are not
yet in adequate shape to meet this challenge at the present date.

Early work attempted to interpret the superconductivity of the fullerides
in terms of Migdal-Eliashberg theory
\cite{AMT,MTA,Schl,vzr,Mazin92,lannoo,Deaven93,Rice94,Gunnarsson92,Antropov}.
The necessity to introduce a comparably large value of the pseudopotential
$\mu^* \approx 0.4$ already suggests that electronic correlations are
exceedingly important.
The large zero-point vibrational energy $\frac 12 \hbar \omega$ associated
to the vibrationally-derived optical modes of these solids also suggests a
breakdown of Migdal's approximation, and that vertex corrections and other
nonadiabatic effects could play a relevant role in the fullerides
\cite{Pietronero92,Pietronero,Cappelluti00,Cappelluti01,Botti02,Paci04}.
However, this kind of theory starts from free electrons in a regular band
metals, hard to accept for the fullerides whose large value of $U$ puts
them close to (if not beyond) the Mott-Hubbard transition.
As anticipated above, for a satisfactory description of the large local
Coulomb correlation and of the band physics on the same footing, a
multi-band Hubbard model, possibly including the JT phonons, and described
within the DMFT \cite{Georges96} or some of its several evolutions
\cite{Lichtenstein01,Parcollet04,Capone04CDMFT} is certainly a more natural
choice.

The simplest $d$-band Hubbard model is written as
\begin{equation} \label{hamiltonian:eq}
\hat{H} = \hat{H}_{\rm hop} + \hat{H}^{\rm e-e}\, .
\end{equation}
Here we assume purely diagonal hoppings between orbitals at nearest
neighbor sites $i$ and $j$:
\begin{equation} \label{Hkin}
\hat{H}_{\rm hop} = - t \sum_{\langle i,j\rangle,\sigma} \sum_{m=1}^d  
\left(\hat{c}^\dagger_{im\sigma} \hat{c}_{jm\sigma} + H.c.\ \right)
+\sum_{i\,m}  \epsilon_m ~ \hat{n}_{im}\,,
\end{equation}
with $d$ bands of the same width $W$, the anisotropic symmetry lowering
entirely embodied in a local diagonal term splitting the on-site orbital
energy $\epsilon_m$.
Of course, these simplifying assumption are not strictly applicable to the
fullerides, where a more complicated $3\times 3$ matrix of overlaps for the
$t_{1u}$ LUMO orbitals at neighboring sites (possibly including merohedral
disorder) should be (and has been) employed
\cite{GunnarssonBook,Han99}.
The essence of the Mott transition in the fullerides should most likely not
be affected qualitatively by the details of the hoppings, but the
quantitative previsions of a model ignoring the correct tight-binding
hopping matrix, including effects of orientational order or merohedral
disorder, should not be taken too literally.

The Coulomb part $\hat{H}^{\rm e-e} = \sum_i \hat{H}_{{\rm e-e}\,i}$, with
the local term defined in Eq.~(\ref{He-ecombined}) for $d=3$.
Another equivalent formulation for the local term goes as follows:
\begin{eqnarray} \label{HUJ:eq}
\hat{H}^{\rm e-e}_i&=&
  (U+J) \sum_{m} \hat{n}_{im\up} \hat{n}_{im\down} 
+ (U-J) \sum_{m<m'\,\sigma} \hat{n}_{im\sigma} \hat{n}_{im'\sigma} 
\\\nonumber
&+& U \sum_{m\neq m'} \hat{n}_{im\up} \hat{n}_{im'\down} 
+ J \sum_{m\neq m'}
	\hat{c}^\dagger_{im'\up} \hat{c}^\dagger_{im\down}
	\hat{c}_{im'\down} \hat{c}_{im\up}\,,
\end{eqnarray}
with the advantage that it applies equally well for any degeneracy
$d=1,2,3$ \cite{Han98}.

The on-site electron-vibration couplings Eq.~(\ref{Hev:eq}) and the ensuing JT
effect, are quite important for many aspects, including superconductivity
and resistivity \cite{GunnarssonBook,MTA,Gunnarsson97,Capone01}, as
discussed in Sec.~\ref{other:sec} below.
While we included it in the single-ion model of Eq.~(\ref{hamiltonian:eq}),
the initial simplification of neglecting the JT coupling in the solid is
useful for a simple approach to the basic physics of the Mott transition.

\subsection{The bare $d=2$ model with anisotropy}
\label{dmft:sec}

For an initial study of the anisotropy effect on the Mott transition, one
can start with the simplest orbitally degenerate Hubbard model, namely
$d=2$ bands, zero JT coupling, zero Hund's rule exchange.
For the anisotropy term we can assume, without loss of generality,
$\epsilon_2=-\epsilon_1=\Delta/2$.
We choose to study filling $n=1$ electron/site: this choice is motivated by
the observation that for $d=2$, half filling $n=d=2$ would yield a trivial
phase diagram, not comparable to the realistic case $n=d=3$.
We study this model at zero temperature, where the Mott transition appears
most clearly, as a function of $\Delta/W$ and $U/W$.

In the $U=0$ limit, the splitting $\Delta>0$ simply shifts band 2 upwards 
with respect to band 1, promoting electron transfer 
from the upper to the lower band.
Above a critical value $\Delta=\Delta_c$, the upper band will be emptied.
For example, with two symmetric bands, above $\Delta_c/W=0.5$ the upper
band is emptied and the lower band remains half filled.
At $U=0$, the transition between the ``two-band metal'' and
``one-band metal'' is continuous.
Because the topology of the Fermi surface changes, this transition is
accompanied by a weak zero-temperature singularity of the total energy first described by
Lifshitz \cite{Lifshitz60}.
When the electron-electron interaction $U$ is turned on, one expects the
emptying of the upper band to take place at smaller values of $\Delta_c$,
owing to the effective band narrowing.
Perturbatively in $U$ one can show that at weak coupling
\begin{equation}
\Delta_c(U) = \Delta_c(0) - \gamma\; U + O(U^2) \,,
\label{smallU}
\end{equation}
where the value of the coefficient $\gamma>0$ depends on details of the
bands.
In addition, electron-electron interactions might modify the nature of the
metal-metal transition singularity relative to the noninteracting case
\cite{Katsnelson00}, a point which we will not further address here.

The $\Delta=0$ and $\Delta\to\infty$ limits reduce then, respectively, to the
(quarter-filled) two-band and (half-filled) single-band Hubbard models,
both possessing a metal-insulator transition as a function of $U$
\cite{Rozenberg97,Rozenberg94,Caffarel94,Bulla99}.
We are not interested here in the weak-coupling antiferromagnetic
instability of the ensuing large-$\Delta$ half-filled band, associated with
the possible presence or absence of nesting characteristic of a specific 
assumed intersite hopping scheme.
Consistently with our neglect of all intrasite multiplet interactions, we 
also ignore for the moment the possibility of charge ordered and/or
superconducting phases. In particular, we leave the local exchange term 
out ($J=0$), with the proviso that this term will later be 
crucial in order to understand the ordered
phases which enrich the phase diagram of the model
\cite{Capone04,Capone01,Capone00,Capone02}.
We assume a genuine Mott transition for the half-filled single-band model
to occur at a finite $U_{\rm c}^{\rm 1-band}>0$ (for $\Delta\to\infty$) and
$U_{\rm c}^{\rm 2-band}>U_{\rm c}^{\rm 1-band}$ (for $\Delta=0$).

The limit of strong interaction, $U\gg W$, is insulating for any value of
$\Delta$.
This limit can be studied by mapping the model (\ref{hamiltonian:eq}) onto
a spin and orbital exchange Hamiltonian which reads \cite{Arovas95}
\begin{equation}
H_{\rm exch} = J_t \sum_{\langle i,j\rangle} \left(
{\bm S}_i\cdot {\bm S}_j +
{\bm T}_i\cdot {\bm T}_j +
4\; {\bm S}_i\cdot {\bm S}_j \ \ {\bm T}_i\cdot {\bm T}_j 
\right)
-\Delta \sum_i T^z_i \;,
\label{Hexch}
\end{equation}
where the true spin operator is ${\bm S}_j=\frac 12 \sum_{m
\nu\nu'} c^\dagger_{j m \nu}\, {\bm \sigma}_{\nu\nu'}\, c_{j m \nu'}$ and 
the pseudospin-1/2 vector operators ${\bm T}_j=\frac 12 \sum_{m
m'\nu} c^\dagger_{j m \nu}\, {\bm \sigma}_{m m'}\, c_{j m' \nu}$ represent
the orbital degrees of freedom, $\bm \sigma$ being the vector of Pauli
matrices, and $J_t=2t^2/U$.
For $\Delta=0$ this model has been studied both in one \cite{Sutherland75}
and two dimensions \cite{Zhang98}, with suggestions that interesting
spin-liquid physics could be realized.
For our purposes, it suffices to note that this model has no ferro-orbital
instability, and has therefore a finite $q=0$ orbital susceptibility.
As a consequence it takes a nonzero value of $\Delta$ to 
fully orbitally polarize the ground state.
Due to the absence of cross-band terms in the kinetic energy, complete
orbital polarization occurs at a finite $\Delta_c\propto J_t=2t^2/U$.
For $\Delta\ge \Delta_c$ the ground state is thus represented 
by a one-band Mott insulator plus a totally empty split-off band.

For the specific case we are interested in ($n=1$ electron in a two-fold
degenerate band) no orbital ordering is present at weak coupling.
On the other hand, the possibility of antiferro-orbital and/or spin ordering
within the Mott insulating phase, and of spin-orbital density waves in the
intermediate $U$ regime, depends crucially on the details of the various 
hopping matrix elements and of lattice coordination.
These are left out of the infinite-dimensional
lattice assumed in the calculations  to be described in the following section.
Therefore, within the standard scheme of single-site DMFT calculations, we can
study the phase diagram of this model restricted to spin and orbital 
paramagnetic states.

\begin{figure}[t]
\centering
\epsfig{file=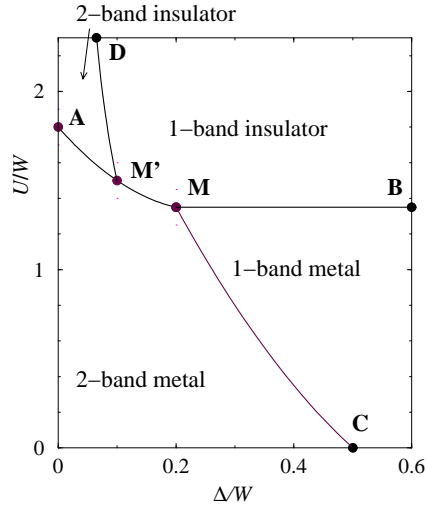,height=7cm}
\caption{\label{phased_th:fig}
Qualitative zero-temperature phase diagram for the two-band Hubbard model
at quarter filling (one electron per site) in the $U$-$\Delta$ plane, 
where $\Delta$ is the anisotropy splitting of the two orbitals.
The various phases and lines are described in the text.
The multicritical points M and M' are not necessarily distinct.
(From Ref.~\cite{splitDMFT}.)
}
\end{figure}

Figure~\ref{phased_th:fig} is a sketch of the zero-temperature phase
diagram of the $d=2$ model in Eq.~(\ref{hamiltonian:eq}) as a function of
$(U,\Delta)$ for $n=1$ electron per site.
The transitions at $\Delta=0$ and $\Delta=\infty$ are located at the values
of $U/W \simeq 1.8$ and $1.5$ as discussed in Sec.~\ref{correlations:sec}.
The descending AB line in Fig.~\ref{phased_th:fig}, representing $U_{\rm
c}$ as a function of $\Delta$ separating metals from Mott insulators,
indicating that $U_{\rm c}^{\rm 2-band}>U_{\rm c}^{\rm 1-band}$.
The DM' line separates the fully orbitally polarized Mott insulator (on the
right) from the two-band insulator, roughly as $U_{\rm c}\sim\Delta^{-1}$
for small $\Delta$.
Similarly, the CM line separates the fully orbitally polarized 
metal from the two-band metal: it starts from point C with a linear slope
$-\gamma^{-1}$, according to Eq.~(\ref{smallU}).
In the region of full orbital polarization the value of $\Delta$ is
irrelevant, and this is the reason why the Mott-transition line MB is
horizontal.
As $\Delta$ increases, for $U<U_{\rm c}^{\rm 1-band}$ the upper-band emptying 
transition takes the two-band metal across the CM line over to a one-band 
metal, while for $U>U_{\rm c}^{\rm 1-band}$ it leads across the AM line to a
Mott insulating state.
The steeply dropping AM line is the main outcome of the calculation,
showing that the two-band Mott transition is heavily affected already at
small $\Delta \ll W$, and not at $\Delta \sim W$, as one might have
expected.
As the DMFT calculations below show, the effective emptying transition
occurs when $\Delta$ increases to reach $\Delta_c(U)\propto z\,W\propto
(U_{\rm c}^{\rm 2-band}-U)$.

\begin{figure}[t]
\centering
\epsfig{file=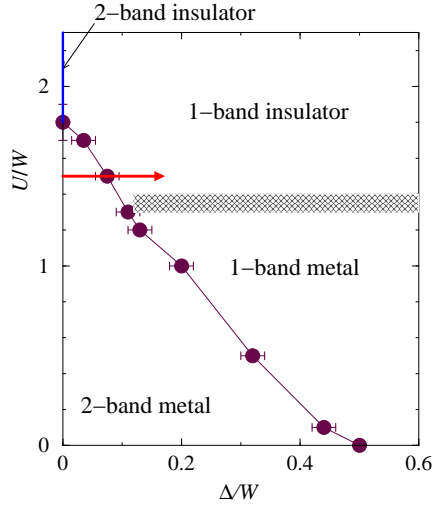,height=7cm}
\caption{\label{phase_diagram:fig}
DMFT zero-temperature phase diagram for the two-band Hub\-bard model
at quarter filling (one electron per site), obtained by the exact 
diagonalization method in the paramagnetic sector. 
%
(From Ref.~\cite{splitDMFT}.)
}
\end{figure}

Further quantitative informations concerning this phase diagram (in
particular in the region of intermediate $U\sim W$ and $\Delta<W$) were
obtained \cite{splitDMFT} by means of DMFT in the exact diagonalization
flavor \cite{Georges96,Caffarel94} on the Bethe lattice.
The results of the DMFT calculations are summarized in 
Fig.~\ref{phase_diagram:fig}.
The DMFT critical $U_{\rm c}^{\rm 1-band}\simeq 1.35\,W$ and $U_{\rm
c}^{\rm 2-band}\simeq 1.8\,W$ are in fair agreement with
corresponding values obtained by other methods
\cite{Rozenberg97,Rozenberg94,Caffarel94,Bulla99}.
The other points in the phase diagram are obtained by following the
stability of the two-band metal for a given value of $U$ and increasing
$\Delta$, marking the emptying transition to the one-band metal or to the
insulator.

A deficiency of this single-site DMFT calculation -- which is restricted as
usual to paramagnetic states only -- is the absence of a two-band
insulating state for $\Delta>0$ (i.e.\ the M' point coincides with A).
In fact, the suppression of the antiferro orbital fluctuations embodied in
the exchange model (\ref{Hexch}) produces a fictitious infinite uniform 
orbital susceptibility which leads to full orbital polarization as soon 
as $\Delta$ is turned on \cite{Georges96}. 
Despite this limitation, the results of the DMFT calculations are
suggestive, revealing the announced sharp reduction of the metal-insulator
$U_{\rm c}(\Delta)$ for small but finite $\Delta$.
Indeed, the DMFT yields a $\Delta_c(U)$ which is roughly proportional to the
quasiparticle residue $z(U)$ of the undistorted ($\Delta=0$) correlated
metal at $U<U_{\rm c}^{\rm 2-band}$,
\begin{equation}
\frac{\Delta_c(U)}{W} \propto \beta z(U) \;, 
\end{equation}
with a proportionality constant $\beta\simeq 0.3$. 
Since $z(U)$ vanishes as $U\to U_{\rm c}^{\rm 2-band}$, most likely
linearly in $(U_{\rm c}^{\rm 2-band}-U)$ \cite{Georges96}, even a
small $\Delta\ll W$ is sufficient to cause a metal-insulator transition in
the strongly correlated metal.
For example, following the bold arrow at $U=1.5\,W$ in 
Fig.~\ref{phase_diagram:fig}, a $\Delta$ value as small as $0.08\, W$ is 
sufficient to cross the transition line from the metal to insulator.

\begin{figure}[t]
\centerline{\epsfig{file=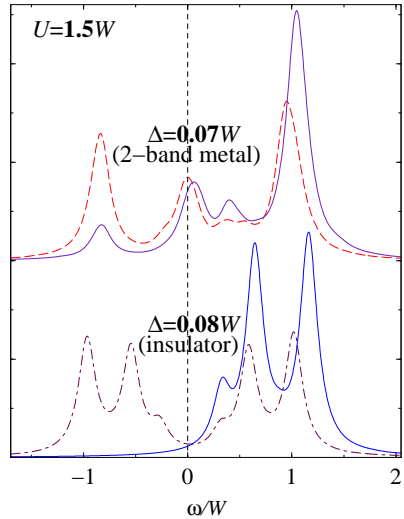,height=7cm}}
\caption{\label{spec_dens:fig}
Spectral density $A_m(\omega)=-\pi^{-1} {\rm Im}\,G_m(\omega)$ 
at $U/W=1.5$ across the Mott-Hub\-bard transition for increasing
anisotropy splitting $\Delta$.
$\omega$ is referred to the Fermi energy.
Solid lines refer to the minority orbital $m=2$.
The multi-peak structures of the high-energy side bands are artifacts of
the finite discretization of the conduction band.
(From Ref.~\cite{splitDMFT}.)
}
\end{figure}

Illustrating further the transition, Fig.~\ref{spec_dens:fig} shows the
behavior of the spectral density
$$ A_m(\omega)=-\pi^{-1}\;{\rm Im}\, G_m(\omega)  \;, $$
$G_m(\omega)$ being the one-particle Green's function of band $m$, on both
sides of the metal-insulator transition.
The asymmetry of the upper-band spectral density $A_2(\omega)$ (solid lines) 
is very pronounced, as this band is nearly ($\Delta/W=0.07$) or
completely ($\Delta/W=0.08$) empty.
As soon as the Kondo-like peaks of the two bands differ enough in energy 
to induce the emptying of band 2, the lower-band spectral density 
$A_1(\omega)$ takes the symmetric shape, characteristic of the half-filled 
one-band Hubbard model.
Here the quasiparticle peak disappears completely, as this value of $U>U_{\rm
c}^{\rm 1-band}$ puts the Hubbard model of band 1 well inside the
insulating regime.
%

\subsection{Noncubic band splitting and correlations in NH$_3\,$K$_3$C$_{60}$}
\label{bands:sec}

The above $d=2$ model calculations show that a small splitting $\Delta
\propto z\,W$ of the orbitally degenerate band can drive the
metal-insulator transition.
We wish to explore the implications that this result -- if assumed to
be more general than the simple model where it was derived -- can have on
the metal-insulator transition which takes place between isoelectronic
K$_3$C$_{60}$ and NH$_3$\,K$_3$C$_{60}$ (the former cubic and the latter
orthorhombic) and on the insulator-metal transition of
NH$_3$\,K$_3$C$_{60}$ itself under pressure.

\begin{figure}[t]
\centering
\epsfig{file=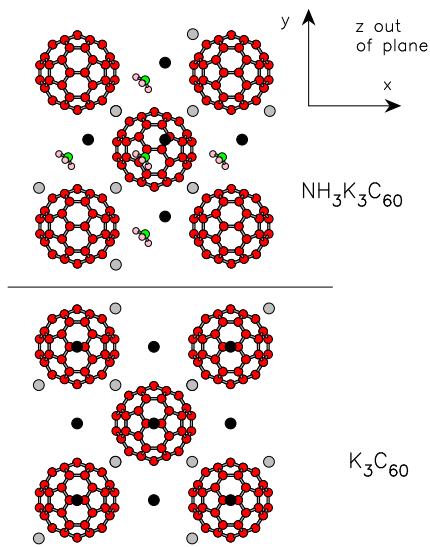,height=8cm}
\caption{\label{geometry_NH3:fig}
Simplified geometry of NH$_3$\,K$_3$C$_{60}$ (top) and K$_3$C$_{60}$
(bottom) used in the DFT-LDA calculation. Octahedral K atoms are indicated
in black, while tetrahedral ones are in gray.
All visible atoms of the central unit cell are shown together with some
atoms in the neighbor unit cells.  In NH$_3$\,K$_3$C$_{60}$ the $c$-axis is
in the $z$ direction.
(From Ref.~\cite{splitDMFT}.)
}
\end{figure}

The effect of ammoniation is apparently twofold. The first effect is a
large volume 
expansion, implying some band narrowing as well as an increase of $U$ 
due to reduced screening. The second effect is a breaking of cubic 
symmetry in the anisotropic lattice structure. The actual strength of
anisotropy in the ammoniated fulleride may be estimated quantitatively from the
DFT-LDA bands of Fig.~\ref{bands_NH3:fig}.
They were obtained in a simplified geometry with a single C$_{60}$ unit
cell with lattice constant $a=14.2$~\AA\ for fcc K$_3$C$_{60}$, and
$a=14.89$~\AA\ for NH$_3$\,K$_3$C$_{60}$ the latter with a centered
tetragonal unit cell with $c/a=0.91$, neglecting the exceedingly small
difference between $a$ and $b$.
Merohedral disorder and the rich antiferro-rotational structure recently
discovered in actual NH$_3$\,K$_3$C$_{60}$ \cite{margadonna} are also
ignored.
The bands indicate that the insertion of the NH$_3$
molecules modifies only weakly the essentially pure K$_3$C$_{60}$ LUMO
conduction band, as expected.
%
%
%
In particular, as apparent from Fig.~\ref{bands_NH3:fig}, the DFT-LDA
bandwidths of NH$_3$\,K$_3$C$_{60}$ and K$_3$C$_{60}$ are very similar,
both $W\sim 0.6$~eV.
The main difference in the two band structures is a splitting of
the threefold-degenerate $t_{1u}$ band of K$_3$C$_{60}$ at the $\Gamma$
point of NH$_3$\,K$_3$C$_{60}$.
The $\Gamma$-point splitting, roughly corresponding to a dimensionless 
ratio $\Delta/W \sim 0.25$, provides a measure of the strength of the non-cubic
crystalline environment seen by the $t_{1u}$ orbital on each fullerene
molecule in the orthorhombic structure of NH$_3$K$_3$C$_{60}$.
Its magnitude is roughly a quarter ($0.2 - 0.3$) of the total bandwidth;
this represents the main result of this DFT calculation.
Now if this compound were an uncorrelated metal, this splitting would have no
real consequences, besides a change of shape of the Fermi surface. 
The consequences can be much more important due to strong correlations,
as anticipated above.

\rem{
The $\Gamma$-point $t_{1u}$ band splitting -- see
Fig.~\ref{bands_NH3:fig}(a) -- can be taken as a crude estimate of the
value of $\Delta \sim 0.15$~eV in NH$_3$K$_3$C$_{60}$, corresponding to a
dimensionless ratio $\Delta/W \sim 0.25$.
This value is of course too small to determine a complete emptying of one
of the bare, uncorrelated bands, and indeed all three bands of
NH$_3$K$_3$C$_{60}$ still cross the Fermi energy $\epsilon_{\rm F}$.
}

\begin{figure}[t]
\centering
\epsfig{file=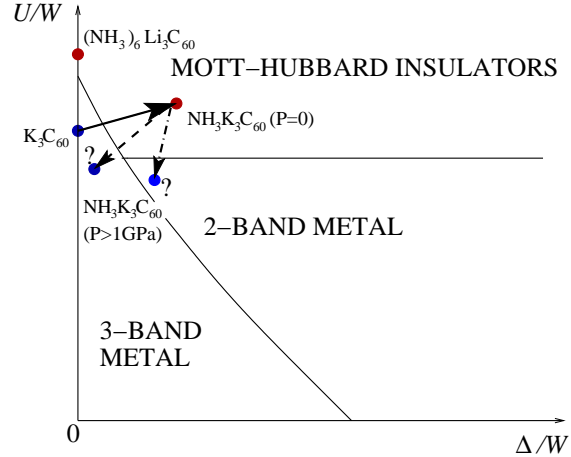,height=6cm}
\caption{\label{phased_NH3:fig}
A schematic $(U,\Delta)$ phase diagram for the $d=3$-bands 
family of compounds K$_3$C$_{60}$ and NH$_3$K$_3$C$_{60}$.
The dashed and dot-dashed arrows indicate two plausible paths of
pressure-induced metallization of NH$_3$K$_3$C$_{60}$.
Cubic insulating (NH$_3$)$_6$Li$_3$C$_{60}$ \cite{Durand03} is also
indicated.
(Adapted from Ref.~\cite{splitDMFT}.)
}
\end{figure}

In fact, as discussed above, the fullerides lie on the metallic side of the
Mott transition in the half-filled $d=3$ band cubic system.
%
%
%
Upon ammoniation of K$_3$C$_{60}$ into
orthorhombic NH$_3$\,K$_3$C$_{60}$, the volume expansion will first of all
increase $U/W$,
while the anisotropy will give rise to a nonzero $\Delta$, 
corresponding to a displacement as indicated by a rightward arrow in
Fig.~\ref{phase_diagram:fig}.
%
%
%
Inspection of Fig.~\ref{phase_diagram:fig} shows that, if $U$ is close
enough to $U_{\rm c}$, even a band splitting $\Delta$ substantially smaller
than what we have estimated for NH$_3$\,K$_3$C$_{60}$ could suffice to
drive that metal insulator transition, even without any appreciable change in
$U/W$.
Conductivity measurements \cite{Kitano02} on the class of compounds
NH$_3$\,K$_{3-x}$Rb$_x$C$_{60}$ supports the possibility that the
orthorhombic distortion could be an important ingredient driving the Mott
transition in these systems.
An interesting -- even if perhaps not practically straightforward -- test
of this overall picture could be obtained by applying uniaxial stress to
cubic superconducting fullerides of the A$_3$C$_{60}$ family.
Contrary to the standard tendency of hydrostatic pressure toward
metallization and lower superconducting $T_c$, the orbital splitting
associated with the appropriate uniaxial strain could drive some of these
compounds towards stronger correlation, thus possibly first toward higher
$T_c$, and eventually Mott insulating.
A similar suggestion was put forward by Koch \cite{Koch02}, although
on rather different grounds.

We must stress here that the scenario sketched above does not 
at all diminish the role, in the metal-insulator transition, 
of the accompanying expansion of the lattice, and of the dependence
of $U/W$ upon volume.
That role is experimentally proven by the observation  of a pressure-driven 
insulator-metal transition, where NH$_3$\,K$_3$C$_{60}$ is transformed 
into a metal (and a superconductor), despite the permanence of the 
orthorhombic structure \cite{Margadonna01}.

\subsection{
Mott transition in other fullerides and other molecular conductors}
\label{related:sec}

The vicinity to the metal-insulator transition is not exclusive of
NH$_3$\,A$_3$C$_{60}$. Other systems of the same family are 
(NH$_3)_6$Li$_3$C$_{60}$ \cite{Durand03},
(NH$_3)_x$\,NaRb$_2$C$_{60}$ ($x\simeq 1.6$) \cite{Ricco00},
(NH$_3)_x$\,NaK$_2$C$_{60}$ ($0.5<x< 1$) \cite{Shimoda96,Ricco01,Ricco03},
and the noncubic expanded alkali fulleride Cs$_3$C$_{60}$ \cite{Palstra95}.
The pressure-driven insulator-superconductor transition observed in
the latter compound further underscores the role of volume expansion in
favoring the Mott insulator.
In fact, (NH$_3$)$_6$Li$_3$C$_{60}$ is an example of a 
compound of the $n=3$ family lying beyond the Mott transition despite
its cubic structure \cite{Durand03},
as illustrated in Fig.~\ref{phased_NH3:fig}.

The $n=4$ alkali fullerides also sit in the vicinity of the metal-insulator
transition, usually on the insulating side.  Indeed, the pressure-induced
insulator to metal transition observed in tetragonal Rb$_4$C$_{60}$
\cite{Kerkoud} is reminiscent of that observed in the C$_{60}^{3-}$ compounds
NH$_3$\,K$_3$C$_{60}$ and Cs$_3$C$_{60}$, but for the lack of
superconductivity.
We shall come back to these C$_{60}^{4-}$ compounds in the final discussion.

\begin{figure}[t]
\centering
\epsfig{file=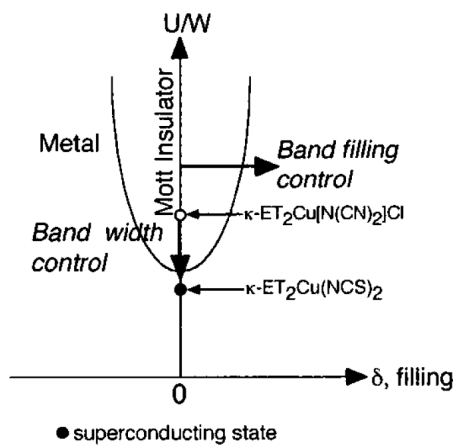,scale=.3333333}
\caption{\label{Mori02:fig}
Figure 1. Electronic phase diagram of organic conductors.
The vertical axis indicates electron correlation $U/W$.
The organic antiferromagnetic Mott insulating state
($\kappa$-ET$_2$Cu[N(CN)$_2$]Cl) transforms into a superconducting one
[$\kappa$-ET$_2$Cu(NCS)$_2$], with a reduction in $U/W$.
The horizontal axis shows the band-filling control.
(From Ref.~\cite{Mori02}.)
}
\end{figure}

In the wider class of molecular conductors, the Mott transition 
is not exclusive of the fullerides; other organic
compounds show a similar phase diagram.
Especially the $\kappa$-ET$_2$ [ET =
bis(ethylene\-di\-thio)tet\-ra\-thia\-fulvalene] compounds show a similar
phenomenology to trivalent fullerides \cite{Mori02}, with a clear Mott
transition at stoichiometry and a superconducting phase all around,
Fig.~\ref{Mori02:fig}.
More recently, insulating metal-phthalocyanine organic films were turned
conducting and metallic by alkali doping \cite{Craciun04, Craciun05}.
A similar scheme to fullerides, and to the metallic organics, has been
proposed \cite{Tosatti04} as a possible scenario for the electron-doped
metal-phthalocyanine conductors. The verification of the existence of
stoichiometric phases, of Mott insulators at integer fillings, and possibly
of superconducting phases in between remains a challenge for future
experiments and theory in these materials.

Next, we should mention 
compounds such as (AsF$_6$)$_2$\,C$_{60}$
\cite{Datars95,Datars96,Panich02,Panich03}, where fullerene is stripped of
two of its $h_u$ valence electrons.
The physics of positive C$_{60}$ ions has been described in
Ref.~\cite{Lueders03}, and is somewhat parallel to that of negative ions,
though with a richer parameter space which implies several complications.
The Hubbard $U$ in the $h_{u}$ HOMO level is slightly larger than in
$t_{1u}$ LUMO.
Cancellation between Hund's rule exchange and JT effect, discussed in
sections \ref{coulomb:sec} and \ref{other:sec}, is present here too. It was
calculated to be nearly exact in the doubly positive ion, with a slight
prevalence of exchange and a marginally stable high-spin ground state
\cite{Lueders03}.
The doubly doped-solid state compounds (AsF$_6$)$_2$C$_{60}$ and
(SbF$_6$)$_2$C$_{60}$ have been reported to be small gap insulators
\cite{Datars95,Datars96,Panich02,Panich03}.
There appears to be no simple way that they could be band insulators, and
the likeliest explanation is that they are Mott insulators.
As for magnetism, some indirect experimental evidence has been suggested,
consistent with high spin in the di-cation.
However the balance between intramolecular exchange and JT is very
marginal, which leaves the possibility for a low-spin hole-doped Mott
insulator wide open. There is even some preliminary indication that this
might in fact be realized in (AsF$_6)_2$C$_{60}$ \cite{riccoPrivate}.
It would be very interesting in this respect to begin an experimental study
of the insulator-metal transition in these compounds, to be obtained for
example by hydrostatic pressure.
Especially if the transition was relatively continuos and not too strongly
first-order, the ensuing metallic phase close to the singlet Mott insulator
should exhibit a pseudogap with strong pair correlations, and possibly
superconductivity, similar to that described in Ref.~\cite{Capone02}.
 
Finally, we note that the pressure- or doping-induced transformation of Mott
insulators into metals and strongly correlated superconductors which we
discuss here for fullerene compounds, is a scenario which shares many
important elements with the high-$T_c$ cuprates. Again, this is an issue to
which we will return later.

\subsection{The crucial role of exchange and of JT on-site interactions } 
\label{other:sec}

While the above bare Hubbard model defines the main backbone of the phase
diagram, the precise details depends strongly on a variety of secondary,
on-site couplings, generally present as permitted by the orbital degeneracy
of the site.
They act as a rule such as to optimize the energetics 
of the isolated site. As such, they generally favor the Mott insulating state, 
where electrons effectively localize on site.
In the metallic state, conversely, kinetic energy washes out much more the
effect of these on-site couplings.
Electrons in the insulating state where on-site occupancy is close
to integer and poorly fluctuating, can take full advantage of these secondary
couplings, whereas travelling electrons in the metallic state cannot.
As a result the overall effect of secondary on-site couplings such as $J$
and $E_{JT}$ is to reduce the critical $U_c$.
On the contrary, we can expect that strongly non\-local/inter\-site couplings
could favor the metallic state, and raise $U_c$.
In the fullerides, intramolecular couplings dominate, while screening may
be assumed to suppress the long-range interactions. In particular, while we
certainly cannot rule out some role for an inter-molecular Coulomb
interaction $V$, we will completely neglect it here.

As has been repeated over and over, the two important intra-site
interactions are the JT coupling, Eq.~(\ref{Hev:eq}), and the Hund's rule
exchange part of Eq.~(\ref{He-ecombined}).
In the DMFT calculation of Sec.~\ref{dmft:sec} however, these terms were
initially left out.
In fact, these on-site interactions, even if small in comparison
with $U$ and $W$, play a crucial role on the Mott
transition in a band-degenerate conductor. Their
presence can actually change nearly everything: the nature of the Mott insulator,
its spin and orbital contents, the nature of the metal phase, its tendency to superconduct
or not, and of course the precise location $U_c$ of the Mott transition
itself.  
Probably the simplest manner to illustrate this physical situation is to
describe the results of DMFT calculations with proper inclusion of on-site
interactions.

\begin{figure}[t]
\centering
\epsfig{file=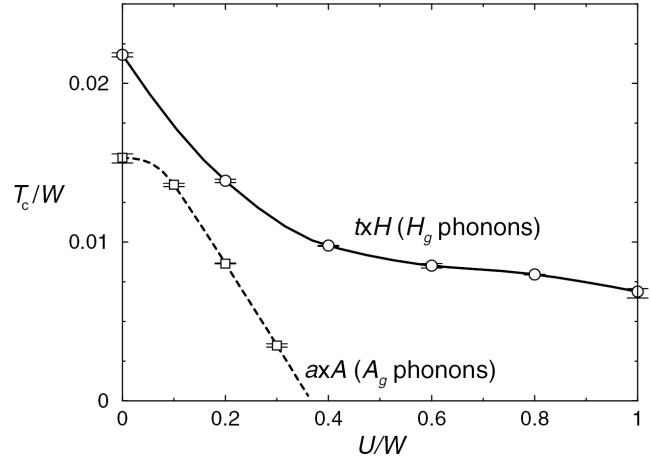,height=6cm}
\caption{\label{Han03:fig}
Superconducting $T_c$ as a function of $U$ for the $t\otimes H$ and
$a\otimes A$ models for the half filled $d=3$ model, treated within the
DMFT.
The parameters are $\lambda= 0.6$ and $\omega/W = 0.25$.
The figure illustrates the important difference between $H_g$ and $A_g$
phonons.
(From Ref.~\cite{Han03}.)
}
\end{figure}

On a given molecular ion, Hund's rule exchange will favor maximum spin and
maximum orbital momentum, while JT will favor just the opposite. Because,
as explained in Sec.~\ref{coulomb:sec}, there is an important cancellation
between the two, neither of them can be safely neglected.
Without entering here any of the technical details, it is difficult to
solve the DMFT problem including exactly both these intra-site couplings,
and some approximation must be made.
Very detailed DMFT work has been done including a realistic representation
of the electron-vibrational JT coupling for the fullerides.
In particular, Refs.~\cite{Han00,Han03,HanKoch00} study a $d=3$
Hubbard-Holstein model including a single JT $H_g$ phonon mode (5
oscillators rather than 40).
The role of different lattice symmetry is investigated in
Ref.~\cite{HanKoch00}. 
These approaches employ very accurate and extensive quantum Monte Carlo
calculations in a finite cluster, including the vibronic coupling with
realistic JT strength $\lambda$ and a realistic band structure.
(An earlier calculation \cite{Han99} even investigated the role of disorder in
these systems).
Reference \cite{Han03} addresses mainly superconductivity in the $d=3$
Hubbard-Holstein-JT model.
As illustrated in Fig.~\ref{Han03:fig}, it was found (like in the $d=2$
model) that the JT coupling to degenerate phonons produces a pairing much
more compatible with strong correlations (large $U$) than coupling to a
single non-JT $A_g$ mode.
In non-JT  $A_g$ electron-phonon coupling superconductivity disappears 
when ${\cal U}_n=U+U^{\rm e-v} \geq 0$, while in JT $H_g$ coupling 
superconductivity survives up to largely repulsive ${\cal U}_n>0$.
On the other hand, for small $U$, the local pairing becomes generally 
less efficient, because charge fluctuations induced by electron hopping disrupt the JT
ground state(s) into uncorrelated electron pairs. Hence the superconducting 
order parameter is always depressed by $U$ at small $U$.
As $U$ is increased, coherent electron hopping is gradually suppressed and
the local pair formation becomes more important.
For realistically large JT coupling $\lambda=\frac56 N(\epsilon_{\rm F})
\sum_k g_k^2 \hbar\omega_k\sim 1$ \cite{GunnarssonBook,Gunnarsson} $d=n=3$
and increasing $U$ the metal phase below $U_c$ is always superconducting,
with $T_c$ a monotonically decreasing function of $U$. In other words,
electron-electron repulsion always disfavors superconductivity for
realistically large JT coupling \cite{Han03}.

However, this approximation is unrealistic as it neglects Hund's rule
exchange, and the large cancellation of JT energetics that it implies.
Inclusion of exchange is difficult, because it leads to a fermion sign
problem in the Monte Carlo DMFT impurity solver. If on the other hand the
impurity is solved by, e.g., Lanczos diagonalization, exchange can be
treated but the phonon ladder gives rise to too many states per each site.
An approximate way out is available very close to the Mott transition.
Here the quasiparticle bandwidth $z W$ tends continuously to zero, and
sufficiently close to $U_c$ it falls below the typical vibrational
frequency $\hbar\omega$. In this regime, the frequency dependence of the
retarded JT coupling (assumed to be weak) can be integrated out, resulting
simply in an additional but inverted Hund's rule like exchange term.
As discussed in Sec.~\ref{coulomb:sec}, for $d=3$ that term is $J^{\rm
e-v}=-\frac 34 \sum_k g_k^2 \hbar\omega_k$.
In this limit -- close enough to the Mott transition, JT coupling
not too strong, relatively high vibration frequencies -- it is therefore 
possible to treat both Hund's rule exchange and JT coupling, by simply 
replacing $J$ with $J_{\rm eff} = J + J^{\rm e-v}$.
This is the route generally followed in recent work by M.\ Capone {\it et
al.\ } \cite{Capone04,Capone01,Capone00,Capone02} and by M.\ Granath and
S.\ \"Ostlund \cite{Granath03}.

In the $t_{1u}$ molecular level of fullerene $J\simeq 50$~meV, and the
DFT-LDA JT couplings of Ref.~\cite{Manini01} yield $J^{\rm e-v}=-57$~meV,
while the couplings obtained from photoemission \cite{Gunnarsson} are
consistent with $J^{\rm e-v}=-127$~meV: with both estimates the net
resulting $J_{\rm eff}$ is negative.
The effective Hund's rule of C$_{60}^{n-}$ is then inverted, the low-spin
states representing the ground state, as dictated by JT coupling.
It is important to underline again here that the massive cancellation
finally leaves fullerides with a {\em weak} on-site spin pairing
interaction, replacing the original stronger couplings, both JT
and Hund. The consequences of this weak residual pairing are, as it turns out,
qualitatively different from either the strong singlet pairing required
by JT alone, or by the strong triplet pairing required by Hund's $J$.
Similarly, in the half filled $e_g$ molecular level of a metal
phthalocyanine ($d=n=2$) it is predicted $J_{\rm eff} = -60$~meV
\cite{Tosatti04}, again inverted, but again small as a result of a strong
cancellation.

Let us consider in detail the consequences of this cancellation
and of the final Hund's rule inversion.
The first consequence is that the nature of the Mott insulator itself is
affected.  Since the isolated molecular ion ground state has low spin (a spin
singlet for even $n$ and a spin doublet for odd $n$), so will the
Mott-insulator sites.
The insulating state is more complicated and intriguing than either the
one-band Mott insulator, or the simple orbitally-ordered cooperative JT
distorted state.  It is rather a Mott Jahn-Teller insulator.  Let us try a
description of that state, in the easy case of $n=2$ electrons/site.
Assume initially zero hopping between sites.  On each site, the JT phonons
are characterized by a 2-dimensional (pseudo)rotor in the trough.
%
Subject to its own quantum fluctuations, this free rotor favors no special
direction and will have its (say) $L=0$ ground state separated by a gap
$\delta$ from its $L=1$ first excited state.
Turning now on a hopping $\propto W$ between sites, this will cause an
intersite interaction between rotors of order $W^2/U$). If that is
strong enough, the rotors will freeze statically in an orbitally ordered
state, similar to a ferroelectric, which is on fact a cooperative JT
distortion \cite{Bersuker,Gehring75,Kaplan95,Dunn04}.
If instead the on-site quantum fluctuations make $\delta$
large enough, then the static cooperative order will quantum mechanically
melt. This quantum melted state, the rotor analogue of a spin singlet, is the
Mott-JT state \cite{Fabrizio97}.
This state bears some resemblance to the quantum paraelectric state
\cite{Martonak}.
It is strongly nonadiabatic in nature, which reflects in a strongly
entangled admixture of electronic and phonon states: all electronic
spectroscopies should observe radically renormalized and vibronically 
broadened ``bands'', even at low temperature.
Other molecular and lattice properties of this state still need to be
worked out and investigated.
This remains a theoretical task for the future.

Further characterization of the Mott-JT state is needed, both theoretically
and experimentally, especially when it comes to its spectroscopical
properties.
The main examples at our disposal are NH$_3$\,K$_3$C$_{60}$ (orthorhombic)
\cite{margadonna} and (NH$_3$)$_6$Li$_3$C$_{60}$ (cubic) \cite{Durand03}
for $n=3$ and a $S=1/2$ on-site ground state, and prominently Rb$_4$C$_{60}$,
K$_4$C$_{60}$ for $n=4$, $S=0$ singlet ground state.
Concerning the $n=3$ Mott insulators, the main evidence so far is their
clear characterization as spin-$1/2$ antiferromagnets. Since there are $3$
electrons/fullerene, that can be explained only by a JT- dominated state.
That state could be at this stage either a statically distorted, collective
JT insulator, or a quantum melted Mott-JT state.
The available crystal and magnetic structure of orthorhombic
NH$_3$\,K$_3$C$_{60}$ \cite{Tou1,margadonna} displays a very rich interplay
of magnetic and orbital order, which is only well analysed but is
apparently compatible with either possibilities.
It should be noted in addition that merohedral disorder in the relative
angular orientation of the C$_{60}$ molecules will frustrate static
collective order, additionally favoring the melted state.
Indeed only in the well ordered, ``discrete'' salts well defined collective
distortions have been observed so far \cite{Reed00review}.
An orbitally-ordered static collective JT state
\cite{Bersuker,Gehring75,Kaplan95,Dunn04} has never been reported in the
alkali fullerides \cite{Kuntscher97}.
On the other hand, in a band-degenerate context, the possibility of
dynamical orbital order should always be considered \cite{Fabrizio97}.
The question whether one or more of the existing Mott insulating fullerides
could be characterized as a dynamical Mott Jahn Teller insulator stands at
this stage as an exciting experimental challenge.

\begin{figure}[t]
\centering
\epsfig{file=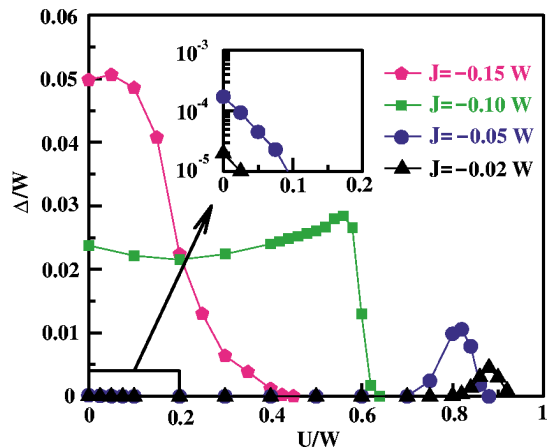,scale=1}
\caption{\label{Capone04:fig}
Superconducting gap in a half-filled $d=2$-band model, as a function of the
on-site repulsion for several (anti-Hund) couplings $J\equiv J_{\rm
eff}=J+J^{\rm e-v}$ (where $J^{\rm e-v}$ accounts antiadiabatically for the
electron-phonon interaction if $J^{\rm e-v} = -\lambda/[2N(\epsilon_{\rm
F})]=-2 g^2 \hbar\omega$, and $g\hbar\omega$ represents the coupling
strength of the $e\otimes E$ JT coupling at each site; the difference with
the relation of Table~\ref{couplings:table} is due to $d=2$ rather than
$d=3$).
Increasing repulsion spoils superconductivity at strong electron-phonon
coupling.  Superconductivity is instead strongly enhanced close to the Mott
transition at weak coupling.
The inset reports the weak-coupling regime on an expanded scale, showing a
much smaller gap at small $U$ compared to $U\simeq U_c$.
(Adapted from Ref.~\cite{Capone04}.)
}
\end{figure}

After this illustration of the important effects of on site interactions on
the mott insulating fullerides, we can now move on to the metallic phase,
obtained at smaller $U$ values.
As we will see, the metallic phase below $U_c$ is even more affected, and
in fact it often becomes superconducting.
This is not in itself totally surprising, since we have after all a small
but definitely attractive $J_{\rm eff}$ reflecting a locally dominant
electron-vibration coupling (even if orders of magnitude smaller than the
repulsive $U$!).
What is more surprising is the behavior of the superconducting gap as a
function of $U$, see Fig.~\ref{Capone04:fig}.
So long as $(-J_{\rm eff})$ is large -- as in the uncompensated JT case
described earlier -- the gap decreases monotonically with increasing
$U$. But when $(-J_{\rm eff})$ is small -- reflecting a strong cancellation
by Hund's rule -- superconductivity only survives near $U = 0$ and near $U
= U_c$.
Superconductivity near the Mott transition was designated ``strongly
correlated superconductivity'' (SCS).
In SCS, the gap near $U_c$ can increase for increasing $U$, and is many
orders of magnitude larger than even the repulsion-free BCS-like
superconducting gap near $U = 0$.
Moreover, close to the Mott insulator, the SCS gap magnitude -- and thus
presumably the superconducting $T_c$ -- has a characteristic bell-shaped
behavior.

\begin{figure}[t]
\centering
\epsfig{file=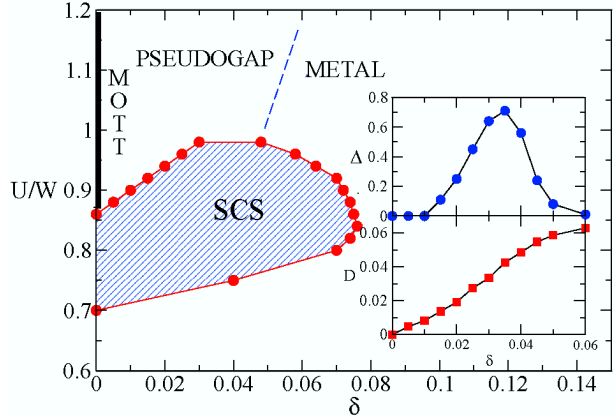,scale=1}
\caption{\label{Capone04phased:fig}
Phase diagram od the $d=2$ model as a function of $U/W$ and doping
$\delta=n-2$ at $J_{\rm eff}=-0.05\,W$. The thick vertical line marks the
singlet Mott insulator.
For $U=0.92\,W$, the inset shows the superconducting gap divided by a
factor 3 and the Drude weight $D$ (i.e.\ the strength of the superfluid
peak normalized to the noninteracting value) as functions of doping.
(From Ref.~\cite{Capone04}.)
}
\end{figure}

The non-superconducting, metallic phase obtained by sligthly doping the
Mott-JT insulator at $U > U_c$ away from stoichiometry is also strongly
unconventional, in that it has a pseudogap.
Even if doping away from stoichiometry is not straightforward in
fullerides, this aspect of the theoretical phase diagram is quite
illuminating.
The presence of the pseudogap is related to $J_{\rm eff} < 0$ and thus to
the slightly prevailing JT effect, but is not at all bipolaronic, since as
illustrated in Ref.~\cite{Capone02} the electron occupancy of each site in
this state is strongly pinned to the mean value $n$, the bipolaronic
disproportionation into $(n-1)$ and $(n+1)$ pairs completely suppressed and
in fact reversed by the large Hubbard $U$.
Upon further doping of the pseudogap metal, the model exhibits SCS, the
maximum gap attained at some optimal doping, then dropping and disappearing
in a final overdoped phase.  as shown in the overall $T=0$ phase diagram of
the $d =2$ $n \simeq 2$ model reported in Fig.~\ref{Capone04phased:fig}.

As one can see, many of these features are very reminiscent of high-$T_c$
cuprates, whereas the model does not describe cuprates at all.
Here we have on-site orbital degeneracy, Hund's rule exchange, and JT
coupling, all elements that are absent or otherwise quite different in the
cuprates.
The strongly correlated superconducting order parameter in the model is, as
in actual fullerides, s-wave and not d-wave as in the cuprates.
We suggest that our models and at least some of the fullerides which they
describe, are new members of a wider family of strongly correlated
superconductors, which includes the cuprates, and all potentially
high-$T_c$ materials.

In this light, it is important to extract and to expose our overall qualitative
understanding out of the model calculations.
The occurrence of superconductivity in our orbitally degenerate molecular
conductors, meant to model the fullerides, can be qualitatively rationalized 
by discussing what happens at integer filling as $U$ is raised to
approach the Mott insulating transition at $U_c$.
First, the quasiparticle bandwidth narrows indefinitely, from $W$ to $zW$ with 
$z \to  0$ as $U \to U_c$. The rest of the single-particle spectral weight 
moves to the incoherent Hubbard bands, far away from the Fermi level.
Second, since all charge fluctuations are gradually frozen out when
approaching the Mott insulator, so are all the charge-related repulsions
(and for that matter, attractions too) between the quasiparticles.  For
example, $U$ is effectively reduced to $\sim zU$, and very near the Mott
transition the propagating quasiparticles tend to infinitely massive 
interacting fermions, with a decreasing absolute repulsion.
Third, thanks to the orbital degeneracy the spin fluctuations are
not frozen out, and so attractions that act in the spin channel are
not renormalized away. Therefore, e.g.\ the effective Hund-rule 
exchange $J$ or $J_{\rm eff}$ retains its bare value even at the Mott
transition, and that can easily overcome the weak repulsion $zU$.

These three elements make the effective quasiparticle Hamiltonian very
similar to an {\em attractive} Hubbard model, with a bandwidth $zW$ and an
on-site attraction $J_{\rm eff} < 0$. The ground state is a weak-coupling
BCS-like superconductor, at least sufficiently below the Mott transition,
so long as $zW \gg |J_{\rm eff}|$.
As $U$ grows, the superconductivity changes from weak to strong coupling,
the maximum gap achieved when $zW \sim |J_{\rm eff}|$.
This point is equivalent to the maximum of the Nozieres-Schmitt-Rink
\cite{Nozieres85,sademelo93} curve of $T_c$ versus $\lambda$ in purely
electron-phonon superconductors and, we believe, also to the optimal doping
point in the cuprates.

If assumed to hold quantitatively, the qualitative mapping onto the
attractive Hubbard model has the additional virtue of even predicting the
maximum 3D critical temperature of our type of model at optimal doping,
by reading it off the attractive Hubbard model studies.
That gives approximately  $k_{\rm B} T_c^{\rm max} \sim 0.07\, |J_{\rm eff}|$ for our models.
In trivalent fullerides the value of $|J_{\rm eff}|$ can be extracted by
equating the observed spin gap $\sim 140$~meV \cite{ZimmerAll,Lukyanchuk95}
to $5\,|J_{\rm eff}|$ (see Table \ref{multiplets:table}), yielding
$T_c^{\rm max} \simeq 0.07\,|J_{\rm eff}|/k_{\rm B} \simeq 23$~K, which is
surprisingly, though probably unfairly, accurate.

\begin{figure}[t]
\centering
\epsfig{file=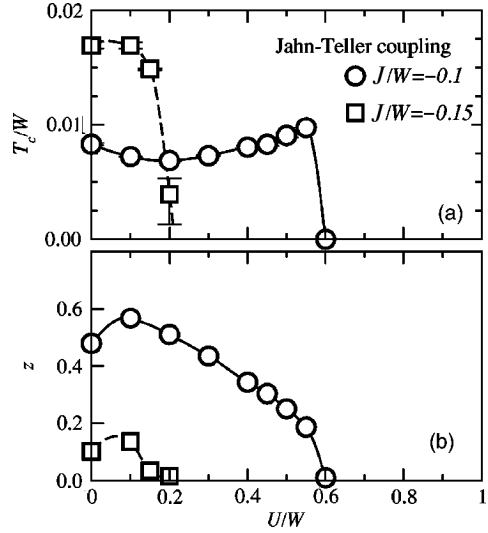,height=7cm}
\caption{\label{Han04_jt:fig}
(a) Superconducting transition temperature $T_c$ vs Coulomb repulsion $U$
in a $d=2$-band model with JT electron-phonon coupling, where the coupling
strength $g\hbar\omega$ is gauged by $J=J^{\rm e-v}= -2 g^2 \hbar\omega$.
Weak JT phonon coupling ($J/ W = -0.1$) produces superconductivity with a
crossover from the conventional superconductivity (small $U$) to the local
pairing regime at large $U$ (a Mott-JT insulator).
(b) Quasiparticle renormalization factor $z$.
(Adapted from Ref.~\cite{Han04}.)
}
\end{figure}

One remaining question is: what is the relationship of this ``island'' of
SCS near the Mott transition, to the more standard BCS-like phonon driven
superconductivity one finds in the same model when there is no
electron-electron repulsion, i.e.\ near $U=0$?
The answer to this question is very instructive, and is obtained by solving
the model for a grid of on-site attractive coupling values -- negative
$J_{\rm eff}$ values \cite{Capone04}.
As shown by the plot of zero temperature superconducting gaps in
Fig.~\ref{Capone04:fig} if $|J_{\rm eff}|$ is large, then the pairing gap
is in fact maximum at $U$ = 0, and decreases monotonically for increasing
$U$, vanishing just at the Mott transition.
This scenario is, we believe, equivalent to that presented by Han {\it et
al.}\ for $d=n=2$ \cite{Han04} and  $d=n=3$ \cite{Han00,Han03,HanKoch00}.
In particular, the monotonic decrease of $T_c$ as $U$ approaches $U_c$ in
Fig.~\ref{Han04_jt:fig} (squares) is equivalent to the large-$|J_{\rm
eff}|$ monotonic gap of Fig.~\ref{Capone04:fig} in a model including
explicitly the JT phonons.
Similarly, the surge in $T_c$ in Fig.~\ref{Han04_jt:fig} (circles) is
essentially identical to that found earlier for small attractive $|J_{\rm
eff}|$.

\begin{figure}[t]
\centering
\epsfig{file=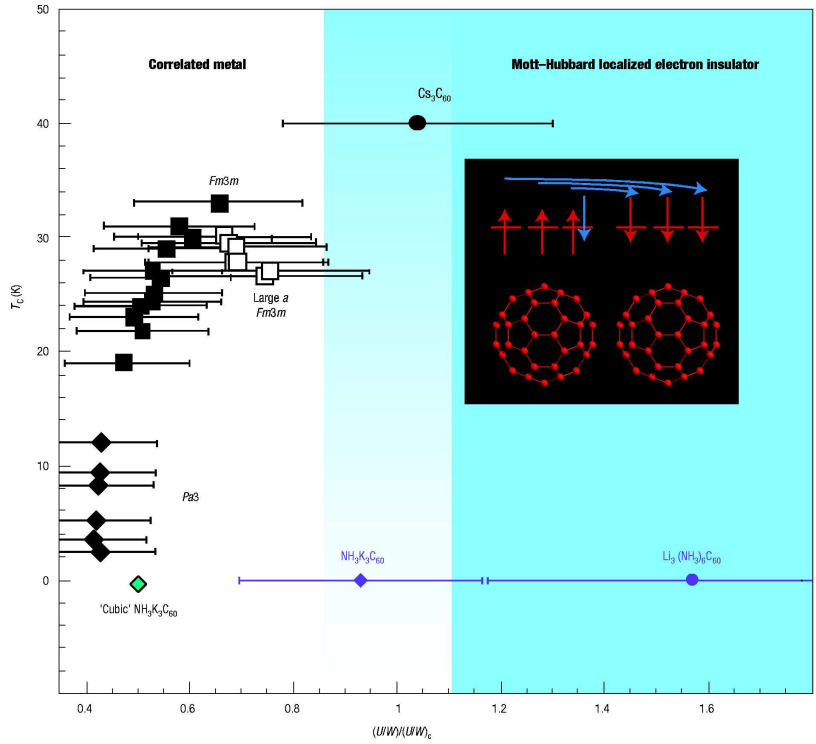,scale=.66666666666667}
\caption{\label{Durand03:fig}
A schematic phase diagram of C$_{60}^{3-}$ fullerides.
The superconducting transition temperature $T_c$ is shown as a function of
the ratio $U/W$ divided by the critical value $U_c/W$ required to produce
electron localization -- $U_c/W$ is estimated to be 2.3 for the f.c.c.\
phases (square and diamond symbols), 1.3 for b.c.c.\ Cs$_3$C$_{60}$ and
Li$_3$\,(NH$_3$)$_6$C$_{60}$ (circles) and 1.1 for the orthorhombic
NH$_3$\,A$_3$C$_{60}$.
The value of $U$ used to calculate $U/W$ was 0.8~eV.
%
%
%
The Mott-insulator region is shaded: uncertainties in the precise numerical
estimates are emphasised by the graded shading of the crossover between
metallic/superconducting and localized.
$W$ is derived for the f.c.c. phases from Ref.~\cite{Satpathy92}, for the
orthorhombic NH$_3$\,K$_3$C$_{60}$ from Ref.~\cite{splitDMFT} and for
b.c.c.\ phases by DFT calculation \cite{Durand03}.
The dramatic effect of lifted orbital degeneracy on the location of
orthorhombic NH$_3$\,K$_3$C$_{60}$ is illustrated by comparison with an
hypothetic ``cubic'' NH$_3$\,K$_3$C$_{60}$) retaining the $t_{1u}$ orbital
degeneracy and located well within the metallic regime.
(From Ref.~\cite{Durand03}.)
}
\end{figure}

Things do change qualitatively when the effective attraction decreases, as
the dominance of JT is more and more eroded and cancelled by Hund's rule
exchange.
For decreasing $|J_{\rm eff}|$, superconductivity remains strongest at $U=0$
and near the Mott transition, but it gradually weakens in between.
Eventually for very small $|J_{\rm eff}|$ one arrives at two separate
superconducting islands, one with a tiny BCS gap near $U =0$, and a second
one, near the Mott transition at $U_c$, with a gap orders of magnitude
larger than BCS. This the SCS island.
At the lower edge of this island (equivalent to the ``overdoped'' regime of
the cuprates), the pairing is again BCS like: but the gap and presumably $T_c$
rises as the electron-electron repulsion increases, rather than the
opposite.
We believe that this regime could describe most of the trivalent
fullerides, and their gigantic $T_c$ and susceptibility increase with
volume expansion.
In fact the whole curve of $T_c$ versus increasing volume as presented,
e.g., by Durand {\it et al.}\ \cite{Durand03} (Fig.~\ref{Durand03:fig})
could in our view be roughly identified with the theoretical bell-shaped
curve of gap versus increasing $U$ of Fig.~\ref{Capone04:fig}.

A corollary is that the highest $T_c$ fullerides should be extremely
strong-coupling materials, comparable to cuprates. This is in agreement
with the extremely large $H_{c2}$ values observed experimentally as well as
with the existence of an irreversibility line in the $H-T$ phase diagram
\cite{Buntar96}. Both features are presently unexplained and both are
similar to cuprates.

The compounds in the descending $T_c$ branch Fig.~\ref{Durand03:fig},
apparently very close to the Mott transition, are the fulleride analog of
the ``underdoped'' cuprates. They should be especially interesting, and
their normal state should for example exhibit strong pseudogap features.

So far the discussion assumed prevalence of JT over Hund's rule exchange,
$J_{\rm eff} < 0$.
Also a hypothetical case of {\em positive} $J_{\rm eff}$ (regular Hund's
rule) would not be without interest.
On-site high-spin states would prevail in this case.
Again, as compared to $J = 0$, the Mott transition shifts to lower $U_{\rm
c}$ \cite{Han04}.
This occurs for two reasons: a reduction of the relevant pair energy ${\cal
U}_n$ proportional to $J$, and the fact that splitting of different Hund's
multiplets opposes the metallic state.
In the DMFT calculation of Ref.~\cite{Han04}, the possibility of triplet
superconductivity was investigated, and indeed it was found that before the
metallic state turns into a magnetic Mott insulator, a superconducting
phase emerges, with triplet pairing, but without p wave, thanks to the
orbital degeneracy.
If for example C$_{60}^{2+}$ was really high spin in (AsF$_6$)$_2$C$_{60}$
or (SbF$_6$)$_2$C$_{60}$, among the metallic phases obtainable under
pressure one could find one such triplet SCS.

In practice a small {\em negative} anti-Hund $J_{\rm eff}$ represents
better the net balance of the positive Coulomb term and a prevailing
electron-phonon attraction (assumed in the antiadiabatic limit) in
C$_{60}^{n-}$ \cite{Lueders03,Leuven02,Lueders02}.
Figure~\ref{Capone04:fig} shows that a strong reduction of $U_{\rm c}$
occurs under the action of an even small negative $J_{\rm eff}$ (as
expected for any local interaction).
In addition, as discussed in Sec.~\ref{coulomb:sec}, a negative $J_{\rm
eff}$ favors local low-spin states, which are spin singlets whenever the
local occupancy is even.
Of course, in the solid, band effects tend to favor local charge
fluctuations and spin admixture, but as $U$ approaches $U_{\rm c}$, the
bandwidth reduces effectively to the width $z\,W$ of the Kondo peak.
When $z\,W$ becomes comparable to $|J|$, the local exchange term promotes
locally paired singlets in a sort of anti-Migdal regime
\cite{Pietronero92,Pietronero,Cappelluti00,Cappelluti01,Botti02,Paci04}
close to the transition, where the average phonon frequencies exceed the
effective bandwidth.
This regime favors superconductivity, and DMFT suggests that a high-$T_c$
($k_{\rm B} T_c$ of order $|J_{\rm eff}|$) superconducting phase should
intrude between the metal and the low-spin Mott insulator
\cite{Capone04,Capone01,Capone00,Capone02}.
As shown in Fig.~\ref{Capone04:fig}, for very small $|J_{\rm eff}|\ll W$
(representative of the cancellation characteristic of C$_{60}^{n-}$), the
very weak superconductivity of the uncorrelated regime $U\simeq 0$ becomes
dominating close to the Mott transition.
An alternative possibility to the intruding superconducting phase is a
discontinuous, first-order transition, jumping directly from the metal to
the insulator, and in fact this is a serious eventuality for the
A$_4$C$_{60}$ fullerides, where no superconductivity is observed.
The pressure-induced reversion from insulator to metal observed in
tetragonal Rb$_4$C$_{60}$ \cite{Kerkoud} is most likely the $n=4$
counterpart of the $n=3$ transition of the ammoniated compounds.

\section{Conclusions}

We have discussed some results and qualitative theoretical ideas on 
the modeling of metal-insulator transition in fullerene compounds.
They have been described as Hubbard models with electron hopping with a
bandwidth $W$ between orbitally degenerate fullerene sites, each of them
supporting a Coulomb repulsion $U$, a Hund's rule exchange $J$, and a
JT distortion with energy gain $E^{\rm e-v}$. The interplay of these
parameters gives rise to a rich phase diagram, comprising Mott insulators,
metals, and superconductors.
For electrons in fullerene, the JT strength marginally prevails over
exchange, a fact which has important implications for all phases.
The Mott insulators can in reality become Mott-JT insulators, whose local
configuration is low spin, with an electron-vibrational entanglement as
intrinsic as that of a superconductor, and possibly displaying other
unexplored features.
The metal phases obtained by doping the even-$n$ Mott insulators can
possess a pseudogap, namely a depression of the density of states at the
Fermi level, and a very reduced susceptibility, constituting a kind of
nonadiabatic semimetals.
The superconductor phases near the Mott insulators arise through pairing of
quasiparticles, these constituting a thin conducting web floating in a
Mott-insulating sea.
These model superconducting phases, although ultimately caused by JT effect
(electron-phonon) and s-wave in character, are shown to share many features
with those of high-$T_c$ cuprates. This possibility calls for further
experimental and theoretical work.

Further work should also be devoted to investigate the detailed nature of
the Mott phase transition.
The alternative possibilities of first-order and second-order transition,
should be distinguised in the $n=3$ and $n=4$ classes of fullerides.
For $n=3$ compounds, the simultaneous disappearence of supercondtucting
order and appearence of magnetic order seem theoretically incompatible with
a continuous transition.

To investigate possible spin/orbital antiferro correlations of the kind
observed in NH$_3$\,K$_3$C$_{60}$ \cite{margadonna}, one should study a
detailed three-dimensional model, including the highly directional hopping
matrix elements between different C$_{60}$ sites (and possibly merohedral
disorder).
Some of these aspects could be addressed by standard mean-field techniques
(of the type applied in a different context \cite{Vernay04}), or quantum
Monte Carlo cluster calculations which unfortunately suffer by significant
finite-size effects.
The DMFT method, based on an infinite lattice is free from these
limitations, but the same aspects are inaccessible to a single-impurity
DMFT model: cluster DMFT methods \cite{Parcollet04,Capone04CDMFT} could
lead to some progress in this direction.

\rem{

Proposal: variation of U with pressure inducing transition.???

fullerides.  A branch of thought focus on the solid-state nature of the
phonons \cite{Chen}, while another line of thought focuses on the molecular

}

\section*{Acknowledgments}
We are indebted to V.\ Brouet, A.\ Goldoni, O.\ Gunnarsson, M.\ Fabrizio,
R.\ Macovez, P.\ Rudolf and G.\ Santoro for useful discussions.
This work was funded in part by the EU's Sixth Framework Programme through
the Nanoquanta Network of Excellence (NMP4-CT-2004-500198).
This work was partly supported by MIUR COFIN2003, MIUR COFIN2004, by FIRB
RBAU017S8R operated by INFM, by MIUR FIRB RBAU017S8 R004, and by CNR-INFM
(Iniziativa trasversale calcolo parallelo).

\bibliographystyle{unsrt}
\bibliography{biblio}

\begin{thebibliography}{100}

\bibitem{Erwin}
{ S.\ C.\ Erwin {\it Buckminsterfullerenes}, edited by W.\ E.\ Billups and M.\
  A.\ Ciufolini (VCH Publishers, New York, 1993), p.\ 217}.

\bibitem{Satpathy92}
{ S.\ Satpathy, V.\ P.\ Antropov, O.\ K.\ Andersen, O.\ Jepsen, O.\ Gunnarsson,
  and A.\ I.\ Liechtenstein, Phys.\ Rev. B {\bf 46}, 1773 (1992)}.

\bibitem{Erwin94}
{ S.\ C.\ Erwin and C.\ Bruder, Physica B {\bf 199-200}, 600 (1994)}.

\bibitem{Benning}
{ P.\ J.\ Benning, F.\ Stepniak, and J.\ H.\ Weaver, Phys.\ Rev.\ B {\bf 48},
  9086 (1993)}.

\bibitem{Martin}
{ M.\ C.\ Martin, D.\ Koller, and L.\ Mihaly, Phys.\ Rev.\ B {\bf 47}, 14607
  (1993)}.

\bibitem{Rosseinsky93}
{ M.\ J.\ Rosseinsky, D.\ W.\ Murphy, R.\ M.\ Fleming, and O.\ Zhou, Nature
  (London) {\bf 364}, 425 (1993)}.

\bibitem{Iwasa95}
{ Y.\ Iwasa and T.\ Kaneyasu, Phys.\ Rev.\ B {\bf 51}, 3678 (1995)}.

\bibitem{Takenobu00}
{ T.\ Takenobu, T.\ Muro, Y.\ Iwasa, and T.\ Mitani, Phys.\ Rev.\ Lett.\ {\bf
  85}, 381 (2000)}.

\bibitem{Obu00}
{ T.\ T.\ Obu, H.\ Shimoda, Y.\ Iwasa, T.\ Mitani, M.\ Kosaka, K.\ U.\
  Tanigaki, C.\ M.\ Brown, and K.\ Prassides, Mol.\ Cryst. Liq.\ Cryst. Sci.\
  Technol., Sect.\ A {\bf 340}, 599 (2000)}.

\bibitem{Tou2}
{ H.\ Tou, Y.\ Maniwa, Y.\ Iwasa, H.\ Shimoda, and T.\ Mitani, Phys.\ Rev. B
  {\bf 62}, R775 (2000)}.

\bibitem{Fleming}
{ R.\ M.\ Fleming, M.\ J.\ Rosseinsky, A.\ P.\ Ramirez, D.\ W.\ Murphy, J.\ C.\
  Tully, R.\ C.\ Haddon, T.\ Siegrist, R.\ Tycko, S.\ H.\ Glarum, P.\ Marsh,
  G.\ Dabbagh, S.\ M.\ Zahurak, A.\ V.\ Makhija, and C.\ Hampton, Nature
  (London) {\bf 352}, 701 (1992)}.

\bibitem{Durand03}
{ P.\ Durand, G.\ R.\ Darling, Y.\ Dubitsky, A.\ Zaopo, and M.\ J.\ Rosseinsky,
  Nature Materials {\bf 2}, 605 (2003)}.

\bibitem{Ramirez}
{ A.\ P.\ Ramirez, Supercond.\ Rev.\ {\bf 1}, 1 (1994)}.

\bibitem{Gelfand}
{ M.\ P.\ Gelfand, Supercond.\ Rev.\ {\bf 1}, 103 (1994)}.

\bibitem{Reed00review}
{ C.\ A.\ Reed and R.\ D.\ Bolskar, Chem.\ Rev. {\bf 100}, 1075 (2000)}.

\bibitem{GunnarssonBook}
{ O.\ Gunnarsson, {\it Alkali-Doped Fullerides: Narrow-Band Solids with Unusual
  Properties} (World Scientific, Singapore, 2004)}.

\bibitem{Schoenherr}
{ E.\ Sch\"onherr, K.\ Matsumoto, and M.\ Wojnowski, J.\ Crystal Growth {\bf
  146}, 227 (1995)}.

\bibitem{David}
{ W.\ I.\ F.\ David, R.\ M.\ Ibberson, J.\ C.\ Matthewman, K.\ Prassides, T.\
  J.\ S.\ Dennis, J.\ P.\ Hare, H.\ W.\ Kroto, R.\ Taylor, and D.\ R.\ M.\
  Walton, Nature (London) {\bf 353}, 147 (1991); W.\ I.\ F.\ David, R.\ M.\
  Ibberson, T.\ J.\ S.\ Dennis, J.\ P.\ Hare, and K.\ Prassides, Europhys.\
  Lett.\ {\bf 18}, 219 (1992)}.

\bibitem{Haddon}
{ R.\ C.\ Haddon, L.\ E.\ Brus, and K.\ Raghavachari, Chem.\ Phys.\ Lett.\ {\bf
  125}, 459 (1986); R.\ C.\ Haddon, T.\ Siegrist, R.\ Tycko, S.\ H.\ Glarum,
  P.\ Marsh, G.\ Dabbagh, S.\ M.\ Zahurak, A.\ V.\ Makhija, and C.\ Hampton,
  Nature (London) {\bf 352}, 701 (1992)}.

\bibitem{Satpathy}
{ S.\ Satpathy, Chem.\ Phys.\ Lett.\ {\bf 130}, 545 (1986)}.

\bibitem{Wilson55}
{ E.\ Bright Wilson, J.\ C.\ Decius, and P.\ C.\ Cross, {\it Molecular
  Vibrations, The Theory of Infrared and Raman Vibrational Spectra}
  (McGraw-Hill, New York, 1955)}.

\bibitem{AMT}
{ A.\ Auerbach, N.\ Manini, and E.\ Tosatti, Phys.\ Rev.\ B {\bf 49}, 12998
  (1994)}.

\bibitem{MTA}
{ N.\ Manini, E.\ Tosatti, and A.\ Auerbach, Phys.\ Rev.\ B {\bf 49}, 13008
  (1994)}.

\bibitem{AssaPRL}
{ A.\ Auerbach, Phys.\ Rev.\ Lett.\ {\bf 70}, 1874 (1994)}.

\bibitem{Delos96}
{ P.\ De Los Rios, N.\ Manini, and E.\ Tosatti, Phys.\ Rev.\ B\ {\bf 54}, 7157
  (1996)}.

\bibitem{Paris97}
{ P.\ De Los Rios and N.\ Manini, {\it Recent Advances in the Chemistry and
  Physics of Fullerenes and Related Materials: Vol.~5}, edited by K.\ M.\
  Kadish and R.\ S.\ Ruoff (The Electrochemical Society, Pennington, NJ, 1997),
  p.\ 468}.

\bibitem{ManiniAErice}
{ N.\ Manini and P.\ De Los Rios, {\it Proceedings of the XIV International
  Symposium on Electron-Phonon Dynamics and Jahn-Teller Effect}, ed. by G.\
  Bevilacqua, L.\ Martinelli and N.\ Terzi (World Scientific, Singapore, 1999),
  p.\ 37}.

\bibitem{noberry}
{N.\ Manini and P.\ De Los Rios, J.\ Phys.: Condens.\ Matter {\bf 10}, 8485
  (1998)}.

\bibitem{hbyh}
{ N.\ Manini and P.\ De Los Rios, Phys.\ Rev. B {\bf 62}, 29 (2000)}.

\bibitem{Moate96}
{ C.\ P.\ Moate, M.\ C.\ M.\ O'Brien, J.\ L.\ Dunn, C.\ A.\ Bates, Y.\ M.\ Liu,
  and V.\ Z.\ Polinger, Phys.\ Rev.\ Lett.\ {\bf 77}, 4362 (1996)}.

\bibitem{Moate97}
{ C.\ P.\ Moate, J.\ L.\ Dunn, C.\ A.\ Bates, and Y.\ M.\ Liu, J.\ Phys.:
  Condens.\ Matter {\bf 9}, 6049 (1997)}.

\bibitem{Manini05}
{ N.\ Manini, Phys.\ Rev. A {\bf 71}, 032503 (2005)}.

\bibitem{Bethune91}
{ D.\ S.\ Bethune, G.\ Meijer, W.\ C.\ Tang, H.\ J.\ Rosen, W.\ G.\ Golden, H.\
  Seki, C.\ A.\ Brown, and M.\ S. de Vries, Chem.\ Phys. Lett.\ {\bf 179}, 181
  (1991)}.

\bibitem{Haymet}
{ A.\ D.\ J.\ Haymet, Chem.\ Phys.\ Lett.\ {\bf 122}, 421 (1985)}.

\bibitem{Negri}
{ F.\ Negri, G.\ Orlandi, and F.\ Zerbetto, Chem.\ Phys.\ Lett.\ {\bf 144}, 31
  (1988)}.

\bibitem{Green96}
{ W.\ H.\ Green Jr., S.\ M.\ Gorun, G.\ Fitzgerald, P.\ W.\ Fowler, A.\
  Ceulemans, and B.\ C.\ Titeca, J.\ Phys.\ Chem. {\bf 100} 14892 (1996)}.

\bibitem{Andreoni}
{ J.\ Kohanoff, W.\ Andreoni, and M.\ Parrinello, Phys.\ Rev.\ B {\bf 46}, 4371
  (1992); J.\ Kohanoff, W.\ Andreoni, and M.\ Parrinello, Chem.\ Phys.\ Lett.\
  {\bf 198}, 472 (1992); B.\ P.\ Feuston, W.\ Andreoni, M.\ Parrinello, and E.\
  Clementi, Phys.\ Rev.\ B {\bf 44}, 4056 (1991)}.

\bibitem{Savina}
{ M.\ R.\ Savina, L.\ L.\ Lohr, and A.\ H.\ Francis, Chem.\ Phys.\ Lett.\ {\bf
  205}, 200 (1993)}.

\bibitem{Tomita05}
{ S.\ Tomita, J.\ U.\ Andersen, E.\ Bonderup, P.\ Hvelplund, B.\ Liu, S.\
  Br{\o}ndsted Nielsen, U.\ V.\ Pedersen, J.\ Rangama, K.\ Hansen, and O.\
  Echt, Phys.\ Rev.\ Lett.\ {\bf 94}, 053002 (2005)}.

\bibitem{Yeretzian}
{ C.\ Yeretzian, K.\ Hansen, and R.\ L.\ Whetten, Science {\bf 260}, 652
  (1993)}.

\bibitem{Wang99}
{ X.\ B.\ Wang, C.\ F.\ Ding, and L.\ S.\ Wang, J.\ Chem.\ Phys. {\bf 110},
  8217 (1999)}.

\bibitem{Yang}
{ C.\ N.\ Yang, Rev.\ Mod. Phys.\ {\bf 34}, 694 (1962)}.

\bibitem{Limbach}
{ P.\ A.\ Limbach, L.\ Schweikhard, K.\ A.\ Cowen, M.\ T.\ McDermott, A.\ G.\
  Marshall, and J.\ V.\ Coe, J.\ Am.\ Chem.\ Soc.\ {\bf 113}, 6795 (1991)}.

\bibitem{Dubois}
{ D.\ Dubois, K.\ M.\ Kadish, S.\ Flanagan, R.\ R.\ Haufler, L.\ P.\ F.\
  Chibante, and L.\ J.\ Wilson, J.\ Am.\ Chem.\ Soc.\ {\bf 113}, 4364 (1991)}.

\bibitem{Heath}
{ G.\ A.\ Heath, J.\ E.\ McGrady, and R.\ L.\ Martin, J.\ Chem.\ Soc., Chem.\
  Commun.\ 1272 (1992)}.

\bibitem{Fullagar}
{ W.\ K.\ Fullagar, I.\ R.\ Gentle, G.\ A.\ Heath, and J.\ W.\ White, J.\
  Chem.\ Soc., Chem.\ Commun.\ 525 (1993)}.

\bibitem{Baumgarten93}
{ M.\ Baumgarten, A.\ G\"ugel, L.\ Gherghel, Adv.\ Mater. {\bf 5}, 458 (1993)}.

\bibitem{Burstein}
{ E.\ Burstein, S.\ C.\ Erwin, M.\ Y.\ Jiang, and R.\ P.\ Messmer, Physica
  Scripta {\bf T42}, 207 (1992)}.

\bibitem{Modesti}
{ S.\ Modesti, S.\ Cerasari, and P.\ Rudolf, Phys.\ Rev.\ Lett.\ {\bf 71}, 2469
  (1993)}.

\bibitem{Allemand}
{ P.\ M.\ Allemand, K.\ C.\ Kemani, A.\ Koch, F.\ Wudl, K.\ Holczer, S.\
  Donovan, G.\ Gr\"uner, and J.\ D.\ Thompson, Science {\bf 253}, 301 (1991);
  K.\ Tanaka, A.\ A.\ Zakhidov, K.\ Yoshizawa, K.\ Okahara, T.\ Yamabe, K.\
  Yakushi, K.\ Kikuchi, S.\ Suzuki, I.\ Ikemoto, and Y.\ Achiba, Phys.\ Lett.\
  A {\bf 164}, 221 (1992)}.

\bibitem{Denisov}
{ V.\ N.\ Denisov, A.\ A.\ Zakhidov, G.\ Ruani, R.\ Zamboni, C.\ Taliani, K.\
  Tanaka, K.\ Yoshizawa, T.\ Okahara, T.\ Yamabe, and Y.\ Achiba, Synth.\
  Metals {\bf 55-57}, 3050 (1993)}.

\bibitem{Hebard}
{ A.\ F.\ Hebard, M.\ J.\ Rosseinsky, R.\ C.\ Haddon, D.\ W.\ Murphy, S.\ H.\
  Glarum, T.\ T.\ M.\ Palstra, A.\ P.\ Ramirez, and A.\ R.\ Kortan, Nature
  (London) {\bf 350}, 600 (1991)}.

\bibitem{Kiefl}
{ R.\ F.\ Kiefl, T.\ L.\ Duty, J.\ W.\ Schneider, A.\ MacFarlane, K.\ Chow, J.\
  W.\ Elzey, P.\ Mendels, G.\ D.\ Morris, J.\ H.\ Brewer, E.\ J.\ Ansaldo, C.\
  Niedermayer, D.\ R.\ Noakes, C.\ E.\ Stronach, B.\ Hitti, and J.\ E.\
  Fischer, Phys.\ Rev.\ Lett.\ {\bf 69}, 2005 (1992)}.

\bibitem{Fisher}
{ J.\ E.\ Fischer and P.\ A.\ Heiney, J.\ Phys.\ Chem.\ Solids {\bf 54}, 1725
  (1993)}.

\bibitem{Chabre}
{ Y.\ Chabre, D.\ Djurado, M.\ Armand, W.\ R.\ Romanons, N.\ Coustel, J.\ P.\
  McCauley Jr, J.\ E.\ Fischer, and A.\ B.\ Smith III, J.\ Am.\ Chem.\ Soc.\
  {\bf 114}, 764 (1992)}.

\bibitem{koh}
{ J.\ Kohanoff, W.\ Andreoni, and M.\ Parrinello, Chem.\ Phys.\ Lett.\ {\bf
  198}, 472 (1992); J.\ Kohanoff, PhD Thesis No. 10079 ETH Zurich (1993); W.\
  Andreoni, {\it Electronic Properties of New Materials: Fullerenes},
  Proceedings of the 1993 Kirchberg Winter School, edited by H.\ Kuzmany, J.\
  Fink, M.\ Mehring, and S.\ Roth (Springer Verlag, Berlin, 1994)}.

\bibitem{Tomita01}
{ S.\ Tomita.\ J.\ U.\ Andersen, C.\ Gottrup.\ P.\ Hvelplund, and U.\ V.\
  Pedersen, Phys.\ Rev.\ Lett.\ {\bf 87}, 073401 (2001)}.

\bibitem{Reed00}
{ C.\ A.\ Reed, K.-C.\ Kim, R.\ D.\ Bolskar, and L.\ J.\ Mueller, Science {\bf
  289}, 201 (2000)}.

\bibitem{Bruno03}
{ C.\ Bruno, I.\ Doubitski, M.\ Marcaccio, F.\ Paolucci, D.\ Paolucci, and A.\
  Zaopo, J.\ Am.\ Chem. Soc.\ {\bf 125}, 15738 (2003)}.

\bibitem{Gasyna}
{ Z.\ Gasyna, L.\ Andrews, and P.\ N.\ Schatz, J.\ Phys.\ Chem.\ {\bf 96}, 1525
  (1992)}.

\bibitem{Datars95}
{ W.\ R.\ Datars and P.\ K.\ Ummat, Solid State Commun.\ {\bf 94}, 649 (1995)}.

\bibitem{Datars96}
{ W.\ R.\ Datars, J.\ D.\ Palidwar, and P.\ K.\ Ummat, J.\ Phys.\ Chem. Solids
  {\bf 57}, 977 (1996)}.

\bibitem{Panich02}
{ A.\ M.\ Panich, P.\ K.\ Ummat, and W.\ R.\ Datars, Solid State Commun.\ {\bf
  121}, 367 (2002)}.

\bibitem{Panich03}
{ A.\ M.\ Panich, H.-M.\ Vieth, P.\ K.\ Ummat, and W.\ R.\ Datars, Physica B
  {\bf 327}, 102 (2003)}.

\bibitem{CeulemansIII}
{ A.\ Ceulemans, P.\ W.\ Fowler, and I.\ Vos, J.\ Chem.\ Phys.\ {\bf 100}, 5491
  (1994)}.

\bibitem{Prassides}
{ K.\ Prassides, T.\ J.\ S.\ Dennis, J.\ P.\ Hare, J.\ Tomkinson, H.\ W.\
  Kroto, R.\ Taylor, and D.\ R.\ M.\ Walton, Chem.\ Phys.\ Lett.\ {\bf 187},
  455 (1991); K.\ Prassides, C.\ Christides, M.\ J.\ Rosseinsky, J.\ Tomkinson,
  D.\ W.\ Murphy, and R.\ C.\ Haddon, Europhys.\ Lett.\ {\bf 19}, 629 (1992);
  R.\ A.\ Jishi and M.\ S.\ Dresselhaus, Phys.\ Rev.\ B {\bf 45}, 2597 (1992);
  M.\ G.\ Mitch, S.\ J.\ Chase, and J.\ S.\ Lannin, Phys.\ Rev.\ B {\bf 46},
  3696 (1992)}.

\bibitem{Zhou}
{ P.\ Zhou, K.\ A.\ Wang, Y.\ Wang, P.\ C.\ Eklund, M.\ S.\ Dresselhaus, G.\
  Dresselhaus, and R.\ A.\ Jishi, Phys.\ Rev.\ B {\bf 46}, 2595 (1992)}.

\bibitem{Onida}
{ G.\ Onida and G.\ Benedek, Europhys.\ Lett.\ {\bf 18}, 403 (1992)}.

\bibitem{Jiang}
{ Q.\ Jiang, H.\ Xia, Z.\ Zhang, and D.\ Tian, Chem.\ Phys.\ Lett.\ {\bf 192},
  93 (1992)}.

\bibitem{Manini01}
{ N.\ Manini, A.\ Dal Corso, M.\ Fabrizio, and E.\ Tosatti, Philos.\ Mag. B
  {\bf 81}, 793 (2001)}.

\bibitem{Balamurugan04}
{ D.\ Balamurugan, M.\ K.\ Harbola, and R.\ Prasad, Phys.\ Rev. A {\bf 69},
  033201 (2004)}.

\bibitem{ChoB97}
{ C.\ C.\ Chancey and M.\ C.\ M.\ O'Brien, {\it The Jahn-Teller Effect in
  C$_{60}$ and Other Icosahedral Complexes} (Princeton Univ.\ Press, Princeton,
  1997)}.

\bibitem{butler81}
{ P.\ H.\ Butler, {\it Point Group Symmetry Applications} (Ple\-num, New York,
  1981)}.

\bibitem{ob69}
{ M.\ C.\ M.\ O'Brien, Phys.\ Rev.\ {\bf 187}, 407 (1969)}.

\bibitem{ob71}
{ M.\ C.\ M.\ O'Brien, J.\ Phys.\ C {\bf 4}, 2524 (1971)}.

\bibitem{Bersuker}
{ I.\ B.\ Bersuker and V.\ Z.\ Polinger, {\it Vibronic Interactions in
  Molecules and Crystals} (Springer-Verlag, Berlin, 1989)}.

\bibitem{Koga}
{ N.\ Koga and K.\ Morokuma, Chem.\ Phys.\ Lett.\ {\bf 196}, 191 (1992)}.

\bibitem{ManiniPhD}
{ N.\ Manini, {\it Electron-Vibron Coupling in Charged Fullerene, Berry Phase,
  and Superconductivity}, PhD Thesis,
  (\verb;http://www.sissa.it/;\-\verb;cm/thesis/1995/manini.ps.gz; SISSA,
  Trieste, 1995)}.

\bibitem{Yabana}
{ K.\ Yabana and G.\ Bertsch, Phys.\ Rev.\ B {\bf 46}, 14263 (1992)}.

\bibitem{Berry}
{ M.\ V.\ Berry, Proc.\ R.\ Soc.\ London A {\bf 392}, 45 (1984)}.

\bibitem{Levi}
{ B.\ Goss Levi, Phys.\ Today {\bf 46}, 17 (1993)}.

\bibitem{Ihm}
{ J.\ Ihm, Phys.\ Rev.\ B {\bf 49}, 10726 (1994)}.

\bibitem{Wilczek}
{ {\it Geometric Phases in Physics}, edited by A.\ Shapere and F.\ Wilczek
  (World Scientific, Singapore, 1989)}.

\bibitem{lh}
{ H.\ C.\ Longuet-Higgins, Adv.\ Spect.\ {\bf 2} 429, (1961), and references
  therein; G.\ Herzberg and H.\ C.\ Longuet-Higgins, Discuss.\ Faraday Soc.\
  {\bf 35}, 77 (1963); H.\ C.\ Longuet-Higgins, Proc.\ Roy.\ Soc.\ London A
  {\bf 344}, 147 (1975); C.\ A.\ Mead and D.\ G.\ Truhlar, J.\ Chem.\ Phys.\
  {\bf 70}, 2284 (1979)}.

\bibitem{Mead}
{ C.\ A.\ Mead, Rev.\ Mod.\ Phys.\ {\bf 64}, 51 (1992)}.

\bibitem{Wolf}
{ J.\ P.\ Wolf, G.\ Delacr\'etaz, and L.\ W\"oste, Phys.\ Rev.\ Lett.\ {\bf
  63}, 1946 (1989); J.\ Blanc, M.\ Broyer, J.\ Chevaleyre, P.\ Dugourd, H.\
  K\"uhling, P.\ Labastie, M.\ Ulbricht, J.\ P.\ Wolf, and L.\ W\"oste, Z.\
  Phys.\ D {\bf 19},7 (1991); P.\ Dugourd, J.\ Chevaleyre, R.\ Antoine, M.\
  Broyer, J.\ P.\ Wolf, and L.\ W\"oste, Chem.\ Phys.\ Lett.\ {\bf 225}, 28
  (1994)}.

\bibitem{Mahan}
{ G.\ D.\ Mahan, {\it Many-Particles Physics} (Ple\-num, New York, 1981)}.

\bibitem{Schl}
{ M.\ Schl\"uter, M.\ Lannoo, M.\ Needels, G.\ A.\ Baraff, and D.\ Tom\'anek,
  Phys.\ Rev.\ Lett.\ {\bf 68}, 526 (1991); J.\ Phys.\ Chem.\ Solids {\bf 53},
  1473 (1992)}.

\bibitem{vzr}
{ C.\ M.\ Varma, J.\ Zaanen, and K.\ Raghavachari, Science {\bf 254}, 989
  (1991)}.

\bibitem{Antropov}
{ V.\ P.\ Antropov, O.\ Gunnarsson, and A.\ I.\ Lichtenstein, Phys.\ Rev.\ B
  {\bf 48}, 7651 (1993)}.

\bibitem{Faulhaber}
{ J.\ C.\ R.\ Faulhaber, D.\ Y.\ K.\ Ko, and P.\ R.\ Briddon, Phys.\ Rev.\ B
  {\bf 48}, 661 (1993)}.

\bibitem{Winter96}
{ J.\ Winter and H.\ Kuzmany, Phys.\ Rev.\ B {\bf 53}, 655 (1996)}.

\bibitem{Gunnarsson}
{ O.\ Gunnarsson, H.\ Handschuh, P.\ S.\ Bechthold, B.\ Kessler, G.\
  Gantef\"or, and W.\ Eberhardt, Phys.\ Rev.\ Lett.\ {\bf 74}, 1875 (1995); O.\
  Gunnarsson, Phys.\ Rev.\ B {\bf 51}, 3493 (1995)}.

\bibitem{ManiniComm03}
{ N.\ Manini and E.\ Tosatti, Phys.\ Rev. Lett.\ {\bf 90}, 249601 (2003)}.

\bibitem{Manini03}
{ N.\ Manini, P.\ Gattari, and E.\ Tosatti, Phys.\ Rev. Lett.\ {\bf 91}, 196402
  (2003)}.

\bibitem{Gattari03}
{ P.\ Gattari, diploma thesis, University Milan (2003),
  \verb;http://;\-\verb;www.mi.infm.it/;\-\verb;manini/theses/gattari.pdf;\,}.

\bibitem{Bruhwiler97}
{ P.\ Br\"uhwiler, A.\ J.\ Maxwell, P.\ Balzer, S.\ Andersson, D.\ Arvanitis,
  L.\ Karlsson, and N.\ M\aa rtensson, Chem.\ Phys. Lett.\ {\bf 279}, 85
  (1997)}.

\bibitem{Canton02}
{ S.\ E.\ Canton, A.\ J.\ Yencha, E.\ Kukk, J.\ D.\ Bozek, M.\ C.\ A.\ Lopes,
  G.\ Snell, and N.\ Berrah, Phys.\ Rev.\ Lett.\ {\bf 89}, 045502 (2002)}.

\bibitem{Bordoni04}
{ A.\ Bordoni and N.\ Manini, {\it Fullerenes and Nanotubes - Materials for the
  New Chemical Frontier - Fullerenes - Vol.~14}, edited by P.\ V.\ Kamat, F.\
  D'Souza, D.\ M.\ Guldi, and S.\ Fukuzumi (The Electrochemical Society,
  Pennington, NJ, 2005), p.\ 118}.

\bibitem{Bergomi}
{ L.\ Bergomi and T.\ Jolicoeur, Comptes Rendus Acad.\ Sci.\ II {\bf 318}, 283
  (1994)}.

\bibitem{Langford99}
{ V.\ S.\ Langford and B.\ E.\ Williamson, J.\ Phys.\ Chem. A {\bf 103}, 6533
  (1999)}.

\bibitem{Gehring75}
{ G.\ A.\ Gehring and K.\ A.\ Gehring, Rep.\ Prog. Phys.\ {\bf 38}, 1 (1975)}.

\bibitem{Kaplan95}
{ M.\ D.\ Kaplan and B.\ G.\ Vekhter, {\it Cooperative Phenomena in Jahn-Teller
  crystals} (Plenum Press, New York, 1995)}.

\bibitem{Reno}
{ N.\ Manini and E.\ Tosatti, {\it Recent Advances in the Chemistry and Physics
  of Fullerenes and Related Materials: Vol.~2}, edited by K.\ M.\ Kadish and
  R.\ S.\ Ruoff (The Electrochemical Society, Pennington, NJ, 1995), p.\ 1017}.

\bibitem{Lueders03}
{ M.\ L\"uders, N.\ Manini, P.\ Gattari, and E.\ Tosatti, Eur.\ Phys. J.\ B
  {\bf 35}, 57 (2003)}.

\bibitem{Leuven02}
{ M.\ Lueders and N.\ Manini, Adv.\ Quantum Chem.\ {\bf 44}, 289 (2003)}.

\bibitem{Agerror:note}
{ The incorrect values of $g_k$ for the $A_g$ modes published in Table~3 of
  Ref.~\cite{Manini01} should be multiplied by a factor 2: in the present work
  we use the correct couplings}.

\bibitem{Ceulemans97}
{ A.\ Ceulemans, L.\ F.\ Chibotaru, and F.\ Cimpoesu Phys.\ Rev. Lett {\bf 78},
  3725 (1997)}.

\bibitem{Baskaran}
{ G.\ Baskaran and E.\ Tosatti, Current Science (Bangalore) {\bf 61}, 33
  (1991); S.\ Chakravarty, M.\ Gelfand, and S.\ Kivelson, Science {\bf 254},
  970 (1991); G.\ N.\ Murthy and A.\ Auerbach, Phys.\ Rev.\ B {\bf 46}, 331
  (1992)}.

\bibitem{Martin93}
{ R.\ L.\ Martin and J.\ P.\ Ritchie, Phys.\ Rev.\ B {\bf 48}, 4845 (1993)}.

\bibitem{Han00}
{ J.\ E.\ Han and O.\ Gunnarsson, Physica B {\bf 292}, 196 (2000)}.

\bibitem{Lueders02}
{ M.\ L\"uders, A.\ Bordoni, N.\ Manini, A.\ Dal Corso, M.\ Fabrizio, and E.\
  Tosatti, Philos.\ Mag. B {\bf 82}, 1611 (2002)}.

\bibitem{Nikolaev02}
{ A.\ V.\ Nikolaev and K.\ H.\ Michel, J.\ Chem.\ Phys. {\bf 117}, 4761
  (2002)}.

\bibitem{Chang91}
{ A.\ H.\ H.\ Chang, W.\ C.\ Ermler, and R.\ M.\ Pitzer, J.\ Phys.\ Chem.\ {\bf
  95}, 9288 (1991)}.

\bibitem{Cowan}
{ R.\ D.\ Cowan, {\it The Theory of Atomic Structure and Spectra} (Univ.\ of
  California Press, Berkeley-CA, 1981)}.

\bibitem{Antropov92}
{ V.\ P.\ Antropov, O.\ Gunnarsson, and O.\ Jepsen, Phys.\ Rev.\ B {\bf 46},
  13647 (1992)}.

\bibitem{Pedersen92}
{ M.\ R.\ Pedersen and A.\ A.\ Quong, Phys.\ Rev.\ B {\bf 46}, 13584 (1992)}.

\bibitem{deCoulon92}
{ V. de Coulon, J.\ L.\ Martins, and F.\ Reuse, Phys.\ Rev.\ B {\bf 45}, 13671
  (1992)}.

\bibitem{Hettic91}
{ R.\ L.\ Hettic, R.\ N.\ Compton, and R.\ H.\ Ritchie, Phys.\ Rev.\ Lett {\bf
  67}, 1242 (1991)}.

\bibitem{gosia03}
{ M.\ Wierzbowska, M.\ L{\"u}ders, and E.\ Tosatti, unpublished}.

\bibitem{Chakravarty92}
{ S.\ Chakravarty, S.\ Kivelson, M.\ I.\ Salkola and S.\ Tewari, Science {\bf
  256}, 1306 (1992); S.\ R.\ White, S.\ Chakravarty, M.\ Gelfand, and S.\
  Kivelson, Phys.\ Rev.\ B {\bf 45} 5062 (1992)}.

\bibitem{Murthy95}
{ N.\ Berdenis and G.\ Murthy, Phys.\ Rev. B {\bf 52}, 3083 (1995)}.

\bibitem{Lin05}
{ F.\ Lin, J. \v{S}makov, E.\ S.\ S\o{}rensen, C.\ Kallin, and A.\ J.\
  Berlinsky, Phys.\ Rev. B {\bf 71}, 165436 (2005)}.

\bibitem{Sookhun03bis}
{ S.\ Sookhun, J.\ L.\ Dunn, and C.\ A.\ Bates, Phys.\ Rev. B {\bf 68}, 235403
  (2003)}.

\bibitem{Dunn05}
{ J.\ L.\ Dunn and H.\ Li, Phys.\ Rev. B {\bf 71}, 115411 (2005)}.

\bibitem{Kerkoud}
{ R.\ Kerkoud, P.\ Auban-Senzier, D.\ Jerome, S.\ Brazovskii, I.\ Luk'yanchuk,
  N.\ Kirova, F.\ Rachdi, and C.\ Goze, J.\ Phys.\ Chem.\ Solids {\bf 57}, 143
  (1996)}.

\bibitem{ZimmerAll}
{ G.\ Zimmer, M.\ Helme, M.\ Mehring, and F.\ Rachdi, Europhys.\ Lett. {\bf
  27}, 543 (1994); G.\ Zimmer, M.\ Mehring, C.\ Goze, and F.\ Rachdi, Phys.\
  Rev. B {\bf 52}, 13300 (1995); G.\ Zimmer, M.\ Mehring, C.\ Goze, and F.\
  Rachdi, {\it Physics and Chemistry of Fullerenes and Derivatives}, edited by
  H.\ Kuzmany, J.\ Fink, M.\ Mehring, and S.\ Roth (World Scientific,
  Singapore, 1995), p.\ 452}.

\bibitem{Lukyanchuk95}
{ I.\ Lukyanchuk, N.\ Kirova, F.\ Rachdi, C.\ Goze, P.\ Molinie, and M.\
  Mehring, Phys.\ Rev. B {\bf 51}, 3978 (1995)}.

\bibitem{Bhyrappa93}
{ P.\ Bhyrappa, P.\ Paul, J.\ Stinchcombe, P.\ D.\ W.\ Boyd, and C.\ A.\ Reed,
  J.\ Am.\ Chem.\ Soc.\ {\bf 115}, 11004 (1993)}.

\bibitem{Bossard93}
{ C.\ Bossard, S.\ Rigaut, D.\ Astruc, M.-H.\ Delville, G.\ Felix, A.\
  Fevrier-Bouvier, J.\ Amiell, S.\ Flandrois, and P.\ Delhaes, J.\ Chem.\ Soc.,
  Chem.\ Commun. 333 (1993)}.

\bibitem{Chen95}
{ J.\ Chen, Z.-E.\ Huang, R.-F.\ Cai, Q.-F.\ Shao, and H.-J.\ Ye, Solid State
  Commun.\ {\bf 95}, 233 (1995)}.

\bibitem{Boyd}
{ P.\ D.\ W.\ Boyd, P.\ Bhyrappa, P.\ Paul, J.\ Stinchcombe, R.\ D.\ Bolskar,
  Y.\ Sun, and C.\ A.\ Reed, J.\ Am.\ Chem.\ Soc.\ {\bf 117}, 2907 (1995)}.

\bibitem{Trulove}
{ P.\ C.\ Trulove, R.\ T.\ Carlin, G.\ R.\ Eaton, and S.\ S.\ Eaton, J.\ Am.\
  Chem.\ Soc.\ {\bf 117}, 6265 (1995)}.

\bibitem{Sun97}
{ Y.\ Sun, PhD Dissertation, University of Southern California (Los Angeles,
  CA, 1997)}.

\bibitem{Brouet99}
{ V.\ Brouet, H.\ Alloul, F.\ Quere, G.\ Baumgartner, and L.\ Forro, Phys.\
  Rev. Lett.\ {\bf 82}, 2131 (1999)}.

\bibitem{Schilder94}
{ A.\ Schilder, H.\ Klos, I.\ Rystau, W.\ Sch\"utz, and B.\ Gotschy, Phys.\
  Rev. Lett.\ {\bf 73}, 1299 (1994)}.

\bibitem{Arovas95}
{ D.\ P.\ Arovas and A.\ Auerbach, Phys.\ Rev. B {\bf 52}, 10114 (1995)}.

\bibitem{Paul94}
{ P.\ Paul, Z.\ Xie, R.\ Bau, P.\ D.\ W.\ Boyd, and C.\ A.\ Reed, J.\ Am.\
  Chem. Soc.\ {\bf 116}, 4145 (1994)}.

\bibitem{Erwin91}
{ S.\ C.\ Erwin and M.\ R.\ Pederson, Phys.\ Rev.\ Lett.\ {\bf 67}, 1610
  (1991)}.

\bibitem{splitDMFT}
{ N.\ Manini, G.\ E.\ Santoro, A.\ Dal Corso, and E.\ Tosatti, Phys.\ Rev. B
  {\bf 66}, 115107 (2002)}.

\bibitem{Gunnarsson98}
{ O.\ Gunnarsson, S.\ C.\ Erwin, E.\ Koch, and R.\ M.\ Martin, Phys.\ Rev.\ B
  {\bf 57}, 2159 (1998)}.

\bibitem{Shirley93}
{ E.\ L.\ Shirley and S.\ G.\ Louie, Phys.\ Rev. Lett.\ {\bf 71}, 133 (1993)}.

\bibitem{WangBrouet03}
{ W.\ L.\ Wang, V.\ Brouet, X.\ J.\ Zhou, H.\ J.\ Choi, S.\ G.\ Louie, M.\ L.\
  Cohen, S.\ A.\ Kellar, P.\ V.\ Bogdanov, A.\ Lanzara, A.\ Goldoni, F.\
  Parmigiani, Z.\ Hussain, and Z.-X.\ Shen, Science {\bf 300}, 303 (2003)}.

\bibitem{GoldoniInSerenaBook}
{ A.\ Goldoni, contribution to present volume}.

\bibitem{DeLeo05}
{ L.\ De Leo and M.\ Fabrizio, Phys.\ Rev. Lett.\ {\bf 94}, 236401 (2005)}.

\bibitem{Capone04}
{ M.\ Capone, M.\ Fabrizio, C.\ Castellani, and E.\ Tosatti, Phys.\ Rev. Lett.\
  {\bf 86}, 5361 (2004)}.

\bibitem{FabrizioPriv}
{ M.\ Fabrizio, private communication}.

\bibitem{Lof}
{ R.\ W.\ Lof, M.\ A. van Veenendaal, B.\ Koopmans, H.\ T.\ Jonkman, and G.\
  A.\ Sawatzky, Phys.\ Rev.\ Lett.\ {\bf 68}, 3924 (1992)}.

\bibitem{Pederson92}
{ M.\ R.\ Pederson and A.\ A.\ Quong, Phys.\ Rev. B {\bf 46}, 13584 (1992)}.

\bibitem{Bruhwiler93}
{ P.\ A.\ Br\"uhwiler, A.\ J.\ Maxwell, A.\ Nilsson, N.\ M\aa rtensson, and O.\
  Gunnarsson, Phys.\ Rev. B {\bf 48}, 18296 (1993)}.

\bibitem{Gunnarsson97}
{ O.\ Gunnarsson, Rev.\ Mod. Phys.\ {\bf 69}, 575 (1997)}.

\bibitem{Rozenberg97}
{ M.\ J.\ Rozenberg, Phys.\ Rev. B {\bf 55}, R4855 (1997)}.

\bibitem{Noack99}
{ R.\ M.\ Noack and F.\ Gebhard, Phys.\ Rev. Lett.\ {\bf 82}, 1915 (1999)}.

\bibitem{Han98}
{ J.\ E.\ Han, M.\ Jarrell, and D.\ L.\ Cox, Phys.\ Rev. B {\bf 58}, R4119
  (1998)}.

\bibitem{Georges96}
{ A.\ Georges, G.\ Kotliar, W.\ Krauth, and M.\ J.\ Rozenberg, Rev.\ Mod.\
  Phys.\ {\bf 68}, 13 (1996)}.

\bibitem{Gunnarsson96}
{ O.\ Gunnarsson, E.\ Koch, and R.\ M.\ Martin, Phys.\ Rev. B {\bf 54}, R11026
  (1996)}.

\bibitem{Koch99}
{ E.\ Koch, O.\ Gunnarsson, and R.\ M.\ Martin, Phys.\ Rev. B {\bf 60}, 15714
  (1999)}.

\bibitem{Rozenberg94}
{ M.\ J.\ Rozenberg, G.\ Kotliar, and X.\ Y.\ Zhang, Phys.\ Rev. B {\bf 49},
  10181 (1994)}.

\bibitem{Caffarel94}
{ M.\ Caffarel and W.\ Krauth, Phys.\ Rev. Lett.\ {\bf 72}, 1545 (1994)}.

\bibitem{Zhou93}
{ O.\ Zhou, R.\ M.\ Fleming, D.\ W.\ Murphy, M.\ J.\ Rosseinsky, A.\ P.\
  Ramirez, R.\ B. van Dover, and R.\ C.\ Haddon, Nature (London) {\bf 362}, 433
  (1993)}.

\bibitem{Blinc96}
{ R.\ Blinc, K.\ Pokhodnia, P.\ Cevc, D.\ Ar\v{c}on, A.\ Omerzu, D.\
  Mihailovi\'c, P.\ Venturini, L.\ Goli\v{c}, Z.\ Trontelj, J.\ Lu\v{z}nik, Z.\
  Jegli\v{c}i\v{c}, and J.\ Pirnat, Phys.\ Rev. Lett.\ {\bf 76}, 523 (1996)}.

\bibitem{Blinc98}
{ D.\ Ar\v{c}on, P.\ Cevc, A.\ Omerzu, and R.\ Blinc, Phys.\ Rev. Lett.\ {\bf
  80}, 1529 (1998)}.

\bibitem{Brouet01}
{ V.\ Brouet, H.\ Alloul, T.\ N.\ Le, S.\ Garaj, and L.\ Forro, Phys.\ Rev.\
  Lett.\ {\bf 86}, 4680 (2001)}.

\bibitem{Robert98}
{ J.\ Robert, P.\ Petit, T.\ Yildirim, and J.\ E.\ Fischer, Phys.\ Rev. B {\bf
  57}, 1226 (1998)}.

\bibitem{Goldoni01}
{ A.\ Goldoni, L.\ Sangaletti, F.\ Parmigiani, G.\ Comelli, and G.\ Paolucci,
  Phys.\ Rev. Lett.\ {\bf 87}, 076401 (2001)}.

\bibitem{Schiessling05}
{ J.\ Schiessling, L.\ Kjeldgaard, T.\ K\"a\"ambre, I.\ Marenne, J.\ N.\
  O'Shea, J.\ Schnadt, C.\ J.\ Glover, M.\ Nagasono, D.\ Nordlund, M.\ G.\
  Garnier, L.\ Qian, J.-E.\ Rubensson, P.\ Rudolf, N.\ M\aa rtensson, J.\
  Nordgren, and P.\ A.\ Br\"uhwiler, Phys.\ Rev. B {\bf 71}, 165420 (2005)}.

\bibitem{Prassides99}
{ K.\ Prassides, S.\ Margadonna, D.\ Ar\v{c}on, A.\ Lappas, H.\ Shimoda, and
  Y.\ Iwasa, J.\ Am.\ Chem. Soc.\ {\bf 121}, 11227 (1999)}.

\bibitem{Tou1}
{ H.\ Tou, N.\ Muroga, Y.\ Maniwa, H.\ Shimoda, Y.\ Iwasa, and T.\ Mitani,
  Physica B {\bf 281}, 1018 (2000)}.

\bibitem{Zhou95}
{ O.\ Zhou, T.\ T.\ M.\ Palstra, Y.\ Iwasa, R.\ M.\ Fleming, A.\ F.\ Hebard,
  P.\ E.\ Sulewski, D.\ W.\ Murphy, and B.\ R.\ Zegarski, Phys.\ Rev. B {\bf
  52}, 483 (1995)}.

\bibitem{Margadonna01}
{ S.\ Margadonna, K.\ Prassides, H.\ Shimoda, Y.\ Iwasa, and M.\ M\'ezouar,
  Europhys.\ Lett.\ {\bf 56}, 61 (2001)}.

\bibitem{Mazin92}
{ I.\ I.\ Mazin, S.\ N.\ Rashkeev, V.\ P.\ Antropov, O.\ Jepsen, A.\ I.\
  Liechtenstein, and O.\ K.\ Andersen Phys.\ Rev. B {\bf 45}, 5114 (1992)}.

\bibitem{lannoo}
{ M.\ Lannoo, G.\ A.\ Baraff, and M.\ Schl\"uter, Phys.\ Rev.\ B {\bf 44},
  12106 (1991)}.

\bibitem{Deaven93}
{ D.\ M.\ Deaven and D.\ S.\ Rokhsar, Phys.\ Rev. B {\bf 48}, 4114 (1993)}.

\bibitem{Rice94}
{ H.\ Y.\ Choi and M.\ J.\ Rice, Phys.\ Rev.\ B {\bf 49}, 7048 (1994)}.

\bibitem{Gunnarsson92}
{ O.\ Gunnarsson and G.\ Zwicknagl, Phys.\ Rev.\ Lett.\ {\bf 69}, 957 (1992)}.

\bibitem{Pietronero92}
{ L.\ Pietronero, Europhys.\ Lett.\ {\bf 17}, 365 (1992)}.

\bibitem{Pietronero}
{ L.\ Pietronero, S.\ Strassler, and C.\ Grimaldi, Physica B {\bf 204}, 222
  (1995)}.

\bibitem{Cappelluti00}
{ E.\ Cappelluti, C.\ Grimaldi, L.\ Pietronero, and S.\ Str\"assler, Phys.\
  Rev. Lett.\ {\bf 85}, 4771 (2000)}.

\bibitem{Cappelluti01}
{ E.\ Cappelluti, C.\ Grimaldi, and L.\ Pietronero, Phys.\ Rev. B {\bf 64},
  125104 (2001)}.

\bibitem{Botti02}
{ M.\ Botti, E.\ Cappelluti, C.\ Grimaldi, and L.\ Pietronero, Phys.\ Rev. B
  {\bf 66}, 054532 (2002)}.

\bibitem{Paci04}
{ P.\ Paci, E.\ Cappelluti, C.\ Grimaldi, L.\ Pietronero, and S.\ Str\"assler,
  Phys.\ Rev. B {\bf 69}, 024507 (2004)}.

\bibitem{Lichtenstein01}
{ A.\ I.\ Lichtenstein, M.\ I.\ Katsnelson, and G.\ Kotliar, Phys.\ Rev. Lett.\
  {\bf 87}, 067205 (2001)}.

\bibitem{Parcollet04}
{ O.\ Parcollet, G.\ Biroli, and G.\ Kotliar, Phys.\ Rev. Lett.\ {\bf 92},
  226402 (2004)}.

\bibitem{Capone04CDMFT}
{ M.\ Capone, M.\ Civelli, S.\ S.\ Kancharla, C.\ Castellani, and G.\ Kotliar,
  Phys.\ Rev. B {\bf 69}, 195105 (2004)}.

\bibitem{Han99}
{ J.\ E.\ Han, O.\ Gunnarsson, and V.\ Eyert, Phys.\ Rev. B {\bf 60}, 6495
  (1999)}.

\bibitem{Capone01}
{ M.\ Capone, M.\ Fabrizio, and E.\ Tosatti, Phys.\ Rev. Lett.\ {\bf 86}, 5361
  (2001)}.

\bibitem{Lifshitz60}
{ I.\ M.\ Lifshitz, Sov.\ Phys.\ JETP {\bf 11}, 1130 (1960)}.

\bibitem{Katsnelson00}
{ M.\ I.\ Katsnelson and A.\ V.\ Trefilov, Phys.\ Rev.\ B {\bf 61}, 1643
  (2000)}.

\bibitem{Bulla99}
{ R.\ Bulla, Phys.\ Rev. Lett.\ {\bf 83}, 136 (1999)}.

\bibitem{Capone00}
{ M.\ Capone, M.\ Fabrizio, P.\ Giannozzi, and E.\ Tosatti, Phys.\ Rev.\ B {\bf
  62}, 7619 (2000)}.

\bibitem{Capone02}
{ M.\ Capone, M.\ Fabrizio, C.\ Castellani, and E.\ Tosatti, Science {\bf 296},
  2364 (2002)}.

\bibitem{Sutherland75}
{ B.\ Sutherland, Phys.\ Rev. B {\bf 12}, 3795 (1975)}.

\bibitem{Zhang98}
{ Y.\ Q.\ Li, M.\ Ma, D.\ N.\ Shi, and F.\ C.\ Zhang, Phys.\ Rev. Lett.\ {\bf
  81}, 3527 (1998)}.

\bibitem{margadonna}
{ S.\ Margadonna, K.\ Prassides, H.\ Shimoda, T.\ Takenobu, and Y.\ Iwasa,
  Phys.\ Rev.\ B {\bf 64}, 132414 (2001)}.

\bibitem{Kitano02}
{ H.\ Kitano, R.\ Matsuo, K.\ Miwa, A.\ Maeda, T.\ Takenobu, Y.\ Iwasa, and T.\
  Mitani, Phys.\ Rev. Lett.\ {\bf 88}, 096401 (2002)}.

\bibitem{Koch02}
{ E.\ Koch, Phys.\ Rev.\ B {\bf 66}, 081401 (2002)}.

\bibitem{Ricco00}
{ M.\ Ricc\`o, R. de Renzi, and A.\ Sartori, Appl.\ Magn. Resonance {\bf 19},
  517 (2000)}.

\bibitem{Shimoda96}
{ H.\ Shimoda, Y.\ Iwasa, and Y.\ Miyamoto Y.\ Maniwa, and T.\ Mitani, Phys.\
  Rev. B {\bf 54}, R15653 (1996)}.

\bibitem{Ricco01}
{ M.\ Ricc\`o, T.\ Shiroka, A.\ Sartori, F.\ Bolzoni, and M.\ Tomaselli,
  Europhys.\ Lett. {\bf 53}, 762 (2001)}.

\bibitem{Ricco03}
{ M.\ Ricc\`o, T.\ Shiroka, E.\ Zannoni, F.\ Barbieri, C.\ Bucci, and F.\
  Bolzoni, Phys.\ Rev. B {\bf 67}, 024519 (2003)}.

\bibitem{Palstra95}
{ T.\ T.\ M.\ Palstra, O.\ Zhou, Y.\ Iwasa, P.\ E.\ Sulewski, R.\ M.\ Fleming,
  and B.\ R.\ Zegarski, Solid State Commun.\ {\bf 93}, 327 (1995)}.

\bibitem{Mori02}
{ H.\ Mori, M.\ Kamiya, M.\ Haemori, H.\ Suzuki, S.\ Tanaka, Y.\ Nishio, K.\
  Kajita, and H.\ Moriyama, J.\ Am.\ Chem. Soc.\ {\bf 124}, 1251 (2002)}.

\bibitem{Craciun04}
{ M.\ F.\ Craciun, S.\ Rogge, M.\ J.\ L. den Boer, S.\ Margadonna, K.\
  Prassides, Y.\ Iwasa, and A.\ F.\ Morpurgo, Adv.\ Mater. {\bf 18}, 320
  (2006)}.

\bibitem{Craciun05}
{ R.\ W.\ I. de Boer, A.\ F.\ Stassen, M.\ F.\ Craciun, C.\ L.\ Mulder, A.\
  Molinari, S.\ Rogge, and A.\ F.\ Morpurgo, Appl.\ Phys. Lett.\ {\bf 86},
  262109 (2005)}.

\bibitem{Tosatti04}
{ E.\ Tosatti, M.\ Fabrizio, J.\ Tobik, and G.\ E.\ Santoro, Phys.\ Rev. Lett.\
  {\bf 93}, 117002 (2004)}.

\bibitem{riccoPrivate}
{ We are grateful to M.\ Ricc\`o for mentioning to us this possibility}.

\bibitem{Han03}
{ J.\ E.\ Han, O.\ Gunnarsson, and V.\ H.\ Crespi, Phys.\ Rev. Lett.\ {\bf 90},
  167006 (2004)}.

\bibitem{HanKoch00}
{ J.\ E.\ Han, E.\ Koch, and O.\ Gunnarsson, Phys.\ Rev. Lett.\ {\bf 84}, 1276
  (2000)}.

\bibitem{Granath03}
{ M.\ Granath and S. \"Ostlund, Phys.\ Rev. B {\bf 68}, 205107 (2003)}.

\bibitem{Dunn04}
{ J.\ L.\ Dunn, Phys.\ Rev. B {\bf 69}, 064303 (2004)}.

\bibitem{Fabrizio97}
{ M.\ Fabrizio and E.\ Tosatti, Phys.\ Rev.\ B {\bf 55}, 13465 (1997)}.

\bibitem{Martonak}
{ R.\ Marton\'ak and E.\ Tosatti, Phys.\ Rev. B {\bf 49}, 12596 (1994); {\it
  ibid.}\ {\bf 54}, 15714 (1996)}.

\bibitem{Kuntscher97}
{ C.\ A.\ Kuntscher, G.\ M.\ Bendele, and P.\ W.\ Stephens, Phys.\ Rev. B {\bf
  55}, R3366 (1997)}.

\bibitem{Nozieres85}
{ P.\ Nozi\`eres and S.\ Schmitt-Rink, J.\ Low Temp.\ Phys. {\bf 59}, 195
  (1985)}.

\bibitem{sademelo93}
{ C.\ A.\ R.\ Sa de Melo, M.\ Randeria, and J.\ R.\ Engelbrecht, Phys.\ Rev.
  Lett.\ {\bf 71}, 3202 (1993)}.

\bibitem{Han04}
{ J.\ E.\ Han, Phys.\ Rev. B {\bf 70}, 054513 (2004)}.

\bibitem{Buntar96}
{ V.\ Buntar, H.\ Weber, Supercond.\ Sci. Technol.\ {\bf 9}, 599 (1996)}.

\bibitem{Vernay04}
{ F.\ Vernay, K.\ Penc, P.\ Fazekas, and F.\ Mila, Phys.\ Rev. B {\bf 70},
  014428 (2004)}.

\end{thebibliography}


\printindex
\end{document}